\documentclass{elsart}

\usepackage{epsfig}

\usepackage{amssymb}

\begin{document}

\begin{frontmatter}



\title{Double-component convection due to different 
boundary conditions with broken reflection symmetry
for a component}


\author{N.~Tsitverblit\thanksref{*}}

\address{School of Mechanical Engineering, Tel-Aviv University,
Ramat-Aviv 69978, Israel}
\thanks[*]{Address for correspondence: 
1 Yanosh Korchak Street, apt. 6, Netanya 42495, Israel; 
e-mail: tsitver@gmail.com}

\begin{abstract}
Onset of two- (2D) and three-dimensional (3D) double-component convection due to different
boundary conditions is studied in a diversely oriented infinite slot with broken symmetry between
the slot conditions for a component. The main focus is on the two compensating background gradients.
Different component conditions at one slot boundary (the distinction boundary) are considered
with such a joint component condition at the other (the similarity boundary) as can be both of the
flux ($\eta=\chi=0$) and of the fixed-value ($\eta=\chi=1$) type. Also examined are such component
conditions at the second boundary (the inverse boundary) as differ from each other inversely to
the distinction boundary ($\eta=0$ and $\chi=1$). In the horizontal slot with inviscid fluid and
oscillatory primary instability for $\eta=\chi$, the most unstable wavelength being infinite at
$\eta=\chi=0$ is rendered finite by flux-component (solute) diffusion at the similarity boundary
when $\eta=\chi=1$.
In the viscous fluid, however, such a diffusion of both components enhances the instability
efficiency compared to $\eta=\chi=0$. With all above types of the broken symmetry, small-amplitude
convection in viscous fluid remains of an oscillatory nature for any slot orientation other than the
inversely-stratified horizontal one. For $\eta=0$ and $\chi=1$, such a universality also involves
various abrupt changes in the marginal-stability curves. These come from respectively identified
mechanisms of switching between dissimilar oscillatory patterns. In inviscid fluid, such changes
emerge with the zero instability threshold. Some of these abrupt changes give rise to new
mechanisms for three-dimensionality of the instability. Such a mechanism arises in viscous fluid for
$\eta=\chi=0$ as well. It comes with multiplicity and isolated existence of as well as hysteresis
between solutions of the linear stability equations. Both the hysteresis region and the other above
abrupt 3D changes are described in terms of analogy between the effect of $G=k_{y}/k$, the ratio of
the 2D and 3D wave numbers, and that of a 2D ratio between two gravity components. Other revealed
3D effects are attributable to new manifestations of their general mechanism identified by
\mbox{Tsitverblit [Ann. Phys. 322 (2007) 1727]}. Apart from dissipation, this mechanism also arises
from solute diffusion at the similarity boundary for $\eta=\chi=1$ and from differential gradient
diffusion at either boundary for $\eta=0$ and $\chi=1$. It also incorporates change of the nature
of instability from steady to oscillatory. In the context of the steady linear instability,
differential gradient diffusion is shown to be more effective at the stress-free slot boundary than
at the no-slip one. Although the mechanism of finite-amplitude steady convection revealed by
\mbox{Tsitverblit [Phys. Lett. A 329 (2004) 445]} is most effective herein for $\eta=\chi=1$,
its manifestation remains well-pronounced for $\eta=\chi<0.5$ as well. Relevance
of this mechanism to abrupt climate change is thus discussed.
\end{abstract}

\end{frontmatter}



\section{\label{s:i}Introduction}
This work addresses manifestations of a broken
symmetry in double-component, buoyancy-driven convection
resulting from the boundary conditions for one component
being different from those for the other. Such convection has
recently been identified as a fundamental class of pattern-forming
hydrodynamic instabilities. The objective of this study is to establish
the understanding of these instabilities for problems where a major
element of previously assumed reflection symmetry is absent: the
distinction between the components coming from one boundary
of the fluid domain is not reflected at the other.

The paradigm of double-component convection in pure fluid first arose in the context
of conventional double-diffusive convection. This is a class of phenomena resulting
from the effects of unequal diffusion coefficients of two density-affecting components
\cite{r:stnf,r:ver,r:stnl}. Among numerous natural science and technology applications
of this subject emphasized in its initial reviews \cite{r:tur}, particular attention has
subsequently been focused on small-scale oceanography \cite{r:scm}, ordinary evolution
of stars \cite{r:sphpc}, geology \cite{r:hs}, geodynamo \cite{r:brgb}, and crystal
growth \cite{r:crsk}. Recently, relevance of double-component convection has
also been highlighted for the dynamics of proto-neutron stars during
core-collapse supernova explosions \cite{r:bmk}, as well
as for colloidal suspensions \cite{r:mceg},
and soap films \cite{r:mw}.

Since the boundary conditions for one component are generically expected to be
different from those for the other, the effects of such different component conditions are 
relevant to all the above areas of application of conventional double-diffusive convection.
However, these effects also apply to an eddy-diffusion description of large-scale environmental
and turbulent processes, where disparity between the component diffusivities may be practically
negligible. Such processes range from \mbox{Langmuir} circulations \cite{r:l77,r:l83th04} to
the global ocean thermohaline circulation \cite{r:stm,r:w,r:r,r:wl,r:qgdm,r:kgmlhr,r:hr}
and associated climate change \cite{r:b,r:a,r:msrc}. In addition, convective
flows are commonly used in fundamental studies of transition to turbulence
\cite{r:br} and nonlinear pattern formation \cite{r:chbk}.

One major subclass of double-component instabilities arising
from the effects of different boundary conditions comprises phenomena
whose nature is conceptually analogous to the classical double-diffusion
\cite{r:stnf,r:ver,r:stnl}. Differential diffusion caused by unequal
component gradients forming in perturbed state due to the different
boundary conditions (differential gradient diffusion) triggers
convection analogously to the effects of disparate diffusivities.
Generalizing the idea in \cite{r:wel}, such analogy has been
introduced in \cite{r:twh,r:tbc,r:tlh,r:trp}
and scrutinized in \cite{r:tap}.

In particular, the nature of an oscillatory instability
highlighted in \cite{r:wel} and analyzed in \cite{r:tap}
is analogous to that in the diffusive regime of the classical
double-diffusion \cite{r:stnf,r:ver}. The viscous problem with
the stratification inverse to that in \cite{r:wel} also gives
rise to a mechanism of steady convection \cite{r:twh,r:tbc} 
that is conceptually analogous to the finger instability in
conventional double-diffusive convection \cite{r:stnf}. This
mechanism generates \mbox{Langmuir} circulations in the presence
of a stable background density stratification \cite{r:l77,r:l83th04}.
For the component conditions being different only at one
boundary and the other boundary being infinitely distant,
the stratification considered in \cite{r:twh,r:tbc} has
been more recently treated in \cite{r:ing}.

For two horizontal component gradients arising in a laterally heated stably
stratified slot, the effect of different boundary conditions \cite{r:tlh}
is also analogous to the classical double-diffusion \cite{r:ths}. As in
conventional double-diffusive convection \cite{r:xqt}, in addition,
steady finite-amplitude instability is triggered from the state of
rest by different sidewall boundary conditions for two compensating
horizontal gradients of the components \cite{r:trp}. Arising without
the linear steady instability of the conduction state \cite{r:rd},
however, such a finite-amplitude manifestation of convection in
\cite{r:trp} also exposes an oscillatory linear instability
whose nature is underlain by differential gradient
diffusion \cite{r:tap}.

Effects of different boundary conditions also extend beyond the
instabilities being due to differential gradient diffusion. As reported
in \cite{r:tpla}, finite-amplitude steady convection arises well before
onset of the respective linear instability in the viscous version of the
problem in \cite{r:wel}. It is then generated by the feedback coming from nonlinear
\mbox{Rayleigh}---\mbox{Benard} convection, despite the stabilizing role
of differential gradient diffusion. Potential relevance of such a
mechanism to abruptly changing global environmental phenomena
\cite{r:stm,r:w,r:kgmlhr,r:hr,r:b,r:a,r:msrc} makes
its examination under more realistic conditions
particularly important.

In the above studies of slot double-component
convection due to different boundary conditions, the
condition for either component at one slot boundary has been
identical to the respective condition at the other. The considered problems
have thus been reflectionally symmetric across the slot. Being an important
initial simplification, the reflection symmetry is however unlikely to
be maintained in real-world applications of the effects of different
boundary conditions. This is also particularly relevant if the
component-dependent forces other than the buoyancy forces
are considered, as suggested in \cite{r:tap,r:tpla}.

Manifestation of physical laws in the absence of certain
symmetries underlying them could make both the laws themselves and their broken
symmetries hardly recognizable. This has been repeatedly illustrated in elementary
particle physics and cosmological theories of unification of fundamental forces
\cite{r:qw}. Another illustration is the irreversibility in statistical
mechanics \cite{r:leb} and its generalized (symmetric) interpretation
\cite{r:skhp}. In geophysics, the global ocean thermohaline circulation
also involves asymmetries \cite{r:r,r:wl,r:qgdm,r:wbfh}. Clarification
of the nature of such asymmetries is viewed as critical for
understanding past and predicting future major
changes of the Earth climate \cite{r:b}.

For classic hydrodynamic instabilities, one can refer to
the structure of multiple steady flows in the \mbox{Taylor}
experiment \cite{r:tay}, where translation invariance is broken
by end walls. Complex as it becomes when the cylinder aspect
ratio is increased \cite{r:m}, this structure is not expected
to transform into that in the translationally invariant problem
even when the aspect ratio tends to infinity \cite{r:bm}. In
addition, if oscillatory instability arises in a flow that
is both reflectionally and translationally symmetric, the
corresponding \mbox{Hopf} bifurcation would give rise
to two respectively symmetric oscillatory branches
\cite{r:gsck}. Elimination of one of the
symmetries from such a system could thus
have a major effect on the structure and
nature of its nonequilibrium flows.

The general objective of the present work has been
to provide a comprehensive insight into the effects of
different boundary conditions in a slot where the previously assumed
boundary conditions symmetry is broken. In a class of such problems, the
component conditions are different only at one slot boundary (hereafter,
the distinction boundary). These problems are addressed for such a joint
component condition at the other (hereafter, the similarity boundary)
as can range from the flux to the fixed-value type. Being referred
to as the inverse boundary, this other boundary is also considered
with such different component conditions as are prescribed
oppositely to the distinction boundary.

Among consequences of the broken symmetry is a universality
of oscillatory manifestation of the effects of different boundary
conditions in viscous fluid. For all above types of the boundary conditions,
the steady linear instability analogous to that in \cite{r:twh,r:tbc} transforms
into an oscillatory one for any slot deviation from the horizontal orientation.
Three-dimensionality and, eventually, an isolated nonlinearity, hysteresis, and
other abrupt changes in such oscillatory-linear-instability curves are underlain
by the respective horizontal-slot steady instabilities. Other new three-dimensional
effects come from abruptly emerging zero thresholds of the inviscid oscillatory instability.
Most pronouncedly manifested herein for the fixed-value similarity boundary, the
mechanism of finite-amplitude steady convection \cite{r:tpla} is
also relevant when the component condition at this boundary
is closer to the flux than to the fixed-value type. 
\section{\label{s:f}The problem formulation and solution procedures}
\subsection{\label{s:fg}The problem and governing equations}
\begin{figure}
\centerline{\psfig{file=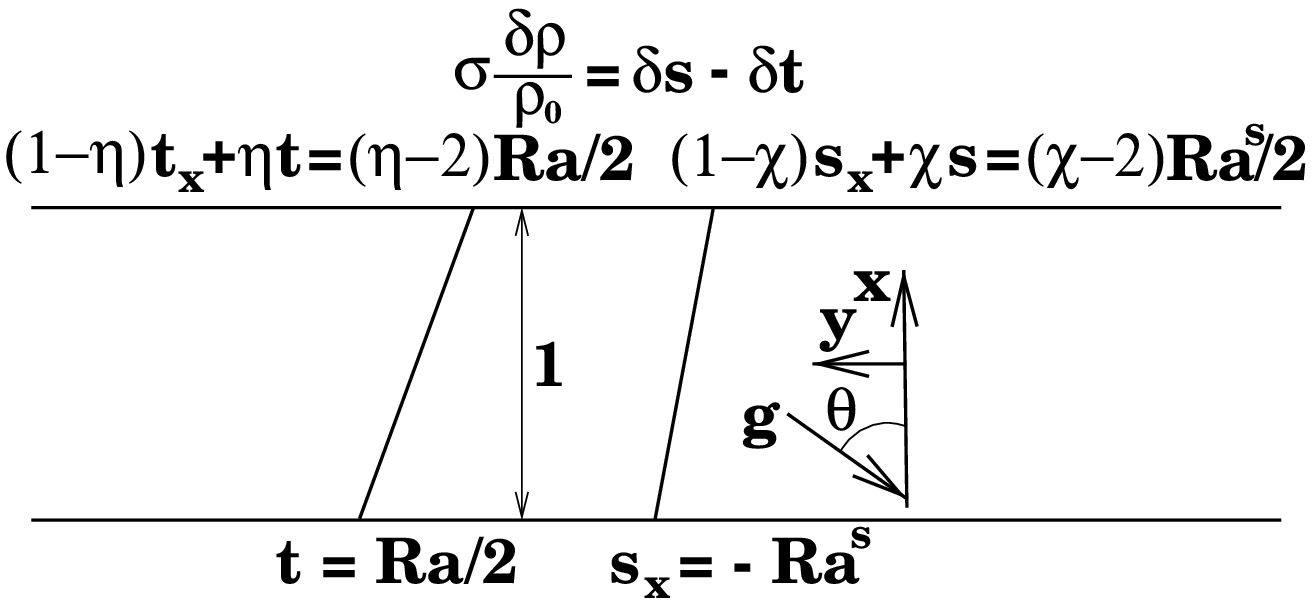,width=11.5cm}}
\vspace*{0.5cm}
\caption{Geometry of the problem. $\delta\rho=\rho-\rho_{0}$ 
is the variation of the (dimensionless) density, $\rho$, due to the variations $\delta s$
and $\delta t$ of solute concentration $s$ and temperature $t$ with respect to their reference
values, at which the density is $\rho_0$; $\sigma=gd^3/\kappa\nu$. $Pr\equiv\nu/\kappa=6.7$,
$Le\equiv\kappa_{T}/\kappa_{S}=1$; $\kappa_{T}$ and $\kappa_{S}$($=\kappa$) are the
component diffusivities. The fluid is of the Boussinesq type.}
\label{f:g}
\end{figure}

A general case of the considered problem is illustrated in
Fig. \ref{f:g}, where $\theta$ ($>0$ in Fig. \ref{f:g}) is the
angle between the direction opposite to the gravity and that of
the across-slot coordinate axis in an infinite slot with pure fluid.
The component gradients in Fig. \ref{f:g} are represented by the
\mbox{Rayleigh} numbers $Ra=g\alpha\Delta \overline{T}d^3/\kappa\nu$ and
$Ra^{s}=-g\beta(\partial\overline{s}/\partial\overline{x})d^4/\kappa\nu\equiv\mu Ra$.
Here, $\overline{x}$ is the (dimensional) across-slot coordinate, $d$ is the width of
the slot, $\Delta\overline{T}$ is the (dimensional) conduction-state difference between
the values of temperature (the component with the fixed-value condition at the distinction
boundary) at the boundaries with smaller and larger across-slot coordinates,
$\partial \overline{s}/\partial\overline{x}$ is the (dimensional) derivative
of solute concentration, the component with the flux condition, at the distinction
boundary, $\alpha$ is the coefficient of thermal expansion, $\beta$ is the coefficient
of the density variation due to the variation of solute concentration, $g$
is the gravitational acceleration, $\nu$ is the kinematic viscosity, and
$\kappa=\kappa_{T}=\kappa_{S}$ is the diffusivity of both components.
The bar means that the respective variable is dimensional. Unless
explicitly stated otherwise, $Ra>0$ and $Ra^{s}>0$ as well as
$\mu=1$ (i.e., the compensating background gradients)
are assumed.

As in \cite{r:twh,r:tbc,r:tlh,r:trp,r:tap,r:tpla}, the
component diffusivities are set equal to eliminate the classical
double-diffusive effects. Such an approach has also been adopted
in most studies of conventional double-diffusive convection, where 
the components with unequal diffusivities were not distinguished
from each other in terms of their boundary conditions. In principle,
equal diffusivities can also be experimentally modeled with two
solutes \cite{r:pmsns}. The \mbox{Prandtl} number, which would
then be significantly different from the present $Pr=6.7$, is
not expected to have a qualitative effect on the main results
and physical interpretations discussed herein. Equal diffusivities
could be interpreted as eddy transport coefficients as well,
as in \cite{r:l77,r:l83th04,r:qgdm}. $Pr=6.7$ is then
also within the range of realistic values for
all diffusion coefficients to be
of the eddy type.

For facilitating comparison of the results for an inclined or vertical
slot with those for a horizontal slot, this study is focused on the
exactly compensating background gradients ($\mu=1$). This eliminates
the along-slot motion arising when the slot orientation differs from
horizontal. In particular, transformation of the oscillatory instabilities
at $\theta\in[0,\pi)$ into the respective steady instabilities at
$\theta=\pi$ can thus be analyzed in the framework of the effects
of different boundary conditions alone. Such compensating gradients
have also been adopted in many studies of conventional
double-diffusive convection \cite{r:xqt,r:bgm}.

The equations describing the two-dimensional (2D) problem
in Fig. \ref{f:g} can be written as follows:
\begin{displaymath}
\frac{\partial\zeta}{\partial\tau}+
\frac{\partial\psi}{\partial x}\frac{\partial\zeta}
{\partial y}-\frac{\partial\psi}{\partial y}
\frac{\partial\zeta}{\partial x}=
\frac{1}{Pr}(\frac{\partial t}{\partial x}-
\frac{\partial s}{\partial x})\sin\theta-    
\end{displaymath}
\begin{equation}
\frac{1}{Pr}(\frac{\partial t}{\partial y}-
\frac{\partial s}{\partial y})\cos\theta+       
\frac{\partial^2\zeta}{\partial x^2}+
\frac{\partial^2\zeta}{\partial y^2},       \label{eq:ns1}
\end{equation}
\begin{equation}
\zeta=\frac{\partial^2\psi}{\partial x^2}+
\frac{\partial^2\psi}{\partial y^2},         \label{eq:ns2}
\end{equation}
\begin{equation}
\frac{\partial \xi_{i}}{\partial\tau}+\frac{\partial\psi}{\partial x}\frac{\partial \xi_{i}}
{\partial y}-\frac{\partial\psi}{\partial y}
\frac{\partial \xi_{i}}{\partial x}=
\frac{1}{Pr}(\frac{\partial^2 \xi_{i}}{\partial x^2}+
\frac{\partial^2 \xi_{i}}{\partial y^2}), \hspace{1cm} i=1,2. \label{eq:dxi}
\end{equation}
Here $\xi_{1}$ and $\xi_{2}$ stand for $t$ and $s$, the
across-slot, $u$, and along-slot, $v$, velocities are
$$u=-\frac{\partial\psi}{\partial y},\;\;\;\;\;\;\;
v=\frac{\partial\psi}{\partial x},$$
vorticity $$\zeta=\frac{\partial v}{\partial x}-
\frac{\partial u}{\partial y},$$
$Pr=\nu/\kappa$ is the \mbox{Prandtl} number, $\tau$ is
the time, $x\in(-1/2,1/2)$, $y\in(-\lambda/2,\lambda/2)$,
and $\lambda=\overline{\lambda}/d$ is the specified
along-slot period.

Eqs. (\ref{eq:ns1})---(\ref{eq:dxi}) are considered along
with wall boundary conditions for $\zeta$ and $\psi$
\begin{equation}
\zeta=\gamma_{\pm}\frac{\partial^{2}\psi}{\partial x^{2}},\;\;\;\;\;
\psi=0\;\;\;\;\; (x=\pm 1/2,\,-\lambda/2<y<\lambda/2), \label{bc:nsl}
\end{equation}
where $\gamma_{\pm}=1$ and $\gamma_{\pm}=0$ stand for the no-slip
and stress-free boundaries, respectively, as well as with
wall boundary conditions for $t$ and $s$
\begin{equation}
t=\frac{Ra}{2}\;\;\;\;\;\;\;\;\;\;\;\;\;\;\;\;\;\;\;\;\;\;\;\;\;\;\;\;\;\;\; 
(x=-1/2,\;-\lambda/2<y<\lambda/2),
\end{equation}
\begin{displaymath}
\frac{\partial s}{\partial x}=-\mu Ra=-Ra^{s}
\;\;\;\;\;(x=-1/2,\;\;-\lambda/2<y<\lambda/2,\;\;y\neq 0),
\end{displaymath}
\begin{equation}
(1-\chi_{0})(s-\frac{Ra^{s}}{2})+\chi_{0}(\frac{\partial s}{\partial x}+Ra^{s})=0
\;\;\;\;\;\;\;(x=-1/2,\;\;y=0),
\label{bc:l}
\end{equation}
\begin{equation}
(1-\eta)(\frac{\partial t}{\partial x}+Ra)+\eta(t+\frac{Ra}{2})=0\;\;\;\;
(x=1/2,\;\;-\lambda/2<y<\lambda/2),
\end{equation}
\begin{displaymath}
\hspace*{-0.9cm}(1-\chi)(\frac{\partial s}{\partial x}+Ra^{s})+\chi(s+\frac{Ra^{s}}{2})=0
\;\;(x=1/2,\;-\lambda/2<y<\lambda/2,\;y\neq 0),
\end{displaymath}
\begin{equation}
(1-\chi_{1})(\frac{\partial s}{\partial x}+\frac{Ra^{s}}{2})+\chi_{1}(s+\frac{Ra^{s}}{2})=0
\;\;\;\;\;\;(x=1/2,\;\;y=0)
\label{bc:r}
\end{equation}
and periodic boundary conditions in the along-slot direction
\begin{displaymath}
\xi(x,\lambda/2)=\xi(x,-\lambda/2),\;\;\;\;\;\;
\frac{\partial\xi(x,\lambda/2)}{\partial y}=
\frac{\partial \xi(x,-\lambda/2)}{\partial y}
\end{displaymath}
\begin{equation}
(-1/2<x<1/2). \label{bc:pbcxi}
\end{equation}
Here $\xi$ stands for $\zeta$, $\psi$, $t$, and $s$, $\chi_{0}=0$
for $\chi<1$ or in all simulations of the linear stability for $\theta=0$ 
and $\chi_{0}=1$ otherwise, $\chi_{1}=1$ for $\theta=\pi/2$ and
$\chi_{1}=\chi$ for $\theta=0$. In the present study, $Pr=6.7$
as well as either $\eta=\chi=0$ or $\eta=\chi=1$ or $\eta=0$
and $\chi=1$ unless $\eta=\chi\in(0,1)$ is specified
for $\theta=0$.

In (\ref{bc:l}) and (\ref{bc:r}), specification of the
middle values of $s$ identifies the solute scale and the solution
phase for $\eta=\chi<1$ at $\theta=0$, symmetrically fixes the solute
scale in the temporal simulations for $\eta=\chi=0$ at $\theta=\pi/2$,
and fixes the solution phase in such simulations for $\eta=\chi=1$ at
$\theta=0$. The phase of a nontrivial steady solution for $\theta=0$ when
$\chi=1$ and either $\eta=1$ or $\eta=0$ was selected by continuation in $\chi$
and $\eta$ from the progenitor of such a solution at $\eta=1$ and $\chi=0$ in
\cite{r:tap,r:tpla}, where the phase and the solute scale were fixed at the periodic
condition boundaries. The latter approach becomes inconsistent with the natural
solute scale specification in the nontrivial steady solutions for $\chi=1$.
It also leads to early spurious oscillatory instabilities of the
background state for $\eta=\chi=0$ at $\theta=0$. It was not
thus employed in the reported results.

In boundary conditions (\ref{bc:nsl})---(\ref{bc:pbcxi}), switching
the distinction and the other boundary designations in Fig. \ref{f:g}
is equivalent to transformation
\begin{equation}
(x,y,\theta,Ra,Ra^{s})^{T}\mapsto(-x,-y,\theta+\pi,-Ra,-Ra^{s})^{T}.
\label{eq:ibc}
\end{equation}
Since Eqs. (\ref{eq:ns1})---(\ref{eq:dxi}) are invariant under (\ref{eq:ibc}),
one can consider only such slot orientation to the gravity as is
depicted in Fig. \ref{f:g}. 

Discretized by central finite differences,
the steady version of Eqs. (\ref{eq:ns1})---(\ref{eq:dxi})
and boundary conditions (\ref{bc:nsl})---(\ref{bc:pbcxi}) was
treated with the \mbox{Euler---Newton} and \mbox{Keller} arclength
\cite{r:kel} continuation algorithms to trace out bifurcating branches
\cite{r:tlh}. These algorithms were based on the Harwell MA32 Fortran routine.
The along-slot period $\lambda=2$ was prescribed. The grid with $33$ nodes
in the across-slot direction was used in the computations [$nx\times\lambda(nx+1)$
with $nx=33$], as in \cite{r:twh,r:tbc,r:tlh,r:trp,r:tap,r:tpla}. As already indicated,
temporal behavior of the linearized version of Eqs. (\ref{eq:ns1})---(\ref{eq:dxi})
and boundary conditions (\ref{bc:nsl})---(\ref{bc:pbcxi}) was also examined near
the onset of oscillatory instability of the conduction state. Such an
examination was conducted with the implicit method and time
step $\delta\tau=0.05$.
\subsection{\label{s:fl}Linear stability calculations}
With the state of rest being the background flow for $\mu=1$, the
\mbox{Fourier} mode of a three-dimensional (3D) marginally unstable
oscillatory perturbation with angular frequency $\omega$ and a wave
number $k=(k_{y}^{2}+k_{z}^{2})^{1/2}$ having $y$ and $z$
components $k_{y}$ and $k_{z}$, respectively,
can be written as 
\begin{equation}
[u'(x),t'(x),s'(x)]^{T} 
e^{i(\omega\tau\pm k_{y}y+k_{z}z)}+cc. \label{eq:p3}
\end{equation}
(The $z$-axis is orthogonal to the $x$--$y$ plane in Fig. \ref{f:g}
and is directed towards the reader.) Here $[u'(x),t'(x),s'(x)]^{T}$
is the \mbox{Fourier}-mode part depending on the across-slot coordinate
alone, the prime near a flow variable denotes such part in the perturbation
of the variable. Expression (\ref{eq:p3}) has been introduced into
the linearized 3D governing equations that were nondimensionalized
consistently with (2D) Eqs. (\ref{eq:ns1})---(\ref{eq:dxi}) and
rewritten in terms of across-slot velocity $u$, 
temperature $t$, and solute concentration $s$.
This leads to:
\begin{displaymath}
(\frac{d^2}{dx^2}-k^2)(\frac{d^2}{dx^2}-k^2-i\omega)\tilde{u}=
\end{displaymath}
\begin{equation}
\mp Ra[i G k\frac{d}{dx}(\tilde{t}-\tilde{s})\sin\theta\pm
k^2(\tilde{t}-\tilde{s})\cos\theta], \label{eq:lsu}
\end{equation}
\begin{equation}
(\frac{d^2}{dx^2}-k^2-i\omega Pr)\tilde{\xi}_{i}=
\tilde{u}, \hspace{1cm} i=1,2, \label{eq:lsts}
\end{equation}
where $G\equiv k_{y}/k$, $\tilde{u}=-u'Pr$, $\tilde{\xi}_{1}\equiv\tilde{t}=t'/Ra$, 
and $\tilde{\xi}_{2}\equiv\tilde{s}=s'/Ra^{s}$ ($Ra^{s}=Ra$).

The variable $\tilde{u}$ is subject to boundary conditions
\begin{equation}
\tilde{u}=(1-\gamma_{\pm})\frac{d^{2}\tilde{u}}{dx^{2}}+\gamma_{\pm}\frac{d\tilde{u}}{dx}=0
\;\;\;\;\;\;\;\; (x=\pm 1/2), \label{bc:uv}
\end{equation}
where $\gamma_{+}=0,1$ and $\gamma_{-}=0,1$. Unless the values of $\gamma_{\pm}$
are explicitly given, the boundaries are assumed to be either both stress-free
($\gamma_{\pm}=0$) or both no-slip ($\gamma_{\pm}=1$). Such boundary
conditions for $\tilde{u}$ are used along with
\begin{equation}
\tilde{t}=\frac{d\tilde{s}}{dx}=0\;\;\;\; (x=-1/2) \label{bc:tsl}
\end{equation}
and
\begin{equation}
(1-\eta)\frac{d\tilde{t}}{d x}+\eta\tilde{t}=
(1-\chi)\frac{d\tilde{s}}{dx}+\chi\tilde{s}=0\;\;\;\; (x=1/2), \label{bc:tsr}
\end{equation}
where either $\eta=\chi=0$ or $\eta=\chi=1$ or $\eta=0$ and $\chi=1$.

Let $[\tilde{u}(x),\tilde{t}(x),\tilde{s}(x)]^{T}$ be a solution of the 
$e^{i(\omega\tau+k_{y}y+k_{z}z)}$-version of Eqs. (\ref{eq:lsu}) and (\ref{eq:lsts})
and boundary conditions (\ref{bc:uv})---(\ref{bc:tsr}) at some $Ra_{c}$ and $\omega_{c}$
for a value of $\theta$ and the orientation of $x$ and $y$ axes as in Fig. \ref{f:g}. If
$(x,y,\theta)^{T}$ is transformed into $(-x,-y,\theta+\pi)^{T}$ in boundary conditions
(\ref{bc:uv})---(\ref{bc:tsr}) (i.e., if the distinction and the other boundary designations
in Fig. \ref{f:g} are switched), $[\tilde{u}(x),\tilde{t}(x),\tilde{s}(x)]^{T}$ would then
be a solution of the $e^{i(\omega\tau-k_{y}y+k_{z}z)}$-version of Eqs. (\ref{eq:lsu})
and (\ref{eq:lsts}) at $-Ra_{c}$ and the same $\omega_{c}$. This symmetry is related
to invariance (\ref{eq:ibc}) of Eqs. (\ref{eq:ns1})---(\ref{eq:dxi}). It allows to
consider only one orientation of the slot boundaries to the gravity, provided
that both signs of $k_{y}$ in (\ref{eq:p3}) are taken into account.

For examination of the linear stability of the conduction state
in inviscid fluid,
\begin{equation}
i\omega(\frac{d^2}{dx^2}-k^2)\tilde{u}=
Ra[\pm i G k\frac{d}{dx}(\tilde{t}-\tilde{s})\sin\theta+
k^2(\tilde{t}-\tilde{s})\cos\theta] \label{eq:lsui}
\end{equation}
was used along with
\begin{equation}
(\frac{d^2}{dx^2}-k^2-i\omega)\tilde{\xi}_{i}=
\tilde{u}, \hspace{1cm} i=1,2, \label{eq:lstsi}
\end{equation}
where the \mbox{Rayleigh} numbers are defined as
$Ra=g\alpha\Delta \overline{T}d^3/\kappa^{2}$ and $Ra^{s}=
-g\beta(\partial\overline{s}/\partial\overline{x})d^4/\kappa^{2}=Ra$.
Here $\omega$ is nondimensionalized with $\kappa/d^2$ as opposed to
$\nu/d^2$ in Eqs.~(\ref{eq:lsu}) and (\ref{eq:lsts}). The definitions
of $Ra$, $Ra^s$, and $\omega$ for inviscid fluid are then different
from the respective definitions for viscous fluid. Such definitions
are thus used below according to the type of fluid in question.

The boundary conditions for inviscid fluid are
\begin{equation}
\tilde{u}=0\;\;\;\; (x=\pm 1/2) \label{bc:ui}
\end{equation}
along with (\ref{bc:tsl}) and (\ref{bc:tsr}).

For $\eta=0$ and $\chi=1$, Eqs. (\ref{eq:lsu}) and (\ref{eq:lsts}) along with
boundary conditions (\ref{bc:uv})---(\ref{bc:tsr}) for $\gamma_{-}=\gamma_{+}$
as well as Eqs. (\ref{eq:lsui}) and (\ref{eq:lstsi}) and boundary conditions
(\ref{bc:tsl}), (\ref{bc:tsr}), and (\ref{bc:ui})
are invariant under transformation
\begin{equation}
[\tilde{u}(x),\tilde{t}(x),\tilde{s}(x),\theta]^{T}\mapsto
[\tilde{u}(-x),\tilde{s}(-x),\tilde{t}(-x),\pi-\theta]^{T}.\label{eq:tt01}
\end{equation}
For $\eta=0$ and $\chi=1$, therefore, $\theta\in[0,\pi/2]$ and $\theta\in[\pi/2,\pi]$
are equivalent in the context of (\ref{eq:tt01}). For $Ra\neq Ra^{s}$ at $\theta=0$
and $\theta=\pi$ as well as for $\gamma_{-}\neq\gamma_{+}$ in viscous fluid, such
an equivalence also implies additional transformations. These are specified
when the respective results are discussed below.

The inviscid problem in a horizontal slot was also examined for the $Ra$ and $Ra^{s}$
being independent of each other. In particular,
\begin{equation}
i\omega(\frac{d^2}{dx^2}-k^2)\tilde{u}=
k^2(Ra\tilde{t}-Ra^{s}\tilde{s})\cos\theta \label{eq:lsuii}
\end{equation}
was considered along with Eqs.~(\ref{eq:lstsi}) and boundary conditions
(\ref{bc:tsl}), (\ref{bc:tsr}), and (\ref{bc:ui}). This was done at given
values of $Ra^{s}$ and $\theta=0$ for any of the pairs of $\eta$ and $\chi$
specified above. As already indicated, such a problem for $\eta=0$ and
$\chi=1$ is equivalent to that with the fixed values
of $Ra$ at $\theta=\pi$. 

For fixed $k_{y}$ and $k_{z}$, $Ra_{c}$ and $\omega_{c}$
are found by searching in the $Ra$--$\omega$ domain for the
smallest $Ra$ at which the complex matrix resulting from the
application of boundary conditions (\ref{bc:uv})---(\ref{bc:tsr})
to the general solution of Eqs. (\ref{eq:lsu}) and (\ref{eq:lsts})
is singular. The same procedure is applied to the general solution
either of Eqs. (\ref{eq:lsui}) and (\ref{eq:lstsi}) or of
Eqs. (\ref{eq:lstsi}) and (\ref{eq:lsuii}) for $\theta=0$
with boundary conditions (\ref{bc:tsl}), (\ref{bc:tsr}),
and (\ref{bc:ui}). NAG Fortran routines were employed
for this purpose. 

Once $Ra_{c}(k_{y}^{0})$ and $\omega_{c}(k_{y}^{0})$ have been found 
for a given $k_{y}^{0}$ and fixed $k_{z}$, the corresponding values of
these parameters, $Ra_{c}(k_{y})$ and $\omega_{c}(k_{y})$, at a nearby 
$k_{y}$ are computed by the \mbox{Euler}---\mbox{Newton} continuation
method. The latter is applied to the solution
of equation
\begin{equation}
F[Ra(k_{y}),\omega(k_{y}),k_{y},k_{z}]=0, \label{eq:sm}
\end{equation}
where $F[Ra(k_{y}),\omega(k_{y}),k_{y},k_{z}]$ stands for the (complex) determinant of the
matrix resulting from the application of boundary conditions (\ref{bc:uv})---(\ref{bc:tsr})
or (\ref{bc:tsl}), (\ref{bc:tsr}), and (\ref{bc:ui}) to the general solution of the
respective set of differential equations. Due to the use of standard Fortran routines,
the Jacobian of $\{Re[F(Ra,\omega,k_{y},k_{z})],Im[F(Ra,\omega,k_{y},k_{z})]\}^{T}$ (with
respect to $Ra$ and $\omega$) and $\partial{F(Ra,\omega,k_{y},k_{z})}/\partial{k_{y}}$
were computed with numerical differentiation. When 3D effects were anticipated,
they were studied by repeating the procedure just described
for different $k_{z}\geq 0$.

Since the linear instability to steady disturbances in viscous fluid
was found to arise only in a horizontal slot, the corresponding problem
is discussed in the 2D framework alone. For the compensating background
gradients, in particular, $\omega=0$ and $k_{y}=k$ are set in
Eqs. (\ref{eq:lsu}) and (\ref{eq:lsts}) to obtain
\begin{equation}
(\frac{d^2}{dx^2}-k^2)^{2}\tilde{u}=
-Ra[ik\frac{d}{dx}(\tilde{t}-\tilde{s})\sin\theta+
k^2(\tilde{t}-\tilde{s})\cos\theta], \label{eq:lssu}
\end{equation}
\begin{equation}
(\frac{d^2}{dx^2}-k^2)\tilde{\xi}_{i}=
\tilde{u}, \hspace{1cm} i=1,2.\label{eq:lssts}
\end{equation}
At $\theta=\pi$, the marginal-stability boundaries
described by Eqs. (\ref{eq:lssu}) and (\ref{eq:lssts}) and boundary conditions 
(\ref{bc:uv})---(\ref{bc:tsr}) were found for any of the above pairs of $\eta$ and $\chi$.
For $\eta=0$ and $\chi=1$, they were found at $\theta=0$ as well, in view of
(\ref{eq:tt01}). For $\theta=0$ and $\theta=\pi$, steady linear instability
was also examined for the $Ra$ and $Ra^{s}$
being independent of each other:
\begin{equation}
(\frac{d^2}{dx^2}-k^2)^{2}\tilde{u}=
-k^2(Ra\tilde{t}-Ra^{s}\tilde{s})\cos\theta \label{eq:lssuu}
\end{equation}
and Eqs. (\ref{eq:lssts}) were used either at $\theta=\pi$ for
any of the pairs of $\eta$ and $\chi$ specified above or at
$\theta=0$ for $\eta=0$ and $\chi=1$ alone.

The general solution either of Eqs. (\ref{eq:lssu}) and (\ref{eq:lssts})
or of Eqs. (\ref{eq:lssts}) and (\ref{eq:lssuu}) is obtained analytically.
Boundary conditions (\ref{bc:uv})---(\ref{bc:tsr}) are then applied to such
a general solution. In the former case, the smallest $Ra$, $Ra_{c}(k)$, at
which the resulting matrix becomes singular is searched for at different
$k$ (and found only for $\theta=\pi$, besides $\theta=0$ when $\eta=0$
and $\chi=1$). In the latter case, such a search yields either the
smallest $Ra^{s}$, $Ra^{s}_{c}(k)$, for a fixed $Ra$ when
$\theta=\pi$ or the smallest $Ra$, $Ra_{c}(k)$,
for a fixed $Ra^{s}$ when $\theta=0$ at
$\eta=0$ and $\chi=1$.
\section{\label{s:roi}Inviscid fluid}
\subsection{\label{s:roih}$\theta=0$}
\subsubsection{\label{s:roih00}$\eta=\chi=0$}
\begin{figure}
\vspace*{-0.5cm}
\centerline{\psfig{file=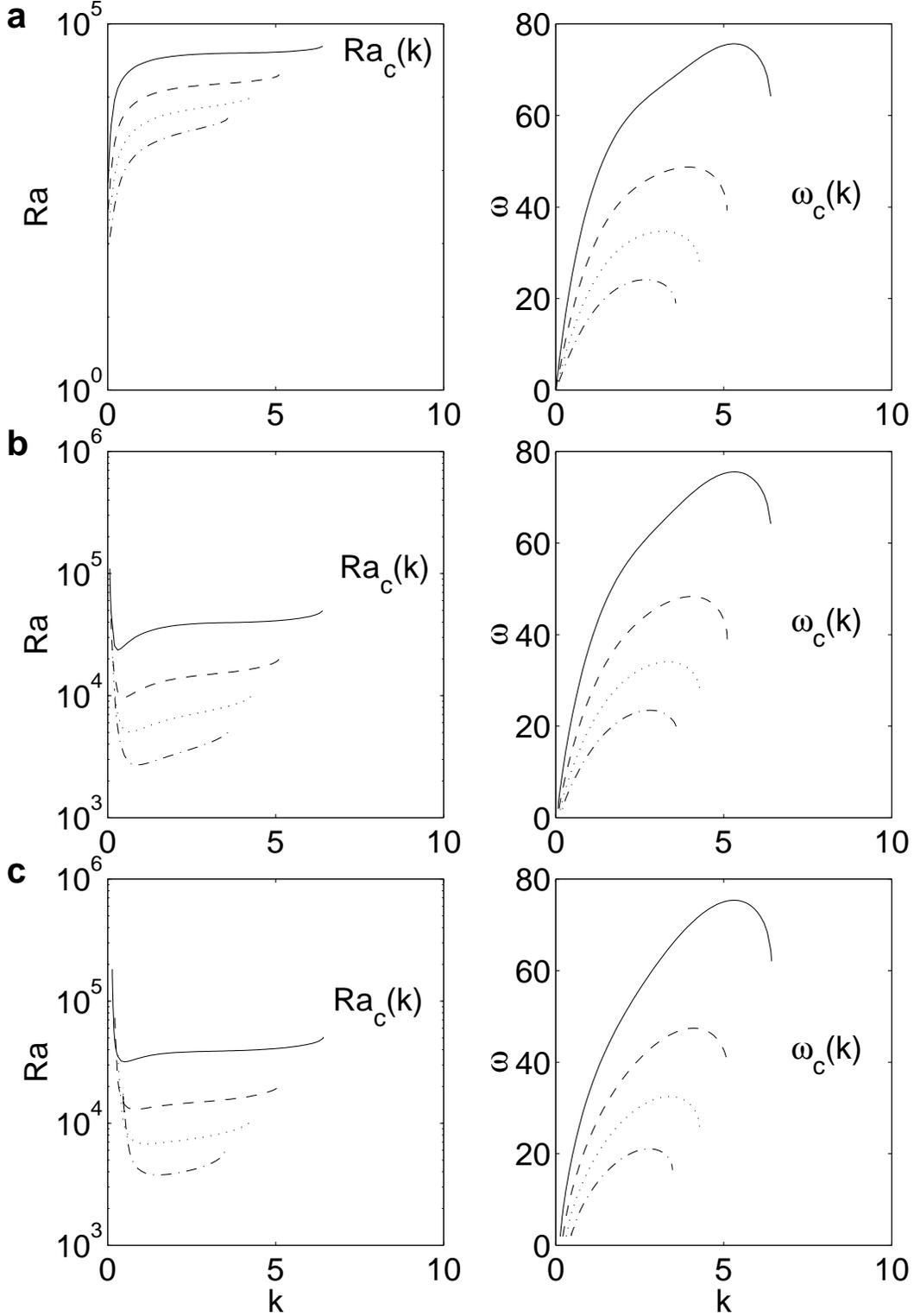,width=14cm}}
\vspace*{-0.4cm}
\caption{$\theta=0$. Inviscid fluid. Curves of the marginal linear stability
to 2D oscillatory disturbances for independently prescribed $Ra^{s}$, $Ra_{c}(k)$
and $\omega_{c}(k)$; $Le=1$. The solid lines: $Ra^{s}=50000$, the dashed lines:
$Ra^{s}=20000$, the dotted lines: $Ra^{s}=10000$, the dash-dot lines: $Ra^{s}=5000$.
(a) $\eta=\chi=0$; (b) $\eta=\chi=1$; (c) $\eta=0$, $\chi=1$ [in view of transformation
(\ref{eq:tt01}), these data apply to $\theta=\pi$ as well if $Ra_{c}\mapsto Ra_{c}^{s}$
and $Ra^{s}\mapsto Ra$]: in particular, $Ra_{c}(k)$ for $Ra^{s}=5000$ exceeds that for
$Ra^{s}=10000$ when $k\leq 0.56$, $Ra_{c}(k)$ for $Ra^{s}=10000$ exceeds that for
$Ra^{s}=20000$ when $k\leq 0.37$, and $Ra_{c}(k)$ for $Ra^{s}=20000$ exceeds
that for $Ra^{s}=50000$ when $k\leq 0.25$.}
\label{f:mscoirss}
\end{figure}

When $\theta=0$, the anticipated linear instability for
$\eta=\chi$ is oscillatory \cite{r:wel,r:tap}. The corresponding inviscid
linear stability problem is generally described by Eqs. (\ref{eq:lstsi})
and (\ref{eq:lsuii}) and boundary conditions (\ref{bc:tsl}), (\ref{bc:tsr}),
and (\ref{bc:ui}). For flux component conditions at the similarity boundary,
the marginal-stability curves, $Ra_{c}(k)$ and $\omega_{c}(k)$, are illustrated
in Fig. \ref{f:mscoirss}(a). The qualitative similarities between them and
such curves in the symmetric ($\eta=1$, $\chi=0$) case [Fig. 2(a) in
\cite{r:tap}] are the result of the same basic physics of
the instability. This physics and its implications in
Fig. \ref{f:mscoirss}(a) are summarized just below.
As explained in \cite{r:tap}, one can consider only
a standing-wave perturbation \cite{r:gsck}.

In the end of a rotation cycle of a perturbation cell, a potential
energy of component perturbation stratifications is generated. Due
to differential gradient diffusion at the distinction boundary, it
is utilized by the cell in the beginning of the cycle of rotation in
the opposite sense. The maximal amount of such energy depends on the
instability horizontal scale. This scale determines the time available
for the two components of a fluid element to diffuse. $Ra_{c}(k)$ thus
decreases with the increase of the horizontal wavelength and becomes
minimal for a given $Ra^{s}$ as $k\rightarrow 0$
[Fig. \ref{f:mscoirss}(a)].

For efficient
utilization of the potential energy, the frequency with which
the marginally unstable cells change their sense of rotation
also has to resonantly match the time for vertical diffusion
naturally specified by the instability wavelength. As
a consequence, $\omega_{c}(k)\rightarrow 0$ as
$k\rightarrow 0$ and $\omega_{c}(k)$ grows
with the increase of $k$ from $0$
[Fig. \ref{f:mscoirss}(a)].

As the wave number is further increased, however, the
wavelength time for the manifestation of differential diffusion
eventually becomes insufficient for the cell oscillation amplitude
to grow. To afford more time for such diffusion, $\omega_{c}(k)$ thus
decreases when certain values of $k$ are exceeded [Fig. \ref{f:mscoirss}(a)].
This leads to an inconsistency between the diffusion time afforded by the
oscillation frequency and that specified by the instability wavelength.
Due to efficiency of the instability mechanism being then reduced, the
instability fails to develop above a critical value of the wave
number. With the increase of $Ra^{s}$ and resulting growth of
$Ra_{c}$, the enhanced gradient disparity in the perturbed
state intensifies the energy transfer to the perturbation
cells. The minimal unstable wavelength thus decreases
as $Ra^{s}$ grows [Fig. \ref{f:mscoirss}(a)].

Compared to $\eta=1$ and $\chi=0$, however, differential
gradient diffusion arises at $\eta=\chi=0$ only from one boundary,
the distinction boundary. More time is thus required for such
process to be effective as that at $\eta=1$ and $\chi=0$. For
this reason, $\omega_{c}(k)$ are smaller at $\eta=\chi=0$
[Fig. \ref{f:mscoirss}(a)] than at $\eta=1$ and $\chi=0$
[Fig. 2(a) in \cite{r:tap}] for all $k>0$. The restriction
of differential gradient diffusion for $\eta=\chi=0$ to a
vicinity of the single boundary also explains why the
respective intervals of unstable $k$ terminate at
slightly smaller wave numbers compared to $\eta=1$
and $\chi=0$. For the same reason, $Ra_{c}(k)$ are
higher at $\eta=\chi=0$ than at $\eta=1$ and
$\chi=0$ for all $k$ outside
a vicinity of $k=0$.

When the instability horizontal scale is large enough,
however, differential gradient diffusion becomes as effective
for development of the instability at one boundary as at both.
Additional temperature diffusion due to the fixed-value condition
at the second boundary for $\eta=1$ and $\chi=0$ then becomes
a stabilizing factor compared to $\eta=\chi=0$. It reduces the
ability of unstable temperature stratification to generate
the potential energy of solute perturbation stratification.
In the vicinity of zero wave number, therefore, $Ra_{c}(k)$
are smaller for $\eta=\chi=0$ than for $\eta=1$ and $\chi=0$.
For the $Ra^{s}$ in the present Fig. \ref{f:mscoirss}(a) and
in Fig. 2(a) of \cite{r:tap}, this takes place when
$k\leq 0.2$ for $Ra^{s}=5000$, $k\leq 0.1$ for
$Ra^{s}=10000$ and $Ra^{s}=20000$, and
$k\leq 0.06$ for $Ra^{s}=50000$.

As for $\eta=1$ and $\chi=0$, the fact that the most unstable wave number
for $\eta=\chi=0$ is zero makes the determination of exact values of $Ra_{c}(0)$
and group velocity $\omega^{c}_{k}(0)\equiv\partial{\omega_{c}(0)}/\partial{k}$
relevant. Using the same long-wavelength expansion as employed for
$\eta=1$ and $\chi=0$ in \cite{r:tap}, one obtains 
\begin{equation}
\omega_{k}^c(0)\equiv\frac{\partial\omega_c(0)}{\partial k}=\omega_0=\sqrt{Ra^{s}/12} \label{eq:omc}
\end{equation}
and then
\begin{equation}
Ra_c(0)=(2Ra^s+5040)/156. \label{eq:rac}
\end{equation}
These expressions differ from the respective expressions for $\eta=1$ and $\chi=0$ \cite{r:tap}
only by the denominator in Eq. (\ref{eq:rac}). The numerical data underlying the marginal-stability
curves in the present Fig. \ref{f:mscoirss}(a) accurately coincide with Eqs. (\ref{eq:omc}) and
(\ref{eq:rac}). For the two compensating gradients, $Ra_{c}(0)=Ra^{s}$, 
\begin{equation}
Ra_c(0)=2520/77,\;\;\; \omega_{k}^{c}(0)=\omega_{0}=\sqrt{210/77}. \label{eq:roc}
\end{equation}
\subsubsection{\label{s:roih11}$\eta=\chi=1$}
For fixed-value component conditions at the similarity boundary, the marginal-stability
curves are illustrated in Fig. \ref{f:mscoirss}(b). They are qualitatively different
from such curves both for the symmetric case [Fig. 2(a) in \cite{r:tap}] and for the
flux similarity boundary [Fig. \ref{f:mscoirss}(a) herein] by the relative stability
of the vicinity of zero wave number. For any $Ra^{s}$, $Ra_{c}(k)$ increases
abruptly when $k$ decreases below a certain value, for which $Ra_{c}(k)$
is minimal. An immediate vicinity of $k=0$ is also stable for any $Ra$.

The abrupt increase of $Ra_{c}(k)$ with decreasing $k$ is associated
with solute diffusion at the similarity boundary. Such diffusion diminishes
the role played by the potential energy of solute perturbation stratification
in the instability mechanism. When the instability wavelength is relatively
short, however, the enhancement of differential gradient diffusion 
with growth of the wavelength still lowers the $Ra_{c}(k)$.

Below a certain $k$, however, neutralization of
the solute perturbation scale forming at the small
$Ra_{c}(k)$ by its diffusion at the similarity boundary becomes
more important than the effect of differential gradient diffusion.
$Ra_{c}(k)$ then grows with the wavelength, to generate a higher
solute perturbation amplitude at the same $Ra^{s}$. Arising from a
fixed $Ra^{s}$, however, the solute perturbation is merely eliminated
by its diffusion at the similarity boundary for any $Ra$ when the
wave number reaches an immediate vicinity of $k=0$. The instability
of such small $k$ thus fails to develop [Fig. \ref{f:mscoirss}(b)].
\subsubsection{\label{s:roih01}$\eta=0$ and $\chi=1$}
For Eqs. (\ref{eq:lstsi}) and (\ref{eq:lsuii}), the present results
would also be applicable to $\theta=\pi$ when transformation (\ref{eq:tt01})
is accompanied by $Ra_{c}\mapsto Ra_{c}^{s}$ and $Ra^{s}\mapsto Ra$.
To avoid confusion, however, they are discussed
only in terms of $\theta=0$.

With the inversely different component conditions at the boundaries,
a horizontal slot combines elements of both steady and oscillatory
instability mechanisms. For $\theta=0$, in particular, the component
stratifications near the distinction boundary give rise to a mechanism
of oscillatory instability of the type discussed above.

Near the inverse boundary for $\theta=0$, however, the component
stratifications correspond to a mechanism of steady instability of
the type reported in \cite{r:twh,r:tbc}. Also arising from differential
gradient diffusion, the mechanism of such an instability leads to
amplitude growth of only such perturbation cells as do not change
their sense of rotation. This mechanism is expected to affect the
oscillatory instability coming from the distinction boundary.

The perturbation cells whose sense of rotation changes periodically in time
would however be located closer to the distinction boundary. Generated by such
cells, the disparity between component perturbation gradients near this boundary
is expected to exceed the opposite one near the inverse boundary. For small and
intermediate wavelengths of the oscillatory perturbation, therefore, the time
available for differential gradient diffusion would prevent such a (relatively
small-gradient-disparity) process near the inverse boundary from being
effective. Largely specified by the distinction boundary, $Ra_{c}(k)$
in Fig. \ref{f:mscoirss}(c) thus decrease with
$k$ for such wavelengths.

Above a critical wavelength, however, the time available for differential 
diffusion becomes sufficient for such a process near the inverse boundary to noticeably 
damp the oscillatory perturbation. The potential energy of component perturbation
stratification generated by differential gradient diffusion at the distinction
boundary is appreciably reduced by such a process at the inverse boundary.
For this reason, $Ra_{c}(k)$ increases with $k$ decreasing below
a certain value [Fig. \ref{f:mscoirss}(c)].

When the wavelength becomes long enough, the combined effects of differential
gradient diffusion near the distinction and inverse boundaries result in larger
$Ra_{c}(k)$ corresponding to the smaller $Ra^{s}$ [Fig. \ref{f:mscoirss}(c)]. 
$Ra^{s}$ is not only a measure of stability near the distinction boundary. Via
the inverse boundary, it also controls the steady opposition to growth of the
oscillatory perturbation. In particular, this opposition is the stronger the more
time is afforded for differential gradient diffusion with the growing wavelength.
As the relative disparity between the diffusion times [represented by the respective
$1/\omega_{c}(k)$, Fig. \ref{f:mscoirss}(c)] for different $Ra^{s}$ also
increases with the wavelength, such effect of the inverse boundary
becomes more important than the effect of the distinction boundary.
Larger $Ra^{s}$ are thus destabilized by the smaller $Ra_{c}(k)$. 

In the immediate vicinity of zero wave number, the (large) time available
for differential gradient diffusion makes the effect of such a process near
the inverse boundary comparable with that near the distinction boundary.
This takes place despite the smaller gradient disparity formed
near the former boundary. As a consequence, the oscillatory
instability fails to develop [Fig. \ref{f:mscoirss}(c)].
\subsection{\label{s:roiv}$\theta\in[0,\pi/2]$}
\subsubsection{\label{s:gen}General}
Oscillatory linear instability in a vertical slot with
viscous fluid is discussed in Sec. \ref{s:rovvim} below. The broken
boundary conditions symmetry then results in only one traveling wave.
Such a disturbance arises from differential gradient diffusion, whether due
to the distinction boundary alone or to both the distinction and the inverse
boundaries. The direction of propagation of such a traveling wave matches the
sign of the background contribution to the density (Fig. \ref{f:g}) at a sidewall
with different boundary conditions from a (component) variable whose flux is prescribed
there by (\ref{bc:l})---(\ref{bc:r}). [This also applies both to the single traveling
wave for $\eta=0$ and $\chi=1$ and to either of the symmetrically counter-propagating
waves for $\eta=1$ and $\chi=0$ (Fig. 7 of \cite{r:tap}).] For the configuration
in Fig. \ref{f:g} ($\theta=\pi/2$), such a traveling wave
has $+k_{y}$ in (\ref{eq:p3}).

Growth of the disturbance with $-k_{y}$ in (\ref{eq:p3}) is inconsistent
with the effect of differential gradient diffusion combined with the
along-slot gravity component. 2D instability to the standing-wave
disturbance at $\theta=0$ cannot thus transform into the respective
instability to the $e^{i(\omega\tau-k_{y}y)}$-mode traveling wave
when $\theta$ is increased from $0$. As a consequence, instability
to the latter traveling wave was found to vanish precipitously
when $\theta$ increases from $0$ for all
types of the boundary conditions.

The vanishing 2D instability is then replaced by the respective
instability to 3D perturbations, since the latter perturbations are
less sensitive to the above asymmetry introduced by the along-slot gravity
component. (In particular, the along-slot gravity does not affect a perturbation
with $k_{y}=0$ and $k_{z}\neq 0$.) Largely driven by the across-slot gravity,
such a 3D instability also has to vanish as $\theta$ increases further. In view
of these considerations, it is only the $e^{i(\omega\tau+k_{y}y+k_{z}z)}$-mode
instability that is discussed herein below both for inviscid
and for viscous fluid. 

For any type of the boundary conditions,
the inviscid $e^{i(\omega\tau+k_{y}y+k_{z}z)}$-mode instability
with certain $k_{z}\geq 0$ possesses intervals of $k_{y}$ for which
$Ra_{c}(k_{y})=\omega_{c}(k_{y})=0$ at some $\theta\in(0,\pi/2]$, as
discussed below. Such a zero $\omega_{c}$, however, does not necessarily
mean that the instability is steady. In particular, steady instabilities
with finite $Ra_{c}$ are absent in viscous fluid for $\theta\in(0,\pi)$ and
the boundary conditions considered herein. The zero-$\omega_{c}$ instability
arising say at $\theta=\pi/2$ from the inviscid equations for oscillatory
marginally unstable perturbation then has no immediate connection with
the instability of a steady origin. It seems thus reasonable to expect
that the perturbation developing at any small $Ra(k_{y})>0$ for some
$\theta\in(0,\pi/2]$ and $k_{z}\geq 0$ would generally have a
small $\omega(k_{y})>0$ as well. This is what is implied
below, although such an effect of along-slot
gravity is referred to as direct.
\subsubsection{\label{s:roiv00}$\eta=\chi=0$}
At $\theta=\pi/2$, the (2D) $e^{i(\omega\tau+k_{y}y)}$-mode 
instability would generally have to be manifested in the
form of traveling cells whose sense of rotation is constant
in time. In inviscid fluid, such a manifestation also has to
take place at any $Ra>0$. In the absence of dissipation, even
an infinitesimal horizontal density difference resulting from
differential gradient diffusion could drive the perturbation.
Eq. (\ref{eq:lsui}) then implies that such instability with
$Ra_{c}(k_{y})=0$ also has to have $\omega_{c}(k_{y})=0$. For
$\theta\in(0,\pi/2)$, however, the effect of across-slot
gravity opposes amplitude growth of the perturbation
cells whose sense of rotation does not
change (Sec. \ref{s:roih00}).

\begin{figure}
\centerline{\psfig{file=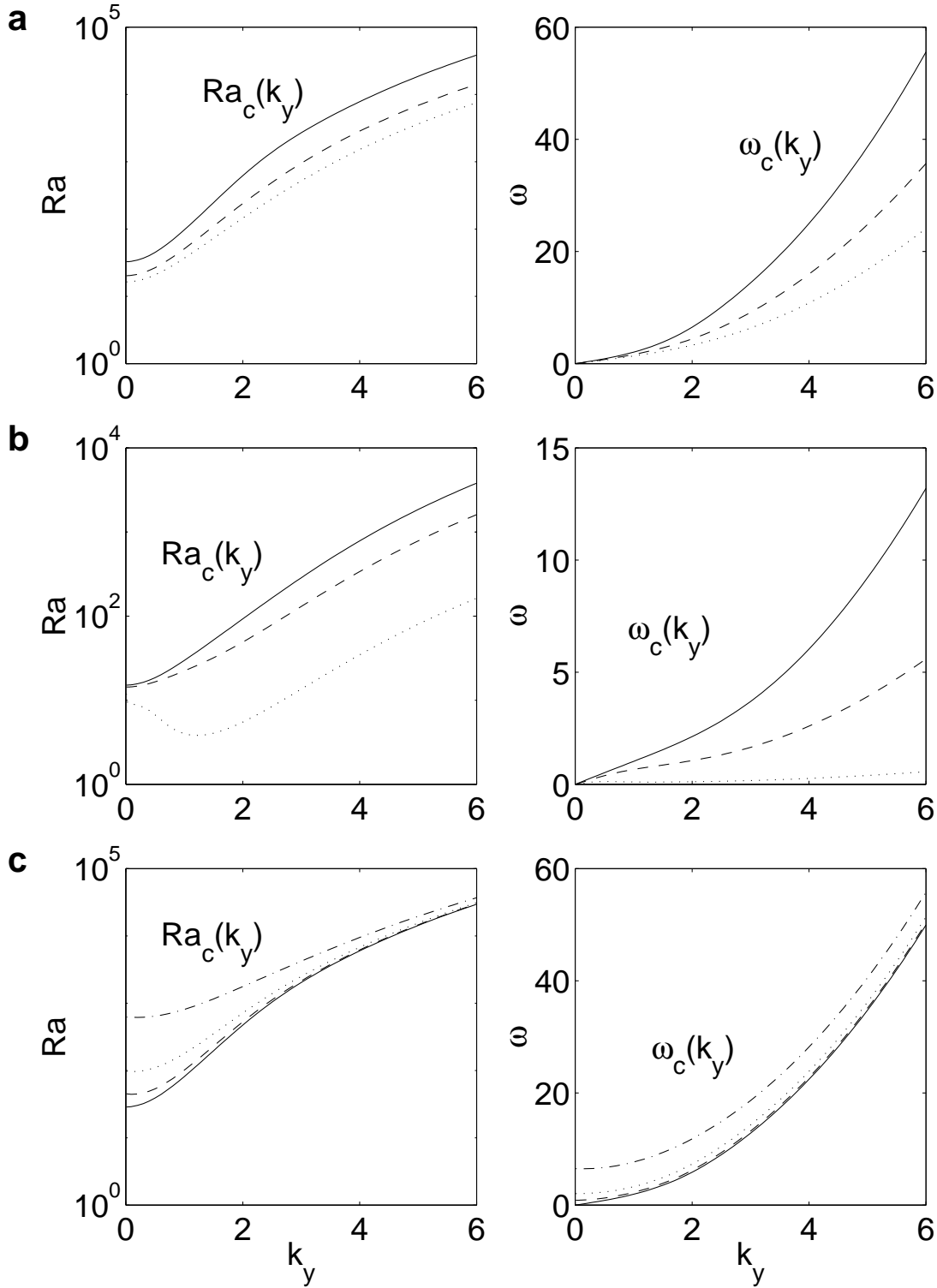,width=14.5cm}}
\caption{Inviscid fluid; $\eta=\chi=0$. 
Curves of the marginal linear stability to 3D ($k_{z}\geq 0$)
oscillatory disturbances for different $k_{z}$, $Ra_{c}(k_{y})$
and $\omega_{c}(k_{y})$; $\mu=1$, $Le=1$. (a) $k_{z}=0$, the solid lines:
$\theta=0$, the dashed lines: $\theta=\pi/8$, the dotted lines:
$\theta=\pi/4$; (b) $k_{z}=0$, the solid lines: $\theta=3\pi/8$,
the dashed lines: $\theta=0.9\pi/2$, the doted lines:
$\theta=0.99\pi/2$; (c) $\theta=0.05\pi/2$, the solid
lines: $k_{z}=0$, the dashed lines: $k_{z}=0.5$, the
dotted lines: $k_{z}=1$, the dash-dot lines:
$k_{z}=2$.}
\label{f:mscoi00}
\end{figure}

The combination of the effects of across-slot
and along-slot gravity thus results in a decrease of
$\omega_{c}(k_{y})$ with $\theta$ increasing from $0$ to $\pi/2$
[Fig. \ref{f:mscoi00}(a) and (b)]. This allows part of the rotation
energy of a perturbation cell to come from the energy directly contributed
by the along-slot gravity component in the current cycle of rotation.
In the end of a cell rotation cycle, the whole rotation energy is
transformed into the potential energy of perturbation stratification
due to the across-slot gravity component. This potential energy
is released in the next rotation cycle. As $\theta$ grows
from $0$, the $Ra_{c}(k_{y})$ thus also decrease.

At the longest 2D wavelengths
for $\eta=\chi=0$, the across-slot gravity component
acts most effectively in opposing the direct contribution
of the along-slot component to the rotation-energy increase.
On the other hand, the shortest scales are least effective
in accommodating the contribution of the along-slot gravity.
It is thus a set of intermediate 2D wavelengths that
becomes most unstable when $\theta=\pi/2$ is approached
[Fig. \ref{f:mscoi00}(b)]. Decreasing for all $k_{y}$
as $\theta\rightarrow\pi/2$, such $Ra_{c}(k_{y})$
and $\omega_{c}(k_{y})$ also become identically
zero at $\theta=\pi/2$.

Upon introduction of the long-wavelength expansion used above
\cite{r:tap} into Eqs. (\ref{eq:lsui}) and (\ref{eq:lstsi}) with $k=k_{y}$,
the $k_{y}^{1}$ order of the $e^{i(\omega\tau+k_{y}y)}$-version of these
equations and boundary conditions (\ref{bc:tsl}), (\ref{bc:tsr}),
and (\ref{bc:ui}) ($\eta=\chi=0$) yields
\begin{equation}
Ra_{c}(0)=24\omega_{0}^{2}/(2\cos\theta+\omega_{0}\sin\theta), \label{eq:roci}
\end{equation}
which is consistent with Eq. (\ref{eq:omc}) for $\theta=0$. The values of
$Ra_{c}(0)$ and $\omega^{c}_{k}(0)=\omega_{0}$ estimated from the numerical
data underlying the (2D) marginal-stability curves in Fig. \ref{f:mscoi00}(a)
and (b) [as well as from such data for $k_{y}<0$ (for which $\omega_{0}<0$ as
well) not reported herein] were found to be fairly consistent
with Eq. (\ref{eq:roci}).
\subsubsection{\label{s:roiv11}$\eta=\chi=1$}
\begin{figure}
\centerline{\psfig{file=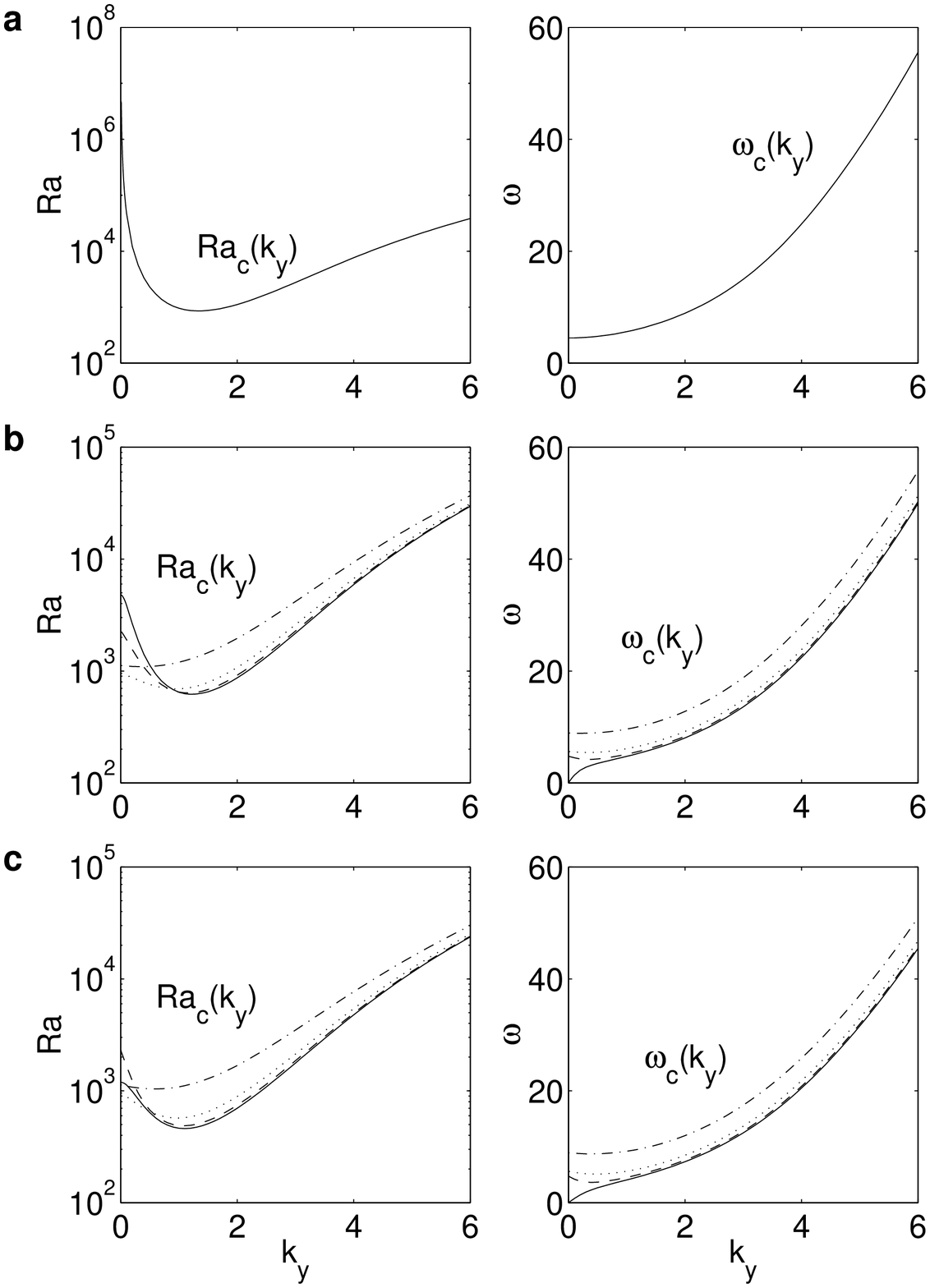,width=14.5cm}}
\caption{Inviscid fluid; $\eta=\chi=1$. 
Curves of the marginal linear stability to 3D ($k_{z}\geq 0$)
oscillatory disturbances for different $k_{z}$, $Ra_{c}(k_{y})$ and
$\omega_{c}(k_{y})$; $\mu=1$, $Le=1$. For (a)---(e), the solid lines:
$k_{z}=0$, the dashed lines: $k_{z}=0.5$, the dotted lines: $k_{z}=1$,
the dash-dot lines: $k_{z}=2$. (a) $\theta=0$; (b) $\theta=0.05\pi/2$;
(c) $\theta=0.1\pi/2$; (d) $\theta=0.15\pi/2$; (e) $\theta=\pi/8$;
(f) $k_{z}=0$, the solid lines: $\theta=\pi/4$, the dashed lines:
$\theta=3\pi/8$, the dotted lines: $\theta=0.9\pi/2$,
the dash-dot lines: $\theta=0.99\pi/2$.}
\label{f:mscoi11}
\end{figure}

At $\theta=\pi/2$ for $\eta=\chi=1$, the inviscid
$e^{i(\omega\tau+k_{y}y)}$-mode instability also has to
arise at any $Ra(k_{y})>0$, and thus $\omega_{c}(k_{y})=0$ for
such instability as well. When only the across-slot gravity
component is present ($\theta=0$), the vicinity of zero
wave number is most stable to the disturbances that
change their sense of rotation periodically in time
[Fig. \ref{f:mscoi11}(a)]. It is therefore this region
of $k_{y}$ that least opposes the direct contribution
of along-slot gravity to the cell rotation energy. As
$\theta$ increases from $0$, 2D $Ra_{c}(k_{y})$ thus
decreases most significantly near $k_{y}=0$ 
[Fig. \ref{f:mscoi11}(a)---(e)]. Closer to
$\theta=\pi/2$, however, the $Ra_{c}(k_{y})$
decreases noticeably and tends to $0$ as
$\theta\rightarrow\pi/2$ for all $k_{y}$ 
[Fig. \ref{f:mscoi11}(f)].

As discussed in \cite{r:tap}, three-dimensionality
of most unstable disturbances could be a consequence
of two general mathematical conditions. One of them
(condition I) is dependence of $Ra_{c}$ only on 
wave number modulus $k$, as in
\begin{equation}
Ra_{c}(k_{y},k_{z})=Ra_{c}[(k^{2}_{y}+k^{2}_{z})^{1/2},0],
\label{eq:kmod}
\end{equation}
at a single value of some parameter ($\theta=0$ in this
case) in whose vicinity $Ra_{c}$ depends on both components of
$\vec{k}$. The other condition (condition II) is the existence of an
interval where $Ra_{c}(k)$ is growing with decreasing $k$ at this value of
the parameter. Three-dimensionality of the instability in a vicinity of the
above parameter value then follows from the assumption that these conditions
would largely apply at nearby values of such a parameter as well.

As discussed in
Sec. \ref{s:roih11} above, condition II holds at $\theta=0$ [Fig. \ref{f:mscoi11}(a)] 
near $k=0$ due to solute diffusion at the similarity boundary. This process
is therefore responsible for the three-dimensionality of instability in
Fig. \ref{f:mscoi11}(b) and (c). With the $Ra_{c}(k_{y})_{|\theta=\pi/2}=0$,
the fast decrease of 2D $Ra_{c}(k_{y})$ near $k_{y}=0$ with $\theta$
growing from $0$ prevents such 3D disturbances from being dominant
for the larger $\theta$. That no 3D most unstable disturbances
are found in Fig. \ref{f:mscoi00}(c) is thus a consequence
both of condition II being then not met for $\theta=0$ 
[Fig. \ref{f:mscoi00}(a)] and of the absence
of another mechanism for 3D instability.
\subsubsection{\label{s:roiv01_2D}2D disturbances for $\eta=0$ and $\chi=1$}
\paragraph{General.}
Transformation (\ref{eq:tt01}) makes
the results of this Sec. \ref{s:roiv01_2D} also
applicable to $\theta\in[\pi/2,\pi]$. Assuming (\ref{eq:tt01}),
they are however discussed below only in terms of $\theta\in[0,\pi/2]$.
As seen from Fig. \ref{f:msco01i2D}, any small increase of $\theta$
from $0$ leads to abrupt changes in the 2D marginal-stability curve
for $\eta=0$ and $\chi=1$. These changes are associated with emergence
of an interval of $k_{y}$ for which $Ra_{c}(k_{y})=\omega_{c}(k_{y})=0$.
Such interval is born with its width and the lower limit tending
to zero as $\theta\rightarrow0$. Both these parameters of
the interval then become finite when $\theta$ increases
from $0$. They also continue to grow with $\theta$
increasing further (see the solid lines in
Fig. \ref{f:msco01i3D}).

In viscous fluid, steady
instability for $\theta=0$ arises at finite viscous
$Ra_{c}(k)$ (Fig. \ref{f:mscsov01}). (It precedes the respective
instability to standing wave.) The inviscid steady $Ra_{c}(k)$ are then
zero for all $k>0$ [Eq. (\ref{eq:lssu}) for $\theta=0$]. As discussed below
(Sec. \ref{s:rovvh}), however, such 2D steady viscous instability transforms
into an oscillatory instability to traveling wave when $\theta$ is increased from
$0$. As $\theta>0$, therefore, the argument just used does not apply. For $\theta>0$,
the inviscid 2D instability with $Ra_{c}(k_{y})=\omega_{c}(k_{y})=0$ still ought to
arise from the type of perturbation that is dominant at $\theta=\pi/2$, where the
largest zero-threshold interval forms [Fig. \ref{f:msco01i3D}(f)]. Generally,
this has to be a traveling wave (whose speed turns infinitesimal with
$\omega_{c}=0$). Such a perturbation is also the first to become
unstable in viscous fluid for $\theta\in(0,\pi/2]$. Its
manifestation would be most convenient to analyze in
the presence of along-slot gravity alone.
\paragraph{Effect of the along-slot gravity.}
In the presence of along-slot gravity alone [Fig. \ref{f:msco01i3D}(f)
for $k_{z}=0$], $Ra_{c}(k_{y})=\omega_{c}(k_{y})=0$ for $k_{y}>k_{y}^{2l}(0)\approx 2.4677$
(Table \ref{t:kc} for $\theta=\pi/2$ and $k_{z}=0$). Just below this value of $k_{y}$,
however, both $Ra_{c}(k_{y})$ and $\omega_{c}(k_{y})$ abruptly increase to finite
magnitudes. Additional calculations for $k_{y}>6$ and $\theta=\pi/2$
also suggest that the upper limit of the interval with zero
$Ra_{c}(k_{y})$ and $\omega_{c}(k_{y})$ is at infinity.

As discussed below for viscous fluid (Sec. \ref{s:rovvim}),
differential gradient diffusion for $\eta=0$ and $\chi=1$ acts in concert
at the vertical-slot boundaries. It gives rise to horizontal density
differences that favor growth of the $e^{i(\omega\tau+k_{y}y)}$-mode
traveling wave. In particular, the density perturbation generated
around a clockwise-(counterclockwise-)rotating cell is largely
specified by negative (positive) solute perturbation at
the distinction boundary and temperature perturbation
at the inverse boundary. This intensifies such
downwards-propagating small-amplitude
convective cells with a steady
sense of rotation. 

In the absence of dissipation, the horizontal density differences
just described would give rise to the instability even when they are
infinitesimal. Such an inviscid vertical-slot instability can thus arise
for any $Ra>0$ no matter how much the effect of differential gradient
diffusion diminishes with decreasing the wavelength. This is
seen in Fig. \ref{f:msco01i3D}(f) ($k_{z}=0$) for
$k_{y}>k_{y}^{2l}(0)\approx 2.4677$.

For $\eta=0$ and $\chi=1$, however, either component
forms a diffusion gradient at one of the boundaries. Such a
diffusion is also the more effective the longer the wavelength
is. When the wavelength exceeds a critical magnitude, therefore,
the perturbation of either component could be neutralized by its
diffusion at one of the boundaries. The infinitesimal horizontal
density differences arising just above $Ra_{c}(k_{y})=\omega_{c}(k_{y})=0$
are thus eliminated by the component diffusion. Both $Ra_{c}(k_{y})$
and $\omega_{c}(k_{y})$ must then increase from zero as $k_{y}$
decreases below the critical value [$k_{y}=k_{y}^{2l}(0)\approx 2.4677$].
This increase is precipitous since such growing $\omega_{c}(k_{y})$
also lowers the efficiency of differential gradient diffusion,
due to inconsistency between the $\omega_{c}(k_{y})$ and
the increasing wavelength.
\paragraph{Combined effects of the along-slot and across-slot gravity.}
The oscillatory instability for $\theta=0$ arises only at finite $Ra_{c}(k_{y})$
(Fig. \ref{f:msco01i2D}). It is characterized by a standing wave, i.e. by convective
cells whose sense of rotation changes periodically in time. Development of
such a perturbation is inconsistent with that of a traveling wave arising
due to the along-slot gravity, for such traveling-wave convective cells
do not change their sense of rotation. For $\theta\in(0,\pi/2)$, the
latter perturbation is also favored and opposed by the effects of
across-slot gravity at the inverse and distinction boundaries,
respectively, particularly when it is manifested with
$Ra_{c}(k_{y})=\omega_{c}(k_{y})=0$. The steadily
rotating cells would thus have to be localized
near the inverse boundary to the extent
$\theta$ is close to $0$.

Due to the expansion of the (steadily rotating) convective cells towards
the distinction boundary with $\theta$ growing from $0$
to $\pi/2$, however, an opposition at this boundary to the steady
sense of cell rotation is generally relevant for any $\theta<\pi/2$.
Its relative role in rotation of a convective cell depends on the
instability wavelength. As such wavelength decreases, in particular,
the effect of across-slot gravity at the distinction boundary (opposing
the steady sense of cell rotation) becomes relatively more pronounced
with respect to that at the inverse boundary (favoring the steadily
rotating cells). In addition, the shorter the wavelength the smaller
the relative portion of streamline particles with across-slot density
differences compared to that with such along-slot differences. The
decreasing (increasing) wavelength thus also enhances (reduces)
the effect of across-slot gravity with respect
to that of along-slot gravity.

The minimal wavelength above which the zero-threshold steadily
rotating cells are dominant then depends on $\theta$. Increasing from $0$
with $\theta$ decreasing from $\pi/2$ [Fig. \ref{f:msco01i3D} for $k_{z}=0$],
it tends to infinity as $\theta\rightarrow 0$ (Fig. \ref{f:msco01i2D}). Just
above such a critical $k_{y}$[$\equiv k_{y}^{2u}(0)$ (Table \ref{t:kc} for
$k_{z}=0$)] for $\theta\in(0,\pi/2)$, the steadily rotating perturbation
cells fail to grow at infinitesimal $Ra$. The $Ra_{c}(k_{y})$ and
$\omega_{c}(k_{y})$ thus increase from $0$.

As such $Ra_{c}(k_{y})$ and $\omega_{c}(k_{y})$ grow from $0$,
however, the effect of across-slot gravity becomes increasingly more
relevant. The efficiency of differential gradient diffusion for the
steadily rotating perturbation cells then decreases. The $Ra_{c}(k_{y})$
and $\omega_{c}(k_{y})$ thus grow precipitously to such $k_{y}$.
Relatively localized near the inverse boundary (to the extent $\theta$ is
close to $0$), the steadily rotating convective cells then also transform
into such perturbation cells as are relatively localized near the distinction
boundary and change their sense of rotation with an adequate frequency.
Control over the instability disturbances is thus largely transferred
to the across-slot gravity component, whose action still remains
affected by the direct contribution from
the along-slot component.

When $\theta=\pi/2$ and
$k_{y}<k_{y}^{2l}(0)_{|\theta=\pi/2}\approx 2.4677$, 
an infinitesimally small $Ra(k_{y})$ is neutralized by diffusion
and thus fails to generate the instability. This takes place when
such a diffusion process is equally active at both boundaries, as
is expected at $\theta=\pi/2$. The closer $\theta$ is to $0$, however,
the more localized the steadily rotating perturbation cells are near
the inverse boundary. Compared to $\theta=\pi/2$, the overall effect
of diffusion for such cells when $\theta\in(0,\pi/2)$ thus becomes
asymmetrically divided between the components. Temperature
diffusion at the distinction boundary is then less active
than solute diffusion at the inverse boundary. For this
reason, $k_{y}^{2l}(0)$ [below which the $Ra_{c}(k_{y})$
abruptly increases from $0$] also decreases with $\theta\in(0,\pi/2)$,
from $k_{y}^{2l}(0)_{|\theta=\pi/2}\approx 2.4677$ to $0$
as $\theta\rightarrow 0$ (Fig. \ref{f:msco01i2D}
and Fig. \ref{f:msco01i3D} for $k_{z}=0$).
\subsubsection{\label{s:roiv01_3D}3D disturbances for $\eta=0$ and $\chi=1$}
\paragraph{\label{s:roiv01_3DG}General.}
Although transformation (\ref{eq:tt01})
makes the results of this Sec. \ref{s:roiv01_3D}
also applicable to $\theta\in[\pi/2,\pi]$, except
for Sec. \ref{s:roiv01_3D3F}, they are discussed
below only in terms of $\theta\in[0,\pi/2]$. 
For any $\theta\in(0,\pi/2)$ in Table \ref{t:kc} and
Fig. \ref{f:msco01i3D}, the 3D lower and upper limits of the above
zero-threshold interval of $k_{y}$, $k_{y}^{2l}(k_{z})$ and $k_{y}^{2u}(k_{z})$, 
decrease from $k_{y}^{2l0}\equiv k_{y}^{2l}(0)$ and $k_{y}^{2u0}\equiv k_{y}^{2u}(0)$,
respectively, with $k_{z}$ increasing from $0$. They also continue to decrease with
$k_{z}$ increasing further. For $\theta=\pi/2$, this formally applies only
to $k_{y}^{2l}(k_{z})$, since $k_{y}^{2u}(k_{z})$ is then at infinity for
any finite $k_{z}\geq 0$. In addition, another zero-threshold area arises
from the vicinity of $k_{y}=0$ when $k_{z}>0$ [see the dashed lines
in Fig. \ref{f:msco01i3D}(a)---(e)]. However, the lower and
upper limits of the latter area, $k_{y}^{3l}(k_{z})$ and
$k_{y}^{3u}(k_{z})$, respectively, increase with
growing $k_{z}$ (Table \ref{t:kc}, $\theta\neq\pi/2$).

At a certain $k_{z}>0$, the interval of finite $Ra_{c}(k_{y})$ and
$\omega_{c}(k_{y})$ between $k_{y}^{3u}(k_{z})$ and $k_{y}^{2l}(k_{z})$
thus vanishes, due to these parameters of $k_{y}$ merging with each other.
A single continuous interval of zero $Ra_{c}(k_{y})$ and $\omega_{c}(k_{y})$
is then formed [the dotted lines in Fig. \ref{f:msco01i3D}(a)---(e)]. Eventually,
such a continuous interval also vanishes when $k_{y}^{3l}(k_{z})$ and
$k_{y}^{2u}(k_{z})$ merge at a still larger $k_{z}$, leaving only
nonzero $Ra_{c}$ and $\omega_{c}$ [the dash-dot lines in
Fig. \ref{f:msco01i3D}(a)---(e)]. Quantitative details
of the behavior just described are reported
in Table \ref{t:kc}.

When 3D perturbations in
Fig. \ref{f:msco01i3D} are dominant, their behavior cannot
be explained only in the framework of the scenario emphasized
in Sec. \ref{s:roiv11} above. Other mechanisms causing such
a three-dimensionality would thus also have to exist. These
mechanisms are associated with the nature of the 2D interval
with $Ra_{c}(k_{y})=\omega_{c}(k_{y})=0$ discussed
in Sec. \ref{s:roiv01_2D} above.
\paragraph{\label{s:roiv01_3D2}$k_{y}^{2l}(k_{z})$ and $k_{y}^{2u}(k_{z})$.}
For a given $\theta\in(0,\pi/2)$ and $k_{z}=0$, the 2D $Ra_{c}(k_{y})$ and
$\omega_{c}(k_{y})$ increase from zero either when $k_{y}$ decreases below
$k_{y}^{2l0}$ or when it increases above $k_{y}^{2u0}$. This takes place
because for the respective wavelength intervals ($\lambda_{y}>2\pi/k_{y}^{2l0}$ and
$\lambda_{y}<2\pi/k_{y}^{2u0}$) at the fixed $\theta\in(0,\pi/2)$, the perturbation
cells with a steady sense of rotation cannot be destabilized by the effects of differential
gradient diffusion at infinitesimal $Ra$ (see Sec. \ref{s:roiv01_2D} above).
Depending on the relative roles of the along-slot and across-slot gravity
components, $k_{y}^{2l0}$ and $k_{y}^{2u0}$ are also the closer
to $0$ the closer $\theta$ is to $0$.

Independent of the orientation of the axis of rotation of a
convective cell, the effect of across-slot gravity does not change
when $k_{z}$ increases from $0$. In this context, one could therefore
refer to $k_{y}^{2l0}$ and $k_{y}^{2u0}$ as the respective critical
values of $k$. With $k_{z}$ growing from $0$, however, the
(relevant) projection of along-slot gravity on the axis
orthogonal to the axis of cell rotation decreases.

Indeed, Eqs. (\ref{eq:lsui}) and (\ref{eq:lstsi}) for $\vec{k}=(k_{y},k_{z})$
are identical to these equations for $\vec{k}=(k_{y},0)$ with that $\sin\theta$
in the latter is replaced by $(k_{y}/k)\sin\theta$ in the former. The relative
role of along-slot gravity is thus diminished with respect to that in the 2D
problem for $\vec{k}=(k_{y},0)$. This is similar to decreasing $\theta$ below
the considered value. With increasing $k_{z}$, therefore, the actual values
of $k_{y}^{2l}(k_{z})$ and $k_{y}^{2u}(k_{z})$ must be smaller than
$q_{y}^{2l}(k_{z})\equiv[(k_{y}^{2l0})^{2}-k_{z}^{2}]^{1/2}$ and
$q_{y}^{2u}(k_{z})\equiv[(k_{y}^{2u0})^{2}-k_{z}^{2}]^{1/2}$,
respectively, as in Table \ref{t:kc} for any
$\theta\neq\pi/2$.

For $\theta=\pi/2$, the relative role of the along-slot gravity component with
respect to that of the (absent) across-slot component would remain infinite for
any finite $k_{z}$. This has to lead to $k_{y}^{2l}(k_{z})=q_{y}^{2l}(k_{z})$,
as is seen from Table \ref{t:kc} ($\theta=\pi/2$). In addition, the infinite
value of $k_{y}^{2u0}$ is retained both by ${q_{y}^{2u}(k_{z})}_{|k_{z}>0}$
and by ${k_{y}^{2u}(k_{z})}_{|k_{z}>0}$ [Fig. \ref{f:msco01i3D}(f)
and Table \ref{t:kc} ($\theta=\pi/2$)].

For $\theta=\pi/2$, $k_{y}^{2l}(k_{z})$ and
$k_{y}^{2u}(k_{z})$ should thus be exactly specified
by $k_{y}^{2l0}$ and $k_{y}^{2u0}$($=\infty$). These
latter have to serve as the critical values of $k$.
Indeed, Eqs. (\ref{eq:lsui}) and (\ref{eq:lstsi})
for $\theta=\pi/2$ suggest that
\begin{equation}
Ra_{c}(k_{y},k_{z})k_{y}/k=Ra_{c}(k,0).
\label{eq:inv}
\end{equation}
This means that ${Ra_{c}(k_{y},k_{z})}_{|\theta=\pi/2}=0
\Longleftrightarrow{Ra_{c}(k,0)}_{|\theta=\pi/2}=0$ for 
\begin{equation}
{k_{y}^{2l0}}_{|\theta=\pi/2}\leq(k_{y}^{2}+k_{z}^{2})^{1/2}\leq
{k_{y}^{2u0}}_{|\theta=\pi/2}=\infty,
\label{eq:invk}
\end{equation}
which is a condition for $k$[$=(k_{y}^{2}+k_{z}^{2})^{1/2}$] alone.

For $\theta=\pi/2$, 3D disturbances are thus most unstable
at least for $k_{y}$ between ${q_{y}^{2l}(k_{z})}_{|\theta=\pi/2}$
($<{k_{y}^{2l0}}_{|\theta=\pi/2}$) and ${k_{y}^{2l0}}_{|\theta=\pi/2}$.
This could cause such a three-dimensionality for small $\pi/2-\theta$ as well.
For any $\theta\in(0,\pi/2)$, however, $k_{y}^{2l}(k_{z})<q_{y}^{2l}(k_{z})$
(because $G=k_{y}/k<1$), as discussed above. The latter inequality is also an
independent cause for dominance of 3D disturbances at $\theta\in(0,\pi/2)$.
Additional causes of perturbation three-dimensionality for $\theta\in(0,\pi/2)$
are associated with another zero-threshold area in the $k_{y}$--$k_{z}$
plane. This area is described in Table \ref{t:kc} by
$k_{y}^{3l}(k_{z})$ and $k_{y}^{3u}(k_{z})$.
\paragraph{\label{s:roiv01_3D3G}General on $k_{y}^{3l}(k_{z})$ and $k_{y}^{3u}(k_{z})$.}
The nature of the area with $Ra_{c}(k_{y},k_{z})=\omega_{c}(k_{y},k_{z})=0$
whose boundaries are described in Table \ref{t:kc} by $k_{y}^{3l}(k_{z})$
and $k_{y}^{3u}(k_{z})$ is fundamentally three-dimensional. First note that
Eq. (\ref{eq:kmod}) is satisfied by $Ra_{c}(k_{y},k_{z})_{|\theta\in(0,\pi/2)}$
for $k_{y}=0$: ${Ra_{c}(0,k_{z})}_{|\theta\in(0,\pi/2)}={Ra_{c}(0,k)}_{|\theta\in(0,\pi/2)}=
{Ra_{c}(k,0)}_{|\theta=0}/\cos\theta$. In a vicinity of $k=0$ [Fig. \ref{f:msco01i2D}],
therefore, ${Ra_{c}(0,k)}_{|\theta\in(0,\pi/2)}$ decreases with increasing $k$.
Conditions I and II discussed in Sec. \ref{s:roiv11} are then met. A region of
small $k_{z}$ is thus expected to be dominated by 3D most unstable
disturbances with small $G=k_{y}/k>0$. This observation, however,
does not unravel the behavior of $k_{y}^{3l}(k_{z})$
and $k_{y}^{3u}(k_{z})$.

Eqs. (\ref{eq:lsui}) and (\ref{eq:lstsi}) for $k_{y}>0$
also suggest that ${k_{y}^{3l}(k_{z})}_{|\theta=\theta_{3}}$
and ${k_{y}^{3u}(k_{z})}_{|\theta=\theta_{3}}$ specify such limits
of the interval of $k$ over which $Ra_{c}(k,0)=\omega_{c}(k,0)=0$ for
$\theta=\theta_{3}$ as are equal to the respective limits of two 2D intervals
of $k_{y}$ over which $Ra_{c}(k_{y},0)=\omega_{c}(k_{y},0)=0$ for two smaller
values of $\theta$. These are such respective $\theta=\theta_{2}$($<\theta_{3}$)
as $\tan\theta_{2}=[k_{3y}(k_{z})/k_{3}(k_{z})]\tan\theta_{3}$, where
$k_{3y}(k_{z})\equiv {k_{y}^{3l}(k_{z})}_{|\theta=\theta_{3}}$,
${k_{y}^{3u}(k_{z})}_{|\theta=\theta_{3}}$
and $k_{3}(k_{z})\equiv[{k^{2}_{3y}(k_{z})}+k_{z}^{2}]^{1/2}$.
For any fixed $\theta\in(0,\pi/2)$, the increase of $G=k_{y}/k$
and $k_{z}$ from $0$ could thus give rise to a 3D area with
$Ra_{c}(k_{y},k_{z})=\omega_{c}(k_{y},k_{z})=0$ analogously
to the emergence of the 2D interval between $k_{y}^{2l0}$
and $k_{y}^{2u0}$ near $k_{y}=0$ when $\theta$ increases
from $0$ (Fig. \ref{f:msco01i2D}). 

Three-dimensionality of the perturbation allows the relative
effects of the two gravity components to be varied independently
of $\theta$, and in particular to mimick the vicinity of $\theta=0$
at any $\theta\in(0,\pi/2)$. As parameters of $k_{y}$ between which
$Ra_{c}(k_{y},k_{z})=\omega_{c}(k_{y},k_{z})=0$, $k_{y}^{3l}(k_{z})$
and $k_{y}^{3u}(k_{z})$ thus not only depend on the relation
between the gravity components. They also
specify this relation.
\paragraph{\label{s:roiv01_3D3Z}$k_{y}^{3l}(k_{z})$
and $k_{y}^{3u}(k_{z})$ for $k_{z}\rightarrow 0$.}
For a fixed $\theta\in(0,\pi/2)$, the effects of across-slot and
along-slot gravity are represented in Figs. \ref{f:msco01i2D} and \ref{f:msco01i3D}(f),
respectively. In view of Eq. (\ref{eq:kmod}), the dependence on $k_{y}$ for $k_{z}=0$ in
Fig. \ref{f:msco01i2D} ($\theta=0$) also represents the dependence on $k$ for $k_{z}>0$. The
dependence on $k_{z}>0$ in Fig. \ref{f:msco01i3D}(f) is then given by Eq. (\ref{eq:inv}). When
$k\rightarrow 0$, $Re[d\tilde{u}(k,0)/dx]\asymp\omega_{c}(k,0)\asymp 1$ both at $\theta=0$
and at $\theta=\pi/2$ [as implied by the scaled continuity and Figs. \ref{f:msco01i2D}
and \ref{f:msco01i3D}(f)]. From Eqs. (\ref{eq:lsui}) and (\ref{eq:lstsi}), the general
solution for $\tilde{u}(x)$ then suggests that ${Ra_{c}(k,0)}_{|\theta=0}\asymp(1/k^{2})$
and ${Ra_{c}(k,0)}_{|\theta=\pi/2}\asymp(1/k)$ as $k\rightarrow 0$.
Defining $R_{\theta_{1}}^{\theta_{2}}(k)\equiv 
{Ra_{c}(k,0)}_{|\theta=\theta_{2}}/{Ra_{c}(k,0)}_{|\theta=\theta_{1}}$, therefore,
${R_{0}^{\pi/2}(k)}_{|k\rightarrow 0}\asymp k$($\rightarrow 0$).

For $(k_{z}/k_{y})<\infty$ [$k_{z}=O(k_{y})$] when $k\rightarrow 0$, therefore,
Eq. (\ref{eq:inv}) implies that the effect of along-slot gravity dominates that of across-slot
gravity. For a finite $\theta\in(0,\pi/2)$, this would be inconsistent with the existence
of an infinitesimal area in the $k_{y}$--$k_{z}$ plane (for infinitesimal $k$) where
$Ra_{c}(k_{y},k_{z})=\omega_{c}(k_{y},k_{z})=0$. Such an area could arise only when
$\varepsilon\equiv(k_{y}/k_{z})\rightarrow 0$ [i.e., $k_{y}=o(k_{z})$] as well.

When $\varepsilon\rightarrow 0$ along with $k\rightarrow 0$
and $R_{c}(k)\equiv R_{0}^{\pi/2}(k)/k$, the necessary balance
between the effects of the two gravity components [analogous
to their 2D balance given by $0<(k_{y}/\tan\theta)<\infty$
as $k_{y}$($=k_{y}^{2u0}$) $\rightarrow 0$
and $\theta\rightarrow 0$],
\begin{equation}
0<\lim_{\stackrel{\scriptstyle{\varepsilon\rightarrow 0}}{\scriptstyle{k\rightarrow 0}}}
{[(k^{2}/k_{y})R_{0}^{\pi/2}(k)/k]}
=\lim_{\stackrel{\scriptstyle{\varepsilon\rightarrow0}}{\scriptstyle{k\rightarrow0}}}
[R_{c}(k)(k^{2}/k_{y})]<\infty,
\label{eq:balr}
\end{equation}
implies that $0<(k^{2}/k_{y})<\infty$ for $k\rightarrow 0$ and $\varepsilon\rightarrow 0$.
In particular, numerical computations show that $R_{c}(k)$ is maximal at
$k\rightarrow 0$ and that 
\begin{equation}
\lim_{k\rightarrow 0}R_{c}(k)\approx 1.055.
\label{eq:rc}
\end{equation}
Using the \mbox{l'Hospital} rule, one could then obtain, in particular,
\begin{displaymath}
(k^{2}/k_{y})\sim(k_{z}^{2}/k_{y})\sim k/\varepsilon\sim k_{z}/\varepsilon
\sim 2 k/(\frac{d k_{y}}{d k})\sim 2 k_{z}/(\frac{d k_{y}}{d k_{z}})\sim
\end{displaymath}
\begin{displaymath}
\frac{d k}{d\varepsilon}\sim\frac{d k_{z}}{d\varepsilon}\sim
[\frac{d(k^{2})}{d(\varepsilon^{2})}]^{1/2}\sim[\frac{d(k_{z}^{2})}{d(\varepsilon^{2})}]^{1/2}
\sim\frac{d k_{y}}{d(\varepsilon^{2})}\sim 1/[\frac{d k_{y}}{d(k^{2})}]\sim
\end{displaymath}
\begin{equation}
1/[\frac{d k_{y}}{d(k_{z}^{2})}]\sim
2/(\frac{d^{2}k_{y}}{d k^{2}})\sim 2/(\frac{d^{2}k_{y}}{d k_{z}^{2}})\;\;\;\;
(k\rightarrow 0,\;\;\varepsilon\rightarrow 0),
\label{eq:balk}
\end{equation}
where $\sim$ denotes the asymptotic equivalence.
With such necessary conditions, the relation between the effects of the two gravity
components is similar to that in the 2D problem when $\theta\rightarrow 0$. For any
finite $\theta\in(0,\pi/2)$, such infinitesimal $k_{y}^{3l}(k_{z})$[$=o(k_{z})$]
and $k_{y}^{3u}(k_{z})$[$=o(k_{z})$] as maintain (\ref{eq:balk}) could
thus arise, $Ra_{c}(k_{y},k_{z})=\omega_{c}(k_{y},k_{z})=0$ for
$k_{y}(k_{z})\in[k_{y}^{3l}(k_{z}),k_{y}^{3u}(k_{z})]$.
\paragraph{\label{s:roiv01_3D3F}$k_{y}^{3l}(k_{z})$ and $k_{y}^{3u}(k_{z})$ for finite $k_{z}$.}
Equivalence relations (\ref{eq:balk}) imply that for
$k_{y}=k_{3y}(k_{z})$[$\equiv k_{y}^{3l}(k_{z})$,$k_{y}^{3u}(k_{z})$],
$dk_{y}/dk_{z}\rightarrow 0$ as $k_{z}\rightarrow 0$ and $\varepsilon\rightarrow 0$.
$k_{3y}(k_{z})$ then has a local minimum in this limit, since $d^{2}k_{y}/dk_{z}^{2}>0$
as $k_{z}\rightarrow 0$ and $\varepsilon\rightarrow 0$ for such a $k_{y}(k_{z})$. One
could thus expect that the respective $dk_{y}/dk_{z}$ be positive within a small
(positive) vicinity of $k_{z}=0$. It is generally shown below that when
$k_{y}=k_{3y}(k_{z})$, $dk_{y}/dk_{z}>0$ for any $k_{z}$($>0$).

As discussed in Sec. \ref{s:roiv01_2D}, $k_{y}^{2l0}$ and $k_{y}^{2u0}$ grow
with increasing (decreasing) $\theta\in(0,\pi/2)$ [$\theta\in(\pi/2,\pi)$], when
the along-slot gravity is enhanced with respect to the across-slot gravity. Eq. (\ref{eq:lsui})
also suggests that the variation of $G(k,k_{z})=(k^{2}-k_{z}^{2})^{1/2}/k$ causes qualitatively
the same effect on the relation between the gravity components as that of $|\tan\theta|$.
$k_{3}(k_{z})$\{$\equiv[k^{2}_{3y}(k_{z})+k_{z}^{2}]^{1/2}$\} thus has to grow with
$G$: $dG(k,k_{z})/dk>0$ for $k=k_{3}(k_{z})$. This yields
\begin{equation}
\frac{dk_{z}}{dk_{3}}<\frac{k_{z}}{k_{3}}\Longleftrightarrow
\frac{dk_{3}}{dk_{z}}>\frac{k_{3}}{k_{z}}.
\label{eq:dk3}
\end{equation}

Since $\varepsilon\rightarrow 0$ as $k_{z}\rightarrow 0$ for $k_{y}=k_{3y}(k_{z})$,
one could expect that $G(k_{3},k_{z})=k_{3y}/k_{3}$ [$\sim k_{y}/k_{z}=\varepsilon$
for infinitesimal $k_{z}$ and $k_{y}=k_{3y}(k_{z})=o(k_{z})$] would increase with
$k_{z}$ within a small vicinity of $k_{z}=0$. Such a behavior is found
to hold for any finite $k_{z}$($>0$) as well: (\ref{eq:dk3}) also
implies that $dG(k,k_{z})/dk_{z}>0$ for $k=k_{3}(k_{z})$.
With $G=k_{y}/(k_{y}^{2}+k_{z}^{2})^{1/2}$
and $k_{y}=k_{3y}(k_{z})$, however,
\begin{equation}
\frac{dG}{dk_{z}}=\frac{k_{z}^{2}}{k^{3}}(\frac{dk_{y}}{dk_{z}}-\varepsilon)>0.
\label{eq:dky}
\end{equation}
This means that $dk_{3y}(k_{z})/dk_{z}>k_{3y}(k_{z})/k_{z}>0$.

$k_{y}^{3l}(k_{z})$ and $k_{y}^{3u}(k_{z})$ thus have to only increase with
growing $k_{z}$. Such a behavior is consistent with the data in Table \ref{t:kc}.
Due to $k_{y}^{2l}(k_{z})$ and $k_{y}^{2u}(k_{z})$ decreasing with growing
$k_{z}$, it also explains both the merging of $k_{y}^{3u}(k_{z})$
with $k_{y}^{2l}(k_{z})$ and that of $k_{y}^{3l}(k_{z})$
with $k_{y}^{2u}(k_{z})$.

Another aspect of the data in Table \ref{t:kc} is that for a fixed
$k_{z}$, $k_{y}^{3l}(k_{z})$ and $k_{y}^{3u}(k_{z})$ are the closer
to $0$ the closer $\theta$ is to $\pi/2$. This is associated with their
ratio between the gravity components, their $G(k_{y},k_{z})\tan\theta$, being a
function of $k$ alone, as suggested by Eq. (\ref{eq:lsui}). Such a function thus has to
be proportional to $1/k$. For $k_{z}$($>0$) being fixed and $k_{y}=k_{3y}(k_{z},\theta)$,
therefore, $d[k_{y}(k_{z},\theta)\tan\theta]/d\theta=0$ implies
$\partial k_{y}(k_{z},\theta)/\partial\theta<0$ [$\partial k_{y}(k_{z},\theta)/\partial\theta>0$]
for $\theta\in(0,\pi/2)$ [$\theta\in(\pi/2,\pi)$]:
\begin{equation}
\frac{\partial k_{y}(k_{z},\theta)}{\partial\theta}=
-\frac{2}{\sin(2\theta)}k_{y}(k_{z},\theta).
\label{eq:kytt}
\end{equation}
Since $k_{y}^{2l}(k_{z})$ and $k_{y}^{2u}(k_{z})$
increase with $\theta$ approaching $\pi/2$, the ultimate vanishing
of all areas with $Ra_{c}(k_{y},k_{z})=\omega_{c}(k_{y},k_{z})=0$, when
$k_{y}^{3l}(k_{z})$ and $k_{y}^{2u}(k_{z})$ merge, takes place at the larger
$k_{z}$ the closer $\theta$ is to $\pi/2$ (Table \ref{t:kc}). For $\theta=\pi/2$,
the absence of across-slot gravity makes the relation between the gravity
components independent of $G$. Eq. (\ref{eq:inv}) thus prohibits
the existence of $k_{y}^{3l}(k_{z})\neq 0$ and
$k_{y}^{3u}(k_{z})\neq 0$ for any $k_{z}$.
\section{\label{s:rv}Viscous fluid}
\subsection{\label{s:rov}Small-amplitude oscillatory convection}
\subsubsection{\label{s:rovh}$\theta=0$}
\paragraph{\label{s:rovhD}Diffusion for $\eta=\chi$.}
Despite dissipation, different component conditions at one boundary give
rise to oscillatory instability at $\theta=0$ both for $\eta=\chi=0$ and for
$\eta=\chi=1$. For the no-slip boundaries, this is illustrated in Fig. \ref{f:mscot0}.
As could be expected, $Ra_{c}(k)$ and $\omega_{c}(k)$ are higher for both values of
$\eta=\chi$ [Fig. \ref{f:mscot0}(a)] than for $\eta=1$ and $\chi=0$ [Fig. 5(a) in
\cite{r:tap}]. Fig. \ref{f:mscot0}(b) also exhibits decaying oscillations of
$\delta Ra_{c}(k)\equiv{Ra_{c}(k)}_{|\eta=\chi=0}-{Ra_{c}(k)}_{|\eta=\chi=1}$
and $\delta\omega_{c}(k)\equiv{\omega_{c}(k)}_{|\eta=\chi=0}-{\omega_{c}(k)}_{|\eta=\chi=1}$.

The behavior of $\delta Ra_{c}(k)$ and $\delta\omega_{c}(k)$
in Fig. \ref{f:mscot0}(b) arises from a more effective
role of the distinction boundary for $\eta=\chi=1$. Compared to
$\eta=\chi=0$, diffusion at the similarity boundary reduces the
component perturbation scales for $\eta=\chi=1$ by the same fraction
(that increases with the wavelength). Specified by the background
scales, the gradient disparity at the distinction boundary then
grows with respect to such a reduced component scale. The
relative amplitude of convective motion thus also grows,
generating a relatively larger gradient disparity for
the next rotation cycle. With such more efficient
mechanism, the small-$k$ instability for
$\eta=\chi=1$ precedes that for
$\eta=\chi=0$.

Figs. \ref{f:perh00}(a),(f) and \ref{f:perh11}(a),(f) illustrate
the stage of potential energy release via differential gradient
diffusion. The ratio of the streamfunction perturbation scale to
that of either component for $\eta=\chi=1$ [Fig. \ref{f:perh11}(a)
and (f)] exceeds this ratio for $\eta=\chi=0$ [Fig. \ref{f:perh00}(a)
and (f)]. Thus relatively more intensive at $\eta=\chi=1$, such a convective
motion gives rise to new component perturbation stratifications
[Figs. \ref{f:perh00}(b)---(d) and \ref{f:perh11}(b)---(d)].
The latter arise from the respective background gradients
with the ones for $\eta=\chi=1$ being smaller.

Via the higher gradient of temperature diffusion at
the distinction boundary, the perturbation stratification
thus formed opposes the current sense of cell rotation
[Figs. \ref{f:perh00}(c),(d) and \ref{f:perh11}(c),(d)].
Due to the higher efficiency of such differential diffusion, this
opposition is more pronounced for the fixed-value similarity boundary
than for the flux one. With respect to the perturbation scale of either
component, in particular, the streamfunction scale for $\eta=\chi=1$
[Fig. \ref{f:perh11}(c),(d)] becomes smaller than that for
$\eta=\chi=0$ [Fig. \ref{f:perh00}(c),(d)]. The relatively unequal
potential energies of perturbation stratification
so generated are then utilized by the reversely
rotating cells [Figs. \ref{f:perh00}(e),(f)
and \ref{f:perh11}(e),(f)].
\paragraph{\label{s:rovhDD}Diffusion and dissipation for $\eta=\chi$.}
A consequence of differential gradient diffusion being more efficient for
$\eta=\chi=1$ than for $\eta=\chi=0$ is thus a relatively larger variation
of the respective velocity scale. For about the same variation time at a
fixed wavelength, this implies steeper spatial velocity gradients for
$\eta=\chi=1$ and thus the respectively greater dissipation.

Of two dissipation mechanisms affecting the
instability \cite{r:tap}, one is merely associated with
damping all motions. Its overall effect is the greater the
shorter the instability wavelength is. The other mechanism causes an
efficiency reduction for the instability feedback. Such feedback links
the component perturbation potential energy generated when a cell rotates
in one sense and the intensity of rotation of such a cell in the opposite
sense. The effect of this mechanism depends on the along-slot part of
dissipation of a convective cell. It is the increasing role of
the latter dissipation mechanism that causes $Ra_{c}(k)$ to rise
infinitely with $k$ decreasing to $0$ for both $\eta=\chi$.
The higher efficiency of differential gradient diffusion
for $\eta=\chi=1$ thus becomes relevant for such $k$.

When the wavelength increases, the growing
destabilizing contribution of diffusion at the
similarity boundary is increasingly opposed by the respective
enhancement of only the second of the two above dissipation
mechanisms. Both these counter effects are commensurately
augmented by the growing wavelength. The enhancement of
overall dissipation for $\eta=\chi=1$ (with respect
to $\eta=\chi=0$) would thus have to remain of
a limited relative significance compared to
the respectively higher efficiency of
differential gradient diffusion.

For sufficiently long waves, therefore, it is the efficiency
of differential gradient diffusion that specifies at which value
of $\eta=\chi$ the instability sets in first: $\delta Ra_{c}(k)>0$
[Fig. \ref{f:mscot0}(b)]. For a fixed wavelength, in addition,
the optimal frequency with which the convective cells change
their sense of rotation has to be largely specified by the
background gradient of the stably stratified component.
For $\mu=1$, therefore, $\delta\omega_{c}(k)>0$
for such $k$.

When the instability wavelength decreases, the
effect of the first dissipation mechanism is enhanced,
due to the across-slot motion being augmented. The role of
the overall dissipation disparity between $\eta=\chi=1$ and
$\eta=\chi=0$ then grows with respect to the (diminishing) effect
of diffusion at the similarity boundary. $|\delta Ra_{c}(k)|$
and $|\delta\omega_{c}(k)|$ thus decay with increasing
$k$ [Fig. \ref{f:mscot0}(b)].

However, the dissipation enhancement for
$\eta=\chi=1$ eventually dominates the respectively
higher efficiency of differential diffusion underlying it.
$\delta Ra_{c}(k)$ thus becomes negative [Fig. \ref{f:mscot0}(b)],
and then so does $\delta\omega_{c}(k)$. The overall efficiency of
the combined effects of diffusion and dissipation then becomes
higher for $\eta=\chi=0$. An additional dissipation arising above
the $k$ where $\delta Ra_{c}(k)=0$ would therefore be greater
for $\eta=\chi=0$ than for $\eta=\chi=1$. When the relative
role of the disparity between such additional dissipations
is augmented sufficiently with $k$ increasing further,
the sign of $\delta Ra_{c}(k)$ changes again.

The oscillatory nature of decay of $\delta Ra_{c}(k)$
and $\delta\omega_{c}(k)$ in Fig. \ref{f:mscot0}(b) is
thus associated with the additional dissipation arising
above a critical value of $k$ where $\delta Ra_{c}(k)=0$
acting against the increase of $|\delta Ra_{c}(k)|$. The
value of $\eta=\chi$ at which the overall efficiency
of diffusion and dissipation becomes higher above
such a critical $k$ is also characterized
by more dissipation.

The values of $k$ where $\delta\omega_{c}(k)=0$ in Fig. \ref{f:mscot0}(b)
slightly exceed the respective $k$ where $\delta Ra_{c}(k)=0$. A comparatively
more time for diffusion and dissipation is thus provided for the value of
$\eta=\chi$ that has just [when $\delta Ra_{c}(k)=0$] become a less
efficient combination of these processes. Over a short interval
of $k$, this outweighs the mismatch between the signs of
$\delta\omega_{c}(k)$ and $\delta Ra_{c}(k)$. 

Unreported herein, the values of $k$ at which $\delta Ra_{c}(k)$ and
$\delta\omega_{c}(k)$ change sign for stress-free boundary conditions
were found to be respectively smaller than those for the no-slip conditions.
This could be associated with the higher no-slip $Ra_{c}(k)$ being more
important for delaying the effects of dissipation than the decrease of
(along-slot) dissipation due to the stress-free boundaries.
\paragraph{\label{s:rovh01}$\eta=0$ and $\chi=1$.}
Oscillatory instability for $\eta=0$ and $\chi=1$ at $\theta=0$
($\theta=\pi$) arises where the fluid is already unstable to steady
disturbances (Fig. \ref{f:mscsov01}). It is however instructive to
comment on its $Ra_{c}(k)$ and $\omega_{c}(k)$ compared to those for
$\eta=\chi$. For $\eta=0$ and $\chi=1$, the oscillatory perturbation
has to be localized near the distinction (inverse) boundary. With
respect to $\eta=\chi$, this implies a higher efficiency of
differential gradient diffusion at this boundary. The relative
portion of moving fluid particles with the generated
horizontal density differences is larger for
$\eta=0$ and $\chi=1$ than for $\eta=\chi$.

For $\eta=0$ and $\chi=1$, however, differential gradient diffusion
at the inverse (distinction) boundary opposes amplitude growth of a
convective cell whose sense of rotation changes periodically. This
opposition is dominant for relatively long wavelengths, whereas the
higher efficiency of oscillatory across-slot motions is more important
for the short waves. In viscous fluid, such a higher efficiency at the
large $k$ for $\eta=0$ and $\chi=1$ is also expected to generate a relatively
higher dissipation. This would however be offset by the across-slot
cell path being shorter for $\eta=0$ and $\chi=1$ than for
$\eta=\chi$. The effects of dissipation thus have to be
relatively little important in the present context.

As $k$ exceeds a certain value, $Ra_{c}(k)$ for
$\eta=0$ and $\chi=1$ thus changes from being larger
to being smaller than that at either of $\eta=\chi$.
For inviscid fluid and for viscous fluid with no-slip
and stress-free boundary conditions, such a value is between
$4.9$ and $5.3$. For $\omega_{c}(k)$, such a value is between
$5.6$ and $5.8$. [The data for most such cases are depicted
in Figs. \ref{f:mscoi00}(a) and \ref{f:mscoi11}(a), partly in
Fig. \ref{f:msco01i2D}, as well as in Figs. \ref{f:mscsov01}
and \ref{f:mscot0}(a).] The larger critical values of $k$ for
$\omega_{c}(k)$ further relatively diminish the effect of
differential gradient diffusion at the inverse (distinction)
boundary. Until then, this effect remains more important
than the match between $\omega_{c}(k)$ and the
respective stable solute stratification.
\subsubsection{\label{s:rovv}$\theta\in[0,\pi/2]$}
\paragraph{\label{s:rovvg}General.}
The marginal-stability curves
in Fig. \ref{f:mscot0}(a) meet conditions I
and II emphasized in Sec. \ref{s:roiv11} above. Here
condition II holds due to the second dissipation mechanism
(Sec. \ref{s:rovhDD}) for both $\eta=\chi$ and also due to
solute diffusion at the similarity boundary for $\eta=\chi=1$.
3D disturbances are thus the first to arise near $k_{y}=0$ for
small $\theta>0$ [Figs. \ref{f:mscof}(a) and \ref{f:mscod}(a)].
Their dominance intervals of $k_{y}$ are shorter than those for
$\eta=1$ and $\chi=0$ \cite{r:tap}. This results from a less
favorable combination of the effects of two gravity components.
The 2D instability for $\eta=1$ and $\chi=0$ sets in at $\theta=0$
before and at $\theta=\pi/2$ after those for $\eta=\chi$. In
particular, the pattern of two counter-traveling waves for
$\eta=1$ and $\chi=0$ at $\theta=\pi/2$ (Fig. 7 in
\cite{r:tap}) generates more dissipation than
such a single-wave pattern for $\eta=\chi$
considered below.

As for $\eta=1$ and $\chi=0$, however, the intervals of
$k_{y}$ with 3D most unstable disturbances vanish for $\eta=\chi$
only when $\theta=\pi/2$ (Figs. \ref{f:mscof} and \ref{f:mscod}). With
$Re[d\tilde{u}(k)/dx]_{|\theta=0}\asymp{\omega_{c}(k)}_{|\theta=0}\asymp 1$ and
$Re[d\tilde{u}(k,0)/dx]_{|\theta=\pi/2}\asymp{\omega_{c}(k,0)}_{|\theta=\pi/2}\asymp 1$
as $k\rightarrow 0$ [Figs. \ref{f:mscot0}(a) and \ref{f:mscoddff}(a)], the general
solution for $\tilde{u}(x)$ from Eqs. (\ref{eq:lsu}) and (\ref{eq:lsts}) suggests
that (Sec. \ref{s:roiv01_3D3Z})
$[R_{0}^{\pi/2}(k)/G]_{|k\rightarrow 0}\asymp(k/G)_{|k\rightarrow 0}>0
\Longleftrightarrow G_{|k\rightarrow 0}=O(k)$($\rightarrow 0$).
Arising from the across-slot gravity alone, the
three-dimensionality is thus retained by a sufficiently
small $G=k_{y}/k$ so long as $\theta<\pi/2$. It could also
be viewed as coming from conditions I and II at $G=0$.
As $k_{y}\rightarrow 0$ for a fixed $k_{z}>0$, in particular, the
$Ra_{c}(k_{y},k_{z})\cos\theta$ and $\omega_{c}(k_{y},k_{z})$ tend
to the respective 2D $Ra_{c}(k_{z})$ and $\omega_{c}(k_{z})$ at $\theta=0$
[Figs. \ref{f:mscot0}(a), \ref{f:mscof}, and \ref{f:mscod}].

In view of transformation (\ref{eq:tt01}),
$Ra_{c}(k_{y},k_{z})$ and $\omega_{c}(k_{y},k_{z})$
for $\eta=0$ and $\chi=1$ when $\theta\in[0,\pi/2)$
are addressed when $\theta\in(\pi/2,\pi]$ is considered
in Sec. \ref{s:rovvh} below. Since Eq. (\ref{eq:inv}) also
applies to viscous fluid, the 2D disturbances must be most
unstable at $\theta=\pi/2$ for any $\eta$ and $\chi$. This
is seen from Figs. \ref{f:mscoddff}(a) and \ref{f:mscodf}(a).
The former figure is independent of $\eta=\chi$. The respective
instability mechanisms for $\eta=\chi$ thus have to be such
as are essentially due only to the distinction boundary.
By virtue of Eq. (\ref{eq:inv}), this could be discussed
in the 2D framework alone. The mechanisms of 2D
instability at $\theta=\pi/2$ for $\eta=\chi$
are thus considered below along with such a
mechanism for $\eta=0$ and $\chi=1$.
\paragraph{\label{s:rovvim}2D instability mechanisms for $\theta=\pi/2$.}
The patterns illustrated in Figs. \ref{f:perv00},
\ref{f:perv11}, and \ref{f:perv01} are traveling waves
propagating in the gravity direction. Their nature can be
clarified if one assumes that in the marginally unstable state,
the speeds of their propagation adequately match the respective
convective velocities oriented downwards. In the reference frame
moving with such a pattern, the component perturbations are
then transported vertically only upwards. This is the
direction where the convective velocities are
augmented by the moving reference frame.

The slot area behind such a propagating clockwise-(counterclockwise-)rotating
cell is supplied with the temperature and solute perturbations from the region
near the left (right) sidewall. The background component values are relatively
high (low) there. Occupied by a counterclockwise-(clockwise-)rotating cell,
such an area is thus largely characterized by the positive (negative)
component perturbations (Figs. \ref{f:perv00}, \ref{f:perv11},
and \ref{f:perv01}).

Due to differential gradient diffusion,
the density near the distinction boundary to the left
is specified mainly by the solute perturbation there. At
the similarity boundary to the right, however, the temperature
and solute isolines behave identically to each other for either
$\eta=\chi$ (Figs. \ref{f:perv00} and \ref{f:perv11}). The density excess
there thus has to be zero. (Due to the fixed solute sidewall values
at $y=0$, this holds only approximately in Fig. \ref{f:perv00}.)
For $\eta=0$ and $\chi=1$ (Fig. \ref{f:perv01}), differential
gradient diffusion at the inverse boundary to the right
results in the density there being specified mainly
by the respective temperature perturbation.

A propagating convective cell that rotates
counterclockwise (clockwise) is thus characterized by positive
(negative) horizontal density differences between the streamline regions
at the left and right sidewalls. This is what drives convective motion
for such a cell. For $\eta=\chi$ (Figs. \ref{f:perv00} and \ref{f:perv11}),
such differences arise mainly from the positive (negative) solute
perturbation at the left sidewall and zero density perturbation at
the right boundary. For $\eta=0$ and $\chi=1$ (Fig. \ref{f:perv01}),
they are mainly due to the positive (negative) solute and temperature
perturbations at the left and right sidewalls, respectively. 

Such a downwards-propagating counterclockwise-(clockwise-)rotating
convective cell also has its component perturbations near the left
(right) sidewall practically steady. The convective-cell velocities
there are largely offset by the speed of propagation. There is thus a steady
horizontal density difference that maintains the downwards propagation of such
a flow pattern. For $\eta=\chi$ (Figs. \ref{f:perv00} and \ref{f:perv11}), this
difference is specified by the positive left-sidewall solute perturbation and
zero right-sidewall density perturbation for a counterclockwise-rotating
cell. For $\eta=0$ and $\chi=1$ (Fig. \ref{f:perv01}), the steady
density difference comes from the cells rotating in both senses.
It is due to the positive (negative) left-(right-) and zero
right-(left-)sidewall solute (temperature) perturbation
for a counterclockwise-(clockwise-)rotating cell. 
\paragraph{\label{s:rovvms}Marginal-stability curves.}
For any pair of $\eta$ and $\chi$ just considered at $\theta=\pi/2$, the
propagating 2D convective pattern gives rise to such a distribution
of the component perturbations as favors its convective motion and
maintains its direction of propagation. This is implemented due to
differential gradient diffusion. Such a process takes place at
the distinction sidewall alone for $\eta=\chi$ and at both
vertical boundaries for $\eta=0$ and $\chi=1$.

For any $\eta$ and $\chi$, the instability mechanism at $\theta=\pi/2$
is underlain by horizontal density differences accompanying
the flow pattern. The independence of 2D $Ra_{c}(k_{y})$ and
$\omega_{c}(k_{y})$ in Fig. \ref{f:mscoddff}(a) of the value of
$\eta=\chi$ is therefore just a manifestation of the component
disparity at the distinction boundary being unaffected by the
orientation of component isolines at the similarity sidewall.
Eq. (\ref{eq:inv}) then suggests that such independence is
retained for $k_{z}>0$ as well. Since the increase of $k_{z}$
for a fixed $k_{y}$ decreases the overall wavelength,
$\omega_{c}(k_{y})$ in Figs. \ref{f:mscoddff}(a) and
\ref{f:mscodf}(a) also increases with $k_{z}$. 

Convective cells driven by such across-slot density
differences between their vertically moving particles make
their along-slot dissipation a part of the instability feedback.
Analogously to the second dissipation mechanism for standing-wave
perturbation at $\theta=0$, discussed in Sec. \ref{s:rovhDD} above,
this causes an infinite growth of 2D $Ra_{c}(k_{y})$ with $k_{y}$
decreasing to $0$ at $\theta=\pi/2$. At $\theta=0$, such a growth for $\eta=\chi=1$
is also caused by solute diffusion at the similarity boundary.
For $\eta=\chi$, the 2D marginal-stability curves at $\theta=0$
[Fig. \ref{f:mscot0}(a)] are thus smoothly transformed into those at $\theta=\pi/2$
[Figs. \ref{f:mscof},\ref{f:mscod}, and \ref{f:mscoddff}(a)]. For $\eta=0$ and
$\chi=1$ at $\theta=\pi/2$, the increase of 2D $Ra_{c}(k_{y})$ near $k_{y}=0$
is also due to neutralization of either component by diffusion. Such a
marginal-stability curve for $\theta$ decreasing from $\pi/2$
to $0$ is discussed in Sec. \ref{s:2tp01} below.

Inversely affecting the component
perturbations at the vertical sidewalls,
differential gradient diffusion for $\eta=0$
and $\chi=1$ still leads to the respective horizontal
density differences augmenting each other by their
superposition. The instability for these $\eta$ and
$\chi$ [Fig. \ref{f:mscodf}(a)] thus sets in substantially
before that for $\eta=\chi$ [Fig. \ref{f:mscoddff}(a)].
Such combination of the effects of differential diffusion
at the boundaries also results in the convection
pattern (Fig. \ref{f:perv01}) without a slope
between its across-slot motion and the
horizontal axis. 

Such horizontality of across-slot motion, however,
makes the overall cell path shorter for $\eta=0$ and
$\chi=1$ than for $\eta=\chi$. With respect to the relative
streamfunction amplitudes, therefore, the maximal horizontal
density differences in Fig. \ref{f:perv01} exceed those in
Figs. \ref{f:perv00} and \ref{f:perv11}. The intensity of
convection in the marginally unstable state has to be matched
by the speed of pattern propagation and thus by the associated
$\omega_{c}(k_{y})$. This explains why $\omega_{c}(k_{y})$ are
respectively smaller for $\eta=0$ and $\chi=1$ [Fig. \ref{f:mscodf}(a)],
where convective motion is relatively weaker, than for $\eta=\chi$
[Fig. \ref{f:mscoddff}(a)]. Also consistent with this argument
is that the difference between such $\omega_{c}(k_{y})$
grows with $k_{y}$.
\subsubsection{\label{s:rovvh}$\theta\in[\pi/2,\pi]$}
\paragraph{\label{s:2tpG}General on 2D disturbances.}
For $\theta=\pi$, the across-slot gravity
is manifested as a steady instability \cite{r:twh,r:tbc}. For the values of $\eta$
and $\chi$ considered in this study, such an instability is discussed in Sec. \ref{s:rsld}
below. Also characterized by steadily rotating convective cells, the 2D traveling
waves arising at $\theta=\pi/2$ thus have to transform into the respective steady
disturbances when $\theta$ increases above $\pi/2$. The speed of flow-pattern
propagation, $\omega_{c}(k_{y})/k_{y}$, then has to decrease to $0$ for all
$k_{y}$. For the reflectionally symmetric pattern of two counter-propagating
waves arising at $\theta=\pi/2$ when $\eta=1$ and $\chi=0$ \cite{r:tap},
such a decrease begins with certain $k_{y}$
at $\theta\approx 1.5\pi/2$. 

For the present values of $\eta$ and $\chi$, however,
the mechanisms described in Sec. \ref{s:rovvim} above have to
retain a flow-pattern propagation for any $\theta\in[\pi/2,\pi)$.
The decrease of $\omega_{c}(k_{y})$ to $0$ for any $k_{y}$($>0$) can thus
take place only when $\theta=\pi$, as in Fig. \ref{f:mscoddff}(b) and (c) and in
Fig. \ref{f:mscodf}(b). The phases of nonzero component perturbations in Figs. \ref{f:perv00},
\ref{f:perv11}, and \ref{f:perv01} also nearly coincide with the streamfunction phases 
of the opposite sign. This is due to the traveling nature of such convective patterns 
(Sec. \ref{s:rovvim}). When $\omega_{c}(k_{y})$ turns $0$ with $\theta$ reaching
$\pi$, however, these relative phases have to become shifted by a quarter of
the wavelength with respect to each other, as in Fig. 2 of \cite{r:tbc}
and in Fig. \ref{f:perhs01}(a). 
\paragraph{\label{s:2tp0011}2D disturbances for $\eta=\chi$.}
For $\theta=\pi/2$, the along-slot part of dissipation reduces efficiency of
the instability feedback. This does not apply to $\theta=\pi$, where only the
first of the two dissipation mechanisms accentuated in Sec. \ref{s:rovhDD} above
is relevant. In the framework of this first mechanism, however, the across-slot
part of dissipation for $\eta=\chi$ is also larger at $\theta=\pi/2$ than at
$\theta=\pi$. This is due to the slope of such an across-slot cell motion at
$\theta=\pi/2$ (Figs. \ref{f:perv00} and \ref{f:perv11}).
For $\eta=\chi=0$, 2D $Ra_{c}(k_{y})$ is thus higher at $\theta\in[\pi/2,\pi)$
than at $\theta=\pi$ for any $k_{y}$ [Fig. \ref{f:mscoddff}(a) and (b)]. The difference
between such $Ra_{c}(k_{y})$ also has to be infinite at any $\theta\in[\pi/2,\pi)$ as
$k_{y}$[$=o(\pi-\theta)$]$\rightarrow 0$ for any ${\omega_{c}(k_{y})}_{|k_{y}\rightarrow 0}$
in Fig. \ref{f:mscoddff}(b), due to the effect of along-slot dissipation. Compared to
${\omega_{c}(k_{y})}_{|k_{y}\rightarrow 0}\rightarrow 0$, however, such effect is
moderated when ${\omega_{c}(k_{y})}_{|k_{y}\rightarrow 0}\asymp 1$.

As $\theta$ exceeds $\pi/2$, in particular, two joined marginal-stability branches
with ${\omega_{c}(k_{y})}_{|k_{y}\rightarrow 0}\rightarrow 0$ (due to the across-slot
gravity) also isolatedly arise from $k_{y}=0$ [Fig. \ref{f:mscoddff}(b), $\theta=1.4\pi/2$].
Their higher $Ra_{c}(k_{y})$ has the slightly higher $\omega_{c}(k_{y})$. Both
their $Ra_{c}(k_{y})$ are smaller and larger than that of the main branch [whose
${\omega_{c}(k_{y})}_{|k_{y}\rightarrow 0}\asymp 1$, due to the along-slot gravity]
at some $k_{y}>0$ and at $k_{y}\rightarrow 0$, respectively.
[${Ra_{c}(k_{y})}_{|k_{y}\rightarrow 0}\rightarrow\infty$ is assumed for the smallest
$\omega_{c}(k_{y})$ as well.] Growing with $\theta$, the higher- and lower-$\omega_{c}$
branches with ${\omega_{c}(k_{y})}_{|k_{y}\rightarrow 0}\rightarrow 0$ meet the main
branch at a finite $k_{y}$ and unfold with its smaller- and larger-$k_{y}$ intervals,
respectively [Fig. \ref{f:mscoddff}(b), $\theta\geq 1.5\pi/2$]. The branch with 
${\omega_{c}(k_{y})}_{|k_{y}\rightarrow 0}\asymp 1$ then exists
only below such a finite $k_{y}$ as decreases
to $0$ with $\pi-\theta$.

For $\eta=\chi=1$, solute diffusion at the similarity boundary increases
2D $Ra_{c}(k_{y})$ infinitely as $k_{y}$ decreases to $0$ at $\theta=\pi$
[Fig. \ref{f:mscoddff}(c)] (Sec. \ref{s:rsld11} below). At $\theta=\pi/2$,
such $Ra_{c}(k_{y})$ is still independent of $\eta=\chi$ [Fig. \ref{f:mscoddff}(a)].
Its infinite increase for $k_{y}\rightarrow 0$ is then due only to the above
role of along-slot dissipation. When $k_{y}$ decreases, however, the relative
portion of streamline particles whose horizontal density differences drive a
convective cell grows at $\theta=\pi/2$ compared to $\theta=\pi$. This efficiency
factor eventually dominates the discrepancy between the growing stabilizing effect
of along-slot dissipation and that of solute diffusion at the similarity
boundary. The 2D $Ra_{c}(k_{y})$ near $k_{y}=0$ thus becomes smaller at
$\theta=\pi/2$ [Fig. \ref{f:mscoddff}(a)] than at $\theta=\pi$
[Fig. \ref{f:mscoddff}(c)]. With $\theta$ increasing from
$\pi/2$, therefore, the $Ra_{c}(k_{y})$ grows for such
very small $k_{y}$ and decreases elsewhere to
transform into its values at $\theta=\pi$.
\paragraph{\label{s:2tp01}2D disturbances for $\eta=0$ and $\chi=1$.}
For $\eta=0$ and $\chi=1$, the across-slot gravity when
$\theta=\pi$ ($\theta=0$) favors and opposes growth of steadily
rotating convective cells at the distinction (inverse) and inverse
(distinction) boundaries, respectively. Its effect is manifested in
terms of respective along-slot density differences arising between the
streamline particles moving across the slot close to these boundaries.
For $\theta\in(\pi/2,\pi)$ [$\theta\in(0,\pi/2)$], it is combined
with the effect of along-slot gravity. The latter effect is
manifested (Fig. \ref{f:perv01}) via across-slot density
differences between the streamline particles moving
along the boundaries. As 2D $k_{y}\rightarrow 0$,
then, $Ra_{c}(k_{y})_{|\theta=\pi/2}<Ra_{c}(k_{y})_{|\theta=\pi,0}$
[Fig. \ref{f:mscodf}(b)]. 

Since the instability for $\theta=\pi/2$ is also due to differential
gradient diffusion at both sidewalls, 2D $Ra_{c}(k_{y})$ for such $\theta$
is smaller than that of steady instability for $\theta=\pi$ ($\theta=0$)
so long as the wavelength is large enough for the effect of diffusion to
be dominant [Fig. \ref{f:mscodf}(b)]. At $\theta=\pi$ ($\theta=0$), however,
the steadily rotating convective cells are localized near the distinction
(inverse) boundary [Fig. \ref{f:perhs01}(a)]. The across-slot part of their
dissipation thus decreases compared to $\theta=\pi/2$. With increasing
$k_{y}$, the relative role of the streamline particles with along-slot
density differences is also enhanced with respect to that with
across-slot density differences. For sufficiently large $k_{y}$
($k_{y}\geq 5.3$), therefore, the $Ra_{c}(k_{y})$ for
$\theta=\pi$ ($\theta=0$) is smaller than that
for $\theta=\pi/2$ [Fig. \ref{f:mscodf}(b)].

As $\theta$ changes from $\pi/2$ to $\pi$ ($0$), the lost contribution
of along-slot gravity to the intensity of the steadily rotating cells
is replaced by the mutually opposing effects of across-slot gravity at
the distinction and inverse boundaries. This weakens such convective
motion. To maintain the remaining effect of along-slot gravity, the
speed of propagation of the flow pattern has to match the relative
intensity of convection. $\omega_{c}(k_{y})$ then decreases. This
shifts the relative phases of component and flow perturbations
with respect to each other. The portion of streamline particles
with favorable across-slot density differences thus decreases,
and the relative effect of along-slot gravity weakens further.
The role of this feedback depends on the wavelength
and $\theta$.

If the wavelength is too short for retaining a sufficiently
long slot interval where the across-slot density differences matter,
an efficient utilization of the along-slot gravity becomes impossible.
As a consequence, the perturbation of such a wavelength occupying the
whole width of the slot cannot persist. It then gives way to a
convective pattern driven mainly by the across-slot gravity. Localized
near the distinction (inverse) boundary, such a pattern has differently
behaving $Ra_{c}(k_{y})$. These just exceed the $Ra_{c}(k_{y})$ at
$\theta=\pi$ ($\theta=0$) for the relatively large $k_{y}$ in
Fig. \ref{f:mscodf}(b). Mainly underlain by the (steady)
effect of across-slot gravity, such a pattern is also
characterized by $\omega_{c}(k_{y})$ [the large-$k_{y}$
$\omega_{c}(k_{y})$ in Fig. \ref{f:mscodf}(b)]
that are substantially smaller than
those at $\theta=\pi/2$.

For sufficiently large $|\theta-\pi/2|$, comparatively abrupt changes in
$\partial Ra_{c}/\partial k_{y}$ and $\partial\omega_{c}/\partial k_{y}$ are
distinguishable in Fig. \ref{f:mscodf}(b). Such changes are a manifestation of
the transition from the (relatively long-wavelength) convective pattern largely
driven by the along-slot gravity to the localized (relatively short-wavelength)
pattern mainly driven by the across-slot gravity. The longer the wavelength is (the
larger is the ratio between the portions of streamline particles with across-slot
and along-slot density differences) the more capable its convective pattern is of
accommodating the effects of across-slot gravity without destroying the mechanism
by means of which the along-slot gravity drives such convection. The abrupt
changes thus arise at the smaller $k_{y}$ the more $|\theta-\pi/2|$ exceeds
$0$. Their $k_{y}$ in Fig. \ref{f:mscodf}(b) also tends to zero
with $\theta$ approaching $\pi$ ($0$).
\paragraph{\label{s:3tp11}3D disturbances for $\eta=\chi=1$.}
For $\eta=\chi=1$, solute diffusion at the
similarity boundary at $\theta=\pi$ results in the $Ra_{c}(k)$
for steady instability rising to infinity with $k$ decreasing to $0$
[Fig. \ref{f:mscoddff}(c)], as also discussed in Sec. \ref{s:rsld11} below. 
Conditions I and II (Sec. \ref{s:roiv11}) for three-dimensionality of the
instability at small $\pi-\theta>0$ are thus met. That the instability at
$\theta<\pi$ is oscillatory is consistent with the corresponding mechanism
for its three-dimensionality. Such oscillatory instability with small
$\omega_{c}>0$ is a perturbation of the steady instability
($\omega_{c}=0$) at $\theta=\pi$ that preserves the
effects of conditions I and II.

3D oscillatory disturbances ($k_{z}>0$) are thus most
unstable near $k_{y}=0$ for small $\pi-\theta>0$ [Fig. \ref{f:ttp11}(a)].
A sufficiently small $G=k_{y}/k$ in Eqs. (\ref{eq:lsu}) and (\ref{eq:lsts}) also
makes the 3D effect of across-slot gravity dominant in the respectively small vicinity
of $k_{y}=0$ so long as $\theta\neq\pi/2$ [Fig. \ref{f:ttp11}(b),(c)]. Indeed, with
$Re[d\tilde{u}(k,0)/dx]_{|\theta=\pi-o(k)}\asymp{\omega_{c}(k,0)}_{|\theta=\pi-o(k)}\asymp 1$
and $Re[d\tilde{u}(k,0)/dx]_{|\theta=\pi/2}\asymp{\omega_{c}(k,0)}_{|\theta=\pi/2}\asymp 1$
as $k\rightarrow 0$ [Fig. \ref{f:mscoddff}(a) and (c)], $G_{|k\rightarrow 0}=O(k)$($\rightarrow 0$)
$\Longleftrightarrow[R_{\pi-o(k)}^{\pi/2}(k)/G]_{|k\rightarrow 0}
\asymp[k/G]_{|k\rightarrow 0}>0$.
The three-dimensionality for any $\theta\in(\pi/2,\pi)$ could be viewed as
coming from conditions I and II at $G=0$ as well. As $k_{y}\rightarrow 0$
for a fixed $k_{z}>0$, in particular, the $Ra_{c}(k_{y},k_{z})|\cos\theta|$
tends to the 2D $Ra_{c}(k_{z})$ at $\theta=\pi$ [Figs. \ref{f:mscoddff}(c)
and \ref{f:ttp11}] with the ${\omega_{c}(k_{y},k_{z})}_{|k_{y}\rightarrow 0}\rightarrow 0$
(Fig. \ref{f:ttp11}). This effect vanishes only when
$\theta=\pi/2$, where Eq. (\ref{eq:inv}) applies.
\paragraph{\label{s:3tp00}3D disturbances for $\eta=\chi=0$.}
For $\eta=\chi=0$, a vicinity of $k_{y}=0$ in
Fig. \ref{f:ttp00} is still dominated by 3D disturbances, despite
condition II being not met at $\theta=\pi$ [Fig. \ref{f:mscoddff}(b)].
The role of 3D disturbances with small $k_{y}$ in Fig. \ref{f:ttp00} is
also enhanced when $\theta$ decreases from $\pi$. Growing with $\pi-\theta$,
the effect of along-slot dissipation on the feedback efficiency increasingly heightens
the 2D $Ra_{c}(k_{y})$ in the vicinity of $k_{y}=0$. The three-dimensionality then comes
from a much more unstable behavior of such $Ra_{c}(k_{y})$ at $\theta=\pi$, since the
latter behavior could be mimicked by $Ra_{c}(k_{y},k_{z})_{|\theta\in(\pi/2,\pi)}$
for small $G=k_{y}/k$.

For any $k_{z}>0$, there have to exist such
$k_{y}$ as make $G$ so small that the along-slot gravity
in Eqs. (\ref{eq:lsu}) and (\ref{eq:lsts}) be negligible. The axes of cell
rotation are then nearly parallel to the along-slot gravity. For these $k_{y}$,
$\omega_{c}(k_{y},k_{z})$ and $Ra_{c}(k_{y},k_{z})|\cos\theta|$ at $\theta\in(\pi/2,\pi)$ are
respectively approximated by $0$ and 2D ${Ra_{c}(k)}_{|\theta=\pi}$. In Fig. \ref{f:mscoddff}(b),
2D ${Ra_{c}(k)}_{|\theta=\pi}$ reaches its minimum, $720$, as $k\rightarrow 0$ \cite{r:tap,r:n}
(Sec. \ref{s:rsld00} below). For such $G$ and small enough $k_{z}$, the
$Ra_{c}(k_{y},k_{z})\approx Ra_{c}(0,k_{z})\approx Ra_{c}(k_{z},0)\approx 720/|\cos\theta|$.
It is thus smaller than the smallest 2D ${Ra_{c}(k_{y})}_{|k_{y}\rightarrow 0}$($\rightarrow\infty$)
at $\theta\in(\pi/2,\pi)$ [Fig. \ref{f:ttp00}(a),(b), Sec. \ref{s:2tp0011}].

The more $\theta$ initially decreases from $\pi$ the greater the smallest 2D
$Ra_{c}(k_{y})$ near $k_{y}=0$ diverges from such $Ra_{c}(k_{y})_{|\theta=\pi}/|\cos\theta|$
[Fig. \ref{f:mscoddff}(b), $\theta\geq 1.5\pi/2$]. However, the neutralization of
along-slot gravity by small $G_{|k_{z}>0}$ also makes the 3D analogue of the 2D branch
with $\omega_{c}(k_{y})_{|k_{y}\rightarrow 0}\asymp 1$ to additionally connect to that of
the higher-$\omega_{c}$ 2D branch with $\omega_{c}(k_{y})_{|k_{y}\rightarrow 0}\rightarrow 0$
[Fig. \ref{f:ttp00}(a) and (b)]. For $k_{z}>0$, closed contours of finite $(Ra_{c},k_{y})$
and $(\omega_{c},k_{y})$ thus form at these $\theta$. Growing with $\pi-\theta$, such
contours collide with the respective smaller 3D $Ra_{c}(k_{y})$ and $\omega_{c}(k_{y})$
around $\theta\approx 1.4\pi/2$ (at the smaller $\pi-\theta$ the smaller $k_{z}$ is,
starting from the reorganization of the 2D branches). Via such a 3D collision, a
structure with two connected limit points unfolds [Fig. \ref{f:ttp00}(c)]. Its
hysteresis reconciles the small- and large-$G$ small-$k_{z}$
branches of disparate 3D nature. 

For small $k_{z}$, in particular, $Ra_{c}(k_{y},k_{z})$
and $\omega_{c}(k_{y},k_{z})$ are approximated by the $Ra_{c}(k_{y},0)$ and
$\omega_{c}(k_{y},0)$ at the same $\theta$. Such are the upper branches in
Fig. \ref{f:ttp00}(c). They fail to exist only for such small $G$ (i.e.,
$k_{y}$ at a fixed $k_{z}$) as make the 3D effect of along-slot gravity
relatively negligible. Largely triggering convection by the across-slot
gravity alone, the small-$G$ nearly-steady lower branches in
Fig. \ref{f:ttp00}(c) fail to exist when the increase of
$G$ (i.e., of $k_{y}$ at a fixed $k_{z}$) transforms
the convective pattern so that it ought to be
driven by the along-slot gravity as well.
Some quantitative details are given
in Table \ref{t:lul}.

With $k$ decreasing along the upper branch, the 2D effect of
along-slot gravity grows with respect to that of across-slot gravity,
according to the relative roles of the across-slot and along-slot density differences.
Neutralization of such along-slot gravity in (3D) Eqs. (\ref{eq:lsu}) and (\ref{eq:lsts})
would thus require $G$ to decrease with $k$ at the upper limit point. With $dG/dk>0$ at
this limit point, $dk_{y}/dk_{z}>k_{y}/k_{z}$, analogously to (\ref{eq:dky}) following
from (\ref{eq:dk3}).

Along the (lower) branch largely specified by $k$ alone, however, 
$\partial G/\partial k>0\Longleftrightarrow dG/dk>0$. Arising due to the across-slot
gravity alone, the lower branch thus reaches its upper bound when $G$ becomes so large that
the latter inequality fails. At this bound then $dG/dk\leq 0$, as discussed in Sec. 4.1.2
of \cite{r:tap} for $|\tan\theta|$ and the exactly steady 2D analogue of such a branch at
$\eta=1$ and $\chi=0$. For the lower limit point, therefore, $dG/dk_{z}\leq 0$ and
$dk_{y}/dk_{z}\leq k_{y}/k_{z}$. This restriction leads to vanishing of the hysteresis
region with increasing $k_{z}$ [Fig. \ref{f:ttp00}(c) and Table \ref{t:lul}]. Such a region
also vanishes as $\theta$ approaches $\pi/2$. For $k_{z}$($>0$) being fixed, in particular,
Eq. (\ref{eq:kytt}) at both limit points implies $\partial k_{y}(k_{z},\theta)/\partial\theta>0$
for the relevant $\theta>\pi/2$ (Table \ref{t:lul}). For $\eta=1$ and $\chi=0$
(Fig. 10 in \cite{r:tap}), the exactly steady analogue of the current
small-$G$ branch also shrinks to $k=0$ as $\theta$ reaches $\pi/2$.
\vspace*{-0.3cm}\paragraph{\label{s:3tp01}3D disturbances for $\eta=0$ and $\chi=1$.}
As discussed in Sec. \ref{s:2tp01} above, the relatively
abrupt changes in 2D $\partial Ra_{c}/\partial k_{y}$ and
$\partial \omega_{c}/\partial k_{y}$ for $\eta=0$ and $\chi=1$
are due to a switching between the primary roles of the two gravity
components in the nature of the perturbation. They are thus specified
by $\theta$ alone. With $k$ substituting for $k_{y}$, the 3D effect of
along-slot gravity decreases with respect to that of across-slot gravity,
due to the emergence of $G$($<1$) in Eqs. (\ref{eq:lsu}) and (\ref{eq:lsts}). The
3D boundary at which the nature of perturbation changes, $k_{y}^{\theta}(k_{z})$, thus
deviates downwards from $q_{y}^{\theta}(k_{z})\equiv[{k_{y}^{\theta}(0)}^{2}-k_{z}^{2}]^{1/2}$.
This is seen from Fig. \ref{f:ttp01}(a) and (b) and from Table \ref{t:jk}. When the change
in 2D $\partial Ra_{c}/\partial k_{y}$ is steep enough [Fig. \ref{f:ttp01}(a)],
such a deviation leads to a short interval of $k_{y}$ with 3D most unstable
disturbances separated from $k_{y}=0$.

Conditions I and II are met for the steady
marginal-stability curve at $\theta=\pi$ ($\theta=0$)
[Figs. \ref{f:mscsov01} and \ref{f:mscodf}(b)]. Here condition
II holds due to the opposition provided by the effect of component
conditions at the inverse (distinction) boundary to a steady cell rotation.
For small $\pi-\theta$ ($\theta$) [Fig. \ref{f:ttp01}(a)], 3D slow-propagating
cells arising from the steady disturbances [localized at the distinction (inverse)
boundary] are thus most unstable near $k_{y}=0$.

With $Re[d\tilde{u}(k,0)/dx]_{|\theta=\pi-o(k)}\asymp Re[d\tilde{u}(k,0)/dx]_{|\theta=o(k)}
\asymp{\omega_{c}(k,0)}_{|\theta=\pi-o(k)}={\omega_{c}(k,0)}_{|\theta=o(k)}\asymp 1$
and $Re[d\tilde{u}(k,0)/dx]_{|\theta=\pi/2}\asymp{\omega_{c}(k,0)}_{|\theta=\pi/2}\asymp 1$
as $k\rightarrow 0$ [Fig. \ref{f:mscodf}(b)], $G_{|k\rightarrow 0}=O(k)$($\rightarrow 0$)
$\Longleftrightarrow[R_{\pi-o(k)}^{\pi/2}(k)/G]_{|k\rightarrow 0}=
[R_{0+o(k)}^{\pi/2}(k)/G]_{|k\rightarrow 0}\asymp[k/G]_{|k\rightarrow 0}>0$.
Such a three-dimensionality then also exists so long as $\theta\neq \pi/2$
[Fig. \ref{f:ttp01}(b) and (c)], since the across-slot gravity is made
dominant by a sufficiently small $G$ at any $\theta\in(\pi/2,\pi)$
[$\theta\in(0,\pi/2)$]. This three-dimensionality could be viewed as
coming from conditions I and II at $G=0$ as well. As $k_{y}\rightarrow 0$
for a fixed $k_{z}>0$, in particular, the $Ra_{c}(k_{y},k_{z})|\cos\theta|$
tends to the 2D $Ra_{c}(k_{z})$ at $\theta=\pi$ ($\theta=0$) with the
${\omega_{c}(k_{y},k_{z})}_{|k_{y}\rightarrow 0}\rightarrow 0$
[Figs. \ref{f:mscsov01}, \ref{f:mscodf}(b),
and \ref{f:ttp01}].

Associated in Table \ref{t:jk} with
$k_{y}^{G}(k_{z})$, the relatively abrupt changes in
$\partial Ra_{c}/\partial k_{y}$ and $\partial \omega_{c}/\partial k_{y}$
arising from $k_{y}=0$ are thus also due to a transition between the
perturbation suited mainly for the across-slot gravity and that
for the along-slot gravity. Such a transition is however
essentially of a 3D nature.

With increasing
$k$, the 2D effect of along-slot gravity decreases with respect to that of
across-slot gravity. At $k_{y}^{G}(k_{z})$ for a fixed $\theta\in(\pi/2,\pi)$
[$\theta\in(0,\pi/2)$], the latter gravity component still gives way to the
former in driving the perturbation. To neutralize the 2D across-slot gravity
in Eqs. (\ref{eq:lsu}) and (\ref{eq:lsts}), $G$ must then grow with such $k$:
$dG/dk>0$ for $k=[{k_{y}^{G}(k_{z})}^2+k_{z}^2]^{1/2}$. As (\ref{eq:dky})
follows from (\ref{eq:dk3}), therefore, $dk_{y}^{G}/dk_{z}>k_{y}^{G}/k_{z}>0$
(Fig. \ref{f:ttp01} and Table \ref{t:jk}). 
With $Re[d\tilde{u}(k,0)/dx]_{|\theta=\pi-o(k)}\asymp Re[d\tilde{u}(k,0)/dx]_{|\theta=o(k)}\asymp 1$
and $Re[d\tilde{u}(k,0)/dx]_{|\theta=\pi/2}\asymp 1$ as $k\rightarrow 0$
[$k_{z}\rightarrow 0$, $k_{y}(k_{z})=k_{y}^{G}(k_{z})=o(k_{z})$],
$[R_{\pi-o(k)}^{\pi/2}(k)k/k_{y}]_{|k\rightarrow 0}=
[R_{0+o(k)}^{\pi/2}(k)k/k_{y}]_{|k\rightarrow 0}\asymp 1$
implies $(k^{2}/k_{y})_{|k\rightarrow 0}\asymp 1$ for $k_{y}(k_{z})=k_{y}^{G}(k_{z})$
[$=o(k_{z})$]. Relations (\ref{eq:balk}) then apply to $k_{y}^{G}(k_{z})$.

For large enough $|\theta-\pi/2|$ [Fig. \ref{f:ttp01}(a) and (b)],
$k_{y}^{\theta}(0)$ is distinguishable within the considered range of $k_{y}$.
Since $k_{y}^{\theta}(k_{z})$ decreases with increasing $k_{z}$, the above increase
of the respective $k_{y}^{G}(k_{z})$ with $k_{z}$ eventually leads to vanishing
of the interval of $k_{y}$ where the instability is mainly due to the along-slot
gravity. This takes place when $k_{y}^{G}(k_{z})$ and $k_{y}^{\theta}(k_{z})$
merge. When $\theta$ is so close to $\pi/2$ that such $k_{y}^{\theta}(k_{z})$
are not found [Fig. \ref{f:ttp01}(c)], the increases of
$Ra_{c}(0,k_{z})=Ra_{c}(k_{z})_{|\theta=0,\pi}/|\cos\theta|$ and of $k_{y}^{G}(k_{z})$
itself with large enough $k_{z}$ make the changes in $\partial Ra_{c}/\partial k_{y}$
and $\partial \omega_{c}/\partial k_{y}$ at $k_{y}^{G}(k_{z})$ indistinguishable.
The slow-propagating cells localized near the distinction (inverse)
boundary then smoothly transform into the traveling
cells occupying the whole slot width.

As discussed above, $k_{y}^{\theta}(k_{z})$ increases when $|\theta-\pi/2|$
decreases for a given $k_{z}$. For $k_{z}$($>0$) being fixed, however,
Eq. (\ref{eq:kytt}) implies $\partial k_{y}^{G}(k_{z},\theta)/\partial\theta>0$
[$\partial k_{y}^{G}(k_{z},\theta)/\partial\theta<0$] for $\theta\in(\pi/2,\pi)$
[$\theta\in(0,\pi/2)$]. The ultimate vanishing of the relatively abrupt changes in
$\partial Ra_{c}/\partial k_{y}$ and $\partial \omega_{c}/\partial k_{y}$ thus has
to take place at the larger $k_{z}$ the closer $\theta$ is to $\pi/2$, as
in Fig. \ref{f:ttp01} and Table \ref{t:jk}.
\subsection{\label{s:rsl}Linear steady instability for $\theta=\pi$}
\subsubsection{\label{s:rslb}Background}
Computed from Eqs. (\ref{eq:lssts}) and (\ref{eq:lssuu}) and boundary conditions
(\ref{bc:uv}), (\ref{bc:tsl}), and (\ref{bc:tsr}), steady marginal-stability
curves for $\theta=\pi$ and prescribed $Ra$ are illustrated in Figs. \ref{f:sp00},
\ref{f:sp11}, and \ref{f:sp01}. Their most unstable wave numbers, $k_{c}$, and the
respective $Ra_{c}^{s}(k_{c})$ are provided in Table \ref{t:km}. Although Fig. \ref{f:sp01}
and its data in Table \ref{t:km} are also relevant to $\theta=0$ (as mentioned in the
captions of Fig. \ref{f:sp01} and Table \ref{t:km}), they are discussed in
Sec. \ref{s:rsld01} below only in terms of $\theta=\pi$. The shape of the
presented curves is specified by the combined effects of across-slot
diffusion and the first dissipation mechanism (Sec. \ref{s:rovhDD}).

For $Ra=0$, such diffusion is stabilizing. Since it is absent for
$\chi=0$, the most unstable wave number is zero, where the minimal
effect of the first dissipation mechanism is achieved. Derivable via
the long-wavelength expansion \cite{r:n,r:sh}, the respective $Ra_{c}^{s}(0)$
(Fig. \ref{f:sp00} and Table \ref{t:km}) depends on the
velocity conditions at the slot boundaries.

For $\chi=1$, the stabilizing effect of diffusion for $Ra=0$ renders the most
unstable wavelength finite. Compared to the stress-free boundary condition, the
enhancement of along-slot dissipation at a no-slip boundary reduces the across-slot
portion of actively moving streamline particles with an along-slot density difference.
Like diffusion, the effect of such disparity in along-slot dissipation is also augmented
with the wavelength. Besides raising $Ra_{c}^{s}(k)$ for all $k$, therefore, the no-slip
effect increases $k_{c}$ as well (Figs. \ref{f:sp11} and \ref{f:sp01} in
Table \ref{t:km} for $Ra=0$) \cite{r:n,r:sh}. The smaller across-slot
portion of streamline particles driving convection then
constitutes a larger part of the overall streamline.

When introduced only at one boundary for $Ra=0$,
the no-slip effect is more pronounced in combination with the flux
solute condition, where the solute perturbation isolines are orthogonal to
the boundary [Figs. \ref{f:sp11}(b) and \ref{f:sp01}(b), and Table \ref{t:km}].
This takes place because for the stress-free condition, the actively moving
streamline particles with along-slot density differences exist infinitesimally
close to the boundary at which there is no diffusion. The (zero-$Ra$) $k_{c}$
in Table \ref{t:km} ($\chi=1$) for $\gamma_{-}=1$ and $\gamma_{+}=0$
($\gamma_{-}=0$ and $\gamma_{+}=1$) is thus closer to its value
for $\gamma_{\pm}=1$ ($\gamma_{\pm}=0$) than to that for
$\gamma_{\pm}=0$ ($\gamma_{\pm}=1$) \cite{r:n,r:sh}.
\subsubsection{\label{s:rsld}$Ra>0$}
\paragraph{\label{s:rsld00}$\eta=\chi=0$.}
When $Ra$ increases from $0$, the effect of
diffusion becomes destabilizing. Being due to differential
gradient diffusion at the distinction boundary, the instability
mechanism is then similar to that described in \cite{r:twh,r:tbc}.
For $\eta=\chi=0$, its effect is maximized at the infinite wavelength,
where any stable fixed-value stratification is fully neutralized by
diffusion. Also minimizing the role of the first dissipation mechanism,
such wavelength thus remains most unstable for any $Ra$ \cite{r:n}. For
$k\rightarrow 0$, the instability parameters in Fig. \ref{f:sp00}
(Table \ref{t:km}) are identical to those for $\eta=1$
and $\chi=0$ \cite{r:twh,r:tbc,r:tap,r:n,r:llm}.

When $k$ increases from $0$, however, quantitative
differences from the respective $Ra_{c}^{s}(k)$ for $\eta=1$
and $\chi=0$ arise, due to differential gradient diffusion taking
place only at one boundary. In Fig. \ref{f:sp00}(b), such differential
diffusion at the distinction boundary is also more effective in
combination with the respective stress-free single boundary
condition ($\gamma_{-}=0$ and $\gamma_{+}=1$) than with
the no-slip one ($\gamma_{-}=1$ and $\gamma_{+}=0$). As
discussed above, the (nondiffusive) flux component
perturbation is affected by the disparity between
the no-slip and stress-free velocity conditions
more than the (diffusive) fixed-value one.
\paragraph{\label{s:rsld11}$\eta=\chi=1$.}
When coupled with the distinction boundary, the
single stress-free condition is more destabilizing
than in its combination with the similarity boundary for
$\eta=\chi=1$ as well. In particular, this explains the relative
location of the solid and dashed lines in Fig. \ref{f:sp11}(b). With
growing $Ra$, in addition, the $Ra_{c}^{s}(k_{c})$ in Table \ref{t:km}
for $\gamma_{-}=0$ and $\gamma_{+}=1$ becomes even much closer to such
a value for $\gamma_{\pm}=0$ than to that for $\gamma_{-}=1$ and
$\gamma_{+}=0$. Likewise, such a $Ra_{c}^{s}(k_{c})$ for
$\gamma_{-}=1$ and $\gamma_{+}=0$ becomes much closer
to its respective value for $\gamma_{\pm}=1$ than to
that for $\gamma_{-}=0$ and $\gamma_{+}=1$.

That the effect of diffusion turns destabilizing when $Ra$
increases from $0$ is also manifested for $\eta=\chi=1$ in
decreasing $k_{c}$ (Fig. \ref{f:sp11} and Table \ref{t:km}).
Due to solute neutralization by its diffusion at the similarity
boundary, however, $k_{c}$ cannot decrease to zero. It is the
balance between the overall differential effect of both
boundaries on component perturbations and the effect
of the similarity boundary on solute perturbation
that specifies the value of $k_{c}$.

As the overall effect of differential diffusion is
augmented with growing $Ra$, however, the ratio of $Ra_{c}^{s}(k)$
to $Ra$ decreases for any $k$. This relatively enhances the effect
of the similarity boundary on solute perturbation in the above balance
specifying $k_{c}$, at which $Ra_{c}^{s}(k)/Ra$ is the smallest for a fixed
$Ra$. When the latter effect turns dominant above certain $Ra$, $k_{c}$
begins to increase. As in Table \ref{t:km}, such an oscillatory behavior
of $k_{c}$ has to be most pronounced for $\gamma_{\pm}=0$, where
diffusion is not restricted by dissipation at the boundaries.
It has to be least pronounced for $\gamma_{\pm}=1$, where
$k_{c}=0.86$ for $Ra=200000$.
\paragraph{\label{s:rsld01}$\eta=0$ and $\chi=1$.}
For $\eta=0$ and $\chi=1$
(Fig. \ref{f:sp01}), the increase of $Ra$ from $0$
introduces differential gradient diffusion that favors
both a steady perturbation at the distinction boundary and a
standing-wave perturbation at the inverse boundary. Such a diffusion
process at the latter boundary thus opposes a steady rotation
of the convective cells. As suggested by Fig. \ref{f:mscsov01},
however, the steady linear instability for $\gamma_{-}=\gamma_{+}$
is expected to precede the oscillatory one. (As indicated in the
captions of Figs. \ref{f:mscsov01}, \ref{f:perhs01}, and \ref{f:sp01},
the results for $\theta=0$ and for $\theta=\pi$ are relevant
to each other.) Whereas both dissipation mechanisms
accentuated in Sec. \ref{s:rovhDD} above damp the
oscillatory (standing-wave) perturbation, only
the first one of them damps the
steady perturbation.

Let the velocity conditions be either
the same at both boundaries ($\gamma_{-}=\gamma_{+}$) or such
as the single stress-free condition is at the distinction boundary
($\gamma_{-}=0$ and $\gamma_{+}=1$). Localization of steady convective
cells near the distinction boundary is thus favored even when $Ra$ is
small. (For $\theta=0$, such a localization at the inverse boundary
is seen in Fig. \ref{f:perhs01}.) The destabilizing effect of the
distinction boundary then dominates the stabilizing effect
of the inverse boundary. Thus $k_{c}$ only decreases
with $Ra$ increasing from $0$ for such
$\gamma_{\pm}$ (Table \ref{t:km}).

When the single stress-free condition is at the inverse
boundary ($\gamma_{-}=1$ and $\gamma_{+}=0$), however, the
effect of differential gradient diffusion at the distinction
boundary may not be dominant. At small $Ra$, in particular, the
onset of steady convection is dominated by the disparity between
the velocity boundary conditions. Indeed, the overall effect of
diffusion on steady convection at $Ra=1000$ remains stabilizing:
${Ra_{c}^{s}(k_{c})}_{|Ra=1000}=1949>Ra+{Ra_{c}^{s}(k_{c})}_{|Ra=0}=1000+817$
(Table \ref{t:km}).

For $\gamma_{-}=1$ and $\gamma_{+}=0$, $k_{c}$ thus first grows with
$Ra$ increasing from $0$ (to at least $1000$, as in Table \ref{t:km}).
Only when the stable fixed-value stratification becomes large enough,
the overall effect of differential gradient diffusion on steady convective
cells becomes destabilizing. With the cells localized near the distinction
boundary, in particular, $k_{c}$ decreases as $Ra$ increases to $5000$ and
above (Table \ref{t:km}). For all pairs of $\gamma_{\pm}$, however, its
decrease to $0$ is still prevented by the stabilizing effect of the
inverse boundary. At the infinite wavelength, such effect becomes
commensurate with the destabilizing role of
the distinction boundary.

Compared to the no-slip boundary condition, the
stress-free condition enhances differential gradient diffusion
at its boundary, as discussed above. Such an enhancement is also
more relevant at the boundary where the convective cells are driven.
In particular, the dashed lines are above the respective solid lines
in Fig. \ref{f:sp01}(b). For growing $Ra$, in addition, $Ra_{c}^{s}(k_{c})$
for $\gamma_{-}=0$ and $\gamma_{+}=1$ remains closer to such a value for
$\gamma_{\pm}=0$ than to that for $\gamma_{-}=1$ and $\gamma_{+}=0$.
The $Ra_{c}^{s}(k_{c})$ for $\gamma_{-}=1$ and $\gamma_{+}=0$ also
remains closer to such a value for $\gamma_{\pm}=1$ than to that
for $\gamma_{-}=0$ and $\gamma_{+}=1$ (Table \ref{t:km} for
Fig. \ref{f:sp01}).
\subsection{\label{s:rsf}Finite-amplitude steady convection for $\theta=0$}
\subsubsection{\label{s:rsf11}$\eta=\chi=1$}
For the component conditions being different only at one boundary and
$\theta=0$, the most pronounced manifestation of the finite-amplitude steady
instability mechanism reported for $\eta=1$ and $\chi=0$ in \cite{r:tpla} takes
place when $\eta=\chi=1$. Its bifurcation diagram is illustrated in Fig. \ref{f:bd}.
When the relative role of stable solute stratification increases with $\mu$, the
linear steady instability is delayed for two reasons. One of them is merely the
stabilizing solute contribution to the overall background stratification. The
other reason is of a double-component nature. It is the opposition provided by
differential gradient diffusion to a steady rotation of convective cells. Such
finite-amplitude convection is however relatively little affected by the
growing solute stratification, due to the nature of its mechanism.
This nature is illustrated in Fig. \ref{f:ff11}.

As the convection amplitude increases 
[Fig. \ref{f:ff11}(a),(b)], a growing number of solute isolines near
the distinction boundary are found to be ''outside'' the flow domain.
This effect is particularly pronounced in the regions of across-slot
motion towards this boundary. The background solute scale is thus reduced
in the convective state. Due to the fixed-value temperature conditions
at both boundaries, however, the isotherms only increase their
density in the direction of across-slot motion near either
boundary. The background temperature scale thus remains
intact even when convection is well-developed
[Fig. \ref{f:ff11}(b)---(e)].

Substantially reducing the across-slot solute scale,
the finite-amplitude perturbation thus generates such
unstable density gradients in the regions of across-slot motion
[Fig. \ref{f:ff11}(a)---(e)] as exceed the linear single-component threshold
[$Ra_{c}(\pi)\approx 1707$ \cite{r:rrh}]. This takes place well before
onset of the double-component linear instability (Fig. \ref{f:bd}). Horizontal
density differences between the respective streamline points are then
formed and give rise to (positive) convective feedback. Coming from
finite-amplitude \mbox{Rayleigh}---\mbox{Benard} convection,
this feedback maintains the disparity between
component gradients.

Such finite-amplitude convection also arises when the overall background
stratification is neutral or stable. For the present formulation, the linear
steady instability is then absent [Fig. \ref{f:ff11}(c)---(e)]. With limit
point $L$ in Fig. \ref{f:bd} moving to higher $Ra$ for growing $\mu$, the
flow amplitude and the disparity between component scales at the same
$Ra$ ($=60000$) still decrease [Fig. \ref{f:ff11}(d),(e)].

Continuation of such a finite-amplitude steady flow for $\mu=1$
in $\theta$ to $\theta>0$ fails even when the step in $\theta$
is close to zero. Such a failure of the continuation procedure
is also experienced when an attempt to continue a supercritical
finite-amplitude steady solution for $\mu=1$ and $\theta=\pi$
to $\theta<\pi$ is made. The latter failure has to be due to the
anticipated transformation of the steady finite-amplitude solutions
into traveling-wave ones, as discussed for the respective small-amplitude
flows say in Sec. \ref{s:2tpG} above. Purely finite-amplitude transformations
of a traveling wave into a steady flow have been reported in binary-fluid
convection \cite{r:blks} and in conventional double-diffusive convection
\cite{r:dkt}. The continuation failure at $\theta=0$ could thus also
be due to a transformation of the steady finite-amplitude
flow into a traveling-wave one. 
\subsubsection{\label{s:rsfc}Comparison with $\eta=1$ and $\chi=0$ and $\eta=\chi<1$}
Differential gradient diffusion at both boundaries delays the linear steady
instability for $\theta=0$ more effectively than this process at the distinction
boundary alone. In the finite-amplitude manifestation of steady convection, the
enhancement of differential diffusion on account of the second boundary also
plays a stabilizing role. Solute stratification is however substantially
reduced in such a convective steady state. Being partly due to differential
gradient diffusion, its (stabilizing) effect on finite-amplitude steady
convection is then much smaller than that on the respective small-amplitude
one. When $\mu<1$, the hysteresis thus has to be more pronounced
for $\eta=1$ and $\chi=0$ than for $\eta=\chi=1$.\linebreak\vspace*{-0.5cm}

In particular, the interval of hysteresis is larger for $\mu=0.6$ in Fig. 2
of \cite{r:tpla} than for $\mu=0.7$ in the present Fig. \ref{f:bd} (see also
Table \ref{t:hys}). Limit point $L$ for $\eta=\chi=1$ still precedes that for
$\eta=1$ and $\chi=0$ at the same $\mu$. Compared to $\eta=\chi=1$, the stabilizing
effect of differential gradient diffusion on finite-amplitude steady convection is 
enhanced for $\eta=1$ and $\chi=0$ more than the efficiency of such finite-amplitude
instability. [Indeed, the convection amplitude in Fig. \ref{f:ff11}(c) is slightly
higher than that in Fig. 4(c) of \cite{r:tpla}.] For this reason, limit point $L$
for $\eta=\chi=1$ continues to precede that for $\eta=1$ and $\chi=0$ at least when
$\mu\in[1,1.5]$, where the linear steady instability sets in at infinite $Ra$ in
both these cases. In particular, this is seen from Fig. 2 (including the
caption) in \cite{r:tpla} and the present Fig. \ref{f:bd} for
$\mu=1$ and $\mu=1.5$.

[Regarding the solute scale and phase specification,
the present numerical formulation for $\eta=\chi=1$ (Sec. \ref{s:fg})
is not identical to that in \cite{r:tpla}. Using the same steady solution
phase and no other restrictions, however, it is most consistent with the
formulation in \cite{r:tpla}. The above relative location of limit point
$L$ for $\eta=\chi=1$, with respect to that for $\eta=1$ and $\chi=0$,
was also found to take place when the solute concentration values are
fixed at the middle points of the slot boundaries
in both cases.]

The finite-amplitude mechanism discussed in
Sec. \ref{s:rsf11} above is underlain by the flux condition
decreasing in the perturbed state the across-slot scale of a component
on which it is imposed at a boundary. The across-slot perturbation
scales of both components then have to be substantially smaller
for $\eta=\chi=0$ than for $\eta=\chi=1$. The resulting smaller
disparity between such scales is thus generally less likely
to trigger finite-amplitude steady convection before the
onset of small-amplitude one. This explains why, for
$\gamma_{\pm}=1$, the hysteresis practically
vanishes when $\eta=\chi$ approaches $0$
at least for $\mu\leq 0.9$.

When $\eta=\chi<1$ is sufficiently above $0$, however, the
hysteresis region persists. In particular, it is indisputably present
at least for $\eta=\chi\approx 0.41$ (Table \ref{t:hys}, $\gamma_{\pm}=1$).
Such a finite-amplitude steady flow is illustrated in Fig. \ref{f:ff11}(f).
As discussed in Sec. \ref{s:rsld} above, differential gradient diffusion is
also more effective at a stress-free slot boundary than at a no-slip one.
Disparity between the small- and finite-amplitude manifestations of such
a process is an important factor for the bifurcation subcriticality.
The stress-free boundary conditions thus ought to have a
quantitative effect on the hysteresis region.

For $\mu\leq 1$ (Table \ref{t:hys}), in particular,
the subcriticality for $\eta=\chi\approx 0.41$ is more
pronounced for $\gamma_{\pm}=0$ than for $\gamma_{\pm}=1$.
(The potentially interesting cases $\gamma_{-}\neq\gamma_{+}$ are 
outside the scope of this work.) Trial computations with $\gamma_{\pm}=0$
at $\mu\in[0.7,0.9]$ also suggested that such a hysteresis then survives
for $\eta=\chi=0$ as well. However, the latter finding could not be
ascertained because of a numerically transcritical manifestation
of the hysteresis at $\eta=\chi=0$. This ought to be due to an
unfolding of the bifurcation diagram by the scale-fixing
condition, to which the formulation for $\eta=\chi=0$
is particularly sensitive.
\subsubsection{\label{s:rsfo}Environmental implications}
Let $Ra$ in Fig. \ref{f:bd} be fixed on branch $A2$
above limit point $L$. This implies a value of $\mu$ where
bifurcation point $B$ is subcritical. For such a flow, let $\mu$ (and $Ra^{s}$)
be varied from a negative value to the (positive) value just defined and
above. Fig. \ref{f:bd} then suggests the existence of hysteresis in $\mu$
as well. One could thus also expect such a hysteresis to arise when $Ra$
and $Ra^{s}$ are independent of each other. In particular, it ought to
take place when $Ra^{s}$ is varied sufficiently for $Ra$ being fixed
on the higher-amplitude branch of the respective limit point.

Viewed as a possible explanation of abrupt climate changes \cite{r:b,r:a},
such a behavior has been reported in numerous model studies of the global
ocean thermohaline circulation and climate \cite{r:kgmlhr,r:hr,r:a,r:msrc}.
In these studies, the fresh-water flux into a North Atlantic ocean
region of deep sinking (deep-water formation) is a parameter
controlling the salt stratification there. 

[Hysteresis was also found for salinity- and temperature-driven
regimes of a box model considered in \cite{r:stm}. The boundary condition for
a component is then specified by the component conductivity at the boundary. Such
hysteresis is viewed as a simplified illustration of multiplicity of the regimes of
global ocean thermohaline circulation and climate \cite{r:w,r:kgmlhr,r:hr,r:b,r:a,r:msrc}.
The boxes in \cite{r:stm} can however represent the upper and lower ocean layers.
The mechanisms of the salinity- and temperature-driven regimes \cite{r:stm} are
then conceptually analogous to the hydrodynamic mechanisms of double-component
convection in \cite{r:twh,r:tbc} (Secs. \ref{s:rsld00} and \ref{s:rsld11}
above) and in \cite{r:tpla} (Secs. \ref{s:rsf11} and \ref{s:rsfc}
above), respectively. 

In the framework of such analogy, in particular, the difference between box
values of either component corresponds to the component gradient in the respective
convective state and the rate of hydraulic flow to the convection amplitude. The
external mechanical mixing is also represented by the (equal) component eddy
diffusivities. (If one disregards the physical dissimilarity between component
boundary conductivities and their fluid diffusivities, such an analogy
could also be respectively extended to the small- \cite{r:stnf}
and finite-amplitude \cite{r:ver} steady instabilities in
conventional double-diffusive convection.)]

A basic constituent of the global ocean
thermohaline circulation is comprised by the so-called Atlantic
Meridional Overturning Circulation (AMOC). Operation of AMOC implies the
existence of a North Atlantic region of deep sinking \cite{r:r,r:kgmlhr,r:b,r:a}.
Such a region also gives rise to penetration of horizontal sea-surface density
differences into the ocean depth, without which AMOC could not be effectively
driven by buoyancy \cite{r:kgmlhr,r:sj,r:wf,r:hg}. In the context of \cite{r:sj}
discussed in \cite{r:kgmlhr,r:wf,r:hg}, therefore, the nature of an AMOC driven
by buoyancy could only be such as suggests no source of the AMOC hysteresis other
than the sinking region. The convection hysteresis, if any, in this region thus
has to be a local factor imposing the solution structure and the parameters
of hysteresis on the entire AMOC (and the respective climate
regimes \cite{r:b,r:a,r:msrc}).

Allowing for transformation (\ref{eq:ibc}), however,
a hysteresis between steady regimes in the convection
region is certain if there is a hydrodynamic formulation
where the temperature condition at a relevant boundary other than the
sea-surface boundary is far enough from the flux type, as discussed above.
Such a formulation could arise for the flow domain just above the bottom
topography. Hysteresis in convection then ought to exist regardless of
its parameters near $\eta=\chi=0$ for the stress-free boundaries if the
eddy diffusivity (defined without convection) within the topographic bottom
layer is not too small compared to that within the upper flow domain itself.
This involves the issue of vertical variation of such an eddy diffusivity.
With many uncertainties, its discussions \cite{r:kgmlhr,r:wf,r:mwg}
still seem to suggest that the lower boundary condition has to be
within the range of existence of a significant hysteresis.
\section{\label{s:c}Summary and concluding remarks}
\subsection{\label{s:cg}General}
This work provides a comprehensive insight into the
manifestation of double-component convection due to different boundary
conditions in a diversely oriented infinite slot when the reflection symmetry
between the slot conditions for a component is broken. In a class of the addressed
problems, different component conditions at one slot boundary (the distinction
boundary) are considered primarily with flux and with fixed-value conditions
for both components at the other (the similarity boundary). Another class of
the problems is such as the component conditions at the second boundary (the
inverse boundary) differ from each other inversely to the distinction boundary.
For elimination of other physical effects, the primary focus is on the
compensating background gradients and equal component diffusivities.
Mainly treating small-amplitude convection, this study also
examines finite-amplitude steady instability.

Being imposed only at one slot boundary, different component
conditions still give rise to double-component convection for any slot
orientation to the gravity. They also do so being inversely prescribed
at the boundaries, although either component then has one fixed-value
and one flux boundary condition. In either of these problem classes,
however, the manifestation of small- and finite-amplitude convection
is substantially disparate from that with the reflectionally
symmetric boundary conditions for a component. 

One aspect of such a disparity can be interpreted as coming
from dissimilar 2D small-amplitude convection patterns and
the ranges of their formation. It is largely a consequence of
only one traveling wave being relevant when the slot orientation
differs from horizontal. Another cause of the disparity is a more
frequent dominance of 3D small-amplitude disturbances, due to
both a previously identified three-dimensionality mechanism
and new ones. Directly underlying these factors, the broken
symmetry also has major implications
for finite-amplitude steady
instability. 
\subsection{\label{s:ci}Inviscid fluid}
When the similarity boundary is of the flux type ($\eta=\chi=0$),
the behavior of an inviscid oscillatory marginal-stability curve in
the horizontal slot for $\theta=0$ is least dissimilar from that under
the reflectionally symmetric component boundary conditions ($\eta=1$ and
$\chi=0$). In particular, such a linear stability boundary is characterized
by the most unstable wave number being zero. For independently prescribed $Ra^{s}$,
$Ra_{c}$ and $\partial\omega_{c}/\partial k$ at $k=0$ obtained from the
long-wavelength expansion also differ from the symmetric case only by
the denominator in the expression for the former parameter. Compared
to $\eta=1$ and $\chi=0$, the lack of temperature diffusion at the
similarity boundary for $\eta=\chi=0$ stabilizes most wavelengths.
However, it also destabilizes such long wavelengths as effectively
neutralize the temperature scale by diffusion
at the distinction boundary alone.

As at $\eta=1$ and $\chi=0$, the zero wave number
at $\eta=\chi=0$ remains most unstable for the compensating
background gradients up until the slot orientation becomes nearly
vertical. For $\theta>0$, however, the reflection asymmetry of
oscillatory perturbation leads to the expression relating 2D $Ra_{c}$
and $\partial\omega_{c}/\partial k_{y}$ at $k_{y}=0$ being entirely
dissimilar from that for $\eta=1$ and $\chi=0$. In addition, zero limit
of the $Ra_{c}(k_{y})$ and $\omega_{c}(k_{y})$ as $\theta\rightarrow\pi/2$
does not necessarily imply that the oscillatory marginal-stability boundary
transforms into a steady one. For any $\theta\in[0,\pi)$, no steady
linear instability arises in viscous fluid at finite $Ra_{c}$
for either $\eta=\chi$. The inviscid instability with
$Ra_{c}=\omega_{c}=0$ for $\theta\geq\pi/2$ could not
thus be underlain by viscous steady instability.
This also applies to other such inviscid
oscillatory-instability zero thresholds
for $\theta\in(0,\pi)$.

Despite the absence of dissipation, the most unstable wave number
for the oscillatory instability at $\theta=0$ becomes finite both
when $\eta=\chi=1$ and when $\eta=0$ and $\chi=1$. For $\eta=\chi=1$,
such a change comes only from solute diffusion at the similarity boundary.
For $\eta=0$ and $\chi=1$, it is due to differential gradient diffusion
at the inverse boundary [or at the distinction boundary for $\theta=\pi$,
allowing for (\ref{eq:tt01})]. When the stable gradient is independently
prescribed, either of such dissimilar diffusion processes still also
prevents a manifestation of the oscillatory instability
in an immediate vicinity of $k=0$.

For $\eta=0$ and $\chi=1$ at $\theta=0$ and $\theta=\pi$,
however, the vanishing of oscillatory instability with decreasing
$k$ is also preceded by a short interval of $k$ where a higher
unstable gradient of one component destabilizes the lower stable
stratification of the other. This is attributed to the growing
relative disparity between the oscillation frequencies at such
different stable stratifications as these frequencies decrease
with the increasing wavelength. Differential diffusion at the
boundary favoring steady convection is thus enhanced relatively
stronger by the lower stable stratification. At small
enough $k$, the unstable gradient is then affected
more by such stabilizing enhancement than by the
effect of the stable stratification at the
boundary favoring oscillatory convection.

When $\theta=0$ in inviscid fluid, solute diffusion at
the similarity boundary for $\eta=\chi=1$ allows to meet (the
second of) two general conditions for three-dimensionality of instability
disturbances formulated in \cite{r:tap}. Due to these conditions, herein
referred to as conditions I and II, such a three-dimensionality then arises
in the vicinity of $k_{y}=0$ for small $\theta>0$. As $\theta$ increases
further to $\pi/2$, however, 2D $Ra_{c}(k_{y})$ and $\omega_{c}(k_{y})$ for
the compensating background gradients decrease to zero when $\eta=\chi=1$
as well. It is also the vicinity of $k_{y}=0$ that is most sensitive to
the direct effect of along-slot gravity, and thus experiences the
fastest decrease of 2D $Ra_{c}(k_{y})$ and $\omega_{c}(k_{y})$.
Unlike other 3D instability manifestations resulting from
conditions I and II, therefore, the present
three-dimensionality is not retained
above small $\theta>0$.

In contrast to the above problems for $\eta=\chi$, an
interval of zero 2D $Ra_{c}(k_{y})$ and $\omega_{c}(k_{y})$ arises
for $\eta=0$ and $\chi=1$ under the compensating background gradients
already when $\theta$ ($\pi-\theta$) is infinitesimal. [Invariance (\ref{eq:tt01})
also makes $\theta\in(0,\pi/2)$ for $\eta=0$ and $\chi=1$ at $\mu=1$ equivalent
to $(\pi-\theta)\in(0,\pi/2)$.] Expanding with growing $\theta$ ($\pi-\theta$),
this interval becomes infinite as $\theta\rightarrow\pi/2$. As its lower limit
also increases with $\theta$ ($\pi-\theta$), however, the $Ra_{c}(k_{y})$
and $\omega_{c}(k_{y})$ below such a limit are left finite for any
$\theta\in(0,\pi/2]$ [$\theta\in[\pi/2,\pi)$]. Neutralizing
either component by diffusion, such relatively long
wavelengths prohibit the 2D instability
manifestation at infinitesimal
$Ra(k_{y})$. 

The upper limit of the interval of zero 2D $Ra_{c}(k_{y})$ and $\omega_{c}(k_{y})$
thus formed for $\eta=0$ and $\chi=1$ when $\theta\in(0,\pi/2]$ [$\theta\in[\pi/2,\pi)$]
is interpreted as resulting from a competition between two dissimilar patterns. Arising
for the relatively longer wavelengths, one of them has steadily rotating cells driven by both
the along-slot gravity and the effect of across-slot gravity at the inverse (distinction) boundary.
It is a traveling wave whose speed turns infinitesimal when $Ra_{c}(k_{y})=\omega_{c}(k_{y})=0$.
The other pattern is featured by cells whose sense of rotation changes periodically in time.
Characterized only by finite $Ra_{c}(k_{y})$ and $\omega_{c}(k_{y})$, it arises from the
effect of across-slot gravity at the distinction (inverse) boundary for the
relatively shorter wavelengths.

Although conditions I and II are met in inviscid fluid at $\theta=0$
($\theta=\pi$) for $\eta=0$ and $\chi=1$ as well, 3D instability then arises for
$\theta\in(0,\pi/2)$ [$\theta\in(\pi/2,\pi)$] largely from more pronounced effects.
Two of these effects are due to the 2D interval with $Ra_{c}(k_{y})=\omega_{c}(k_{y})=0$.
One of them is underlain by an invariance [Eq. (\ref{eq:inv})] of the linear stability
equations at $\theta=\pi/2$. With the 2D area of $Ra_{c}(k)=0$, this invariance leads
to an interval of $k_{y}$ where 3D disturbances are most unstable for $\theta=\pi/2$.
Such an interval thus arises at small $|\pi/2-\theta|$ as well. For any
$\theta\in(0,\pi/2)$ [$\theta\in(\pi/2,\pi)$], however, 3D disturbances
are also dominant due to their unequal effect on the two gravity
components. Relatively reducing the along-slot gravity by
$G=k_{y}/k$, this effect shifts the 3D zero-threshold
interval of $k_{y}$ below such (2D) interval
specified by $\theta$ alone.

3D most unstable disturbances in inviscid
fluid for $\eta=0$ and $\chi=1$ also arise from another type of
3D areas with zero $Ra_{c}(k_{y},k_{z})$ and $\omega_{c}(k_{y},k_{z})$.
Such a 3D area has no 2D progenitor. Taking place at any $\theta\in(0,\pi/2)$
[$\theta\in(\pi/2,\pi)$], such a perturbation three-dimensionality is still consistent
with conditions I and II when they are considered at $G=k_{y}/k=0$. The nature of
the zero-threshold area underlying such a 3D instability is interpretable in
terms of the analogy between the effect of $G$ in the 3D marginal-stability
equations for $k$ and the effect of $|\tan\theta|$ in the 2D equations for
$k_{y}$. With this analogy, differential behavior of the boundaries of
such an area with respect to $k_{z}$ and $\theta$ is consistent with
the respective numerical data. It also matches the asymptotic
behavior of the area boundaries for $k_{z}\rightarrow 0$
obtained by an independent approach.
\subsection{\label{s:cv}Viscous fluid}
For all the considered combinations of
$\eta$ and $\chi$, oscillatory linear instability also
arises in viscous fluid. When $\theta=0$, in particular, the
marginal-stability curve for either $\eta=\chi$ is characterized by
an infinite increase of $Ra_{c}(k)$ with $k$ decreasing to $0$. Both
such increases come from reduction of the instability feedback efficiency
by along-slot dissipation. First identified for $\eta=1$ and $\chi=0$
in \cite{r:tap}, this mechanism of efficiency reduction is herein
referred to as the second dissipation mechanism. For
$\eta=\chi=1$, however, the long wavelengths are
also so stabilized by solute diffusion
at the similarity boundary.

The viscous instability at $\theta=0$ for $\eta=\chi=1$
still precedes that for $\eta=\chi=0$ so long as $k$ is not too large:
$\delta Ra_{c}(k)>0$ for such $k$. Compared to $\eta=\chi=0$, diffusion
of both components at the similarity boundary then leads to efficiency of
the instability mechanism for $\eta=\chi=1$ being higher. Such higher efficiency,
however, generates steeper velocity gradients and thus more dissipation. Relatively
increasing with $k$, this effect of dissipation eventually dominates the
effect of diffusion at the similarity boundary. $\delta Ra_{c}(k)$ thus
changes its sign. With $k$ increasing further, it also exhibits
decaying oscillations. The oscillatory behavior results from 
an additional dissipation arising above a value of $k$ where
$\delta Ra_{c}(k)=0$ acting against the increase of
$|\delta Ra_{c}(k)|$. The additional dissipation
is greater for such $\eta=\chi$ as has the
higher efficiency of the combined effects
of diffusion and dissipation
at the given $k$.

The second dissipation mechanism is the reason why oscillatory linear
instability for $\eta=0$ and $\chi=1$ at $\theta=0$ and $\theta=\pi$ is preceded by
the steady one. For sufficiently long wavelengths, such oscillatory instability at
$\theta=0$ is preceded by those for $\eta=\chi$ as well. The stabilizing effect of
the inverse boundary for $\eta=0$ and $\chi=1$ is then more important than the cell
localization near the distinction boundary. As the wave number increases, however,
the role of this stabilizing effect diminishes. Compared to $\eta=\chi$, the cell
localization for $\eta=0$ and $\chi=1$ also leads to a faster growth of the
relative portion of fluid particles with along-slot density differences,
and thereby to the faster enhancement of efficiency of the instability
mechanism. Above certain $k$, therefore, the oscillatory instability
for $\eta=0$ and $\chi=1$ at $\theta=0$ precedes
those for $\eta=\chi$. 

In viscous fluid, conditions I and II are met at $\theta=0$ for either
$\eta=\chi$. In both cases, 3D disturbances are thus most unstable in a
vicinity of $k_{y}=0$ when $\theta>0$ is small. For either $\eta=\chi$,
however, 3D disturbances with small $k_{y}$ also remain the first to arise
so long as $\theta<\pi/2$. This takes place because a sufficiently small
$G=k_{y}/k$ renders the effect of across-slot gravity dominant at any
$\theta\in(0,\pi/2)$. Compared to the inviscid fluid for $\eta=\chi=1$,
this is possible due to the respective viscous 2D effects of
along-slot gravity being near $k=0$ sufficiently weak
relatively to those of across-slot gravity.

In viscous fluid, the broken symmetry between boundary conditions
for a component is manifested most commonly at $\theta=\pi/2$.
Oscillatory instability is then characterized by a sequence of 2D
counter-rotating convective cells propagating with a nonzero speed
in the gravity direction. For any pair of considered $\eta$ and $\chi$,
the respective propagating pattern gives rise to such a distribution of
the component perturbations as favors its convective motion and maintains
the direction of its propagation. This direction is sustained by a nearly
steady horizontal density difference generated between the traveling
convective pattern and slot boundaries. Such a mechanism
of traveling-wave instability is underlain by
differential gradient diffusion.

Although differential gradient diffusion for $\eta=0$ and $\chi=1$ acts
inversely at the slot boundaries, its sidewall effects at $\theta=\pi/2$ match
each other in the framework of the above features of the instability pattern.
Such inversely symmetric effects of differential diffusion at $\theta=\pi/2$,
however, result in horizontality of across-slot convective motion, in contrast
to $\eta=\chi$. The corresponding reduction of the across-slot cell path
then leads to a smaller relative convection amplitude of the
instability. As a consequence, the marginal-stability
frequencies are respectively lower for $\eta=0$ and
$\chi=1$ than for $\eta=\chi$. 

Horizontal density differences comprise the
key quantitative feature of the vertical-slot
instability mechanism. Their independence of orientation
of the isolines of both components at the similarity boundary is
thus the reason why the marginal-stability curves at $\theta=\pi/2$
are independent of $\eta=\chi$. The effect of along-slot dissipation on
the efficiency of instability feedback at $\theta=\pi/2$ is also similar
to that of the second dissipation mechanism at $\theta=0$. Although the
marginal-stability curve for $\eta=\chi=1$ at $\theta=0$ is affected by
component diffusion at the similarity boundary as well, this effect
does not introduce a qualitative change. Its role also gradually
vanishes as $\theta=\pi/2$ is approached. Continuous
transformation of either 2D marginal-stability curve
for $\eta=\chi$ thus takes place as $\theta$
increases from $0$ to $\pi/2$.

The nature of vertical-slot instability to the single
gravity-directed traveling wave suggests a universal consequence of the
broken symmetry in viscous fluid. In particular, such a traveling-wave character
of the instability is maintained for any considered pair of $\eta$ and $\chi$
up until $\theta$ increases to $\pi$, where the instability becomes steady. Any
of these 2D transformation scenarios is also accompanied by a manifestation of
3D oscillatory disturbances. Such a manifestation is underlain by different
features of the respective steady instability at $\theta=\pi$. Embodied
at $\theta<\pi$ in a small-frequency behavior of the respective 3D
oscillatory-instability boundary, these dissimilar features thus
make the specific 3D scenario of transformation depend
on the values of $\eta$ and $\chi$.

The simplest 3D scenario of the above transformation 
takes place for $\eta=\chi=1$. It is specified by the nature of 3D
disturbances being merely due to conditions I and II at $\theta=\pi$.
In this region of $\theta$, however, such conditions imply that the
instability is generally of an oscillatory nature with only $\theta=\pi$
being specifically characterized by $\omega_{c}(k)=0$. The small value
of $\omega_{c}>0$ is thus a part of the perturbation introduced by
the deviation of $\theta$ from $\pi$. An interval of $k_{y}$ where
3D disturbances are dominant is then retained by a properly decreased
$G=k_{y}/k$ when $\theta$ is close to $\pi/2$ as well. This
is permitted by the appropriate relative behavior of
the respective 2D effects of the along-slot and
across-slot gravity components near $k=0$.

For $\eta=\chi=0$, condition II is not met at $\theta=\pi$.
However, a vicinity of $k_{y}=0$ is then substantially more unstable
to 2D disturbances at $\theta=\pi$ than at $\theta\in(\pi/2,\pi)$. Mimicking
at small $G=k_{y}/k$ the behavior of the 2D steady marginal-stability boundary
at $\theta=\pi$, such a 3D oscillatory-instability curve at $\theta\in(\pi/2,\pi)$
thus gives rise to 3D most unstable disturbances. With growing $\pi-\theta$, the 
large-$G$ behavior of the small-$k_{z}$ 3D linear-instability curve increasingly
diverges from the small-$G$ one. Isolated contours of solutions of the 3D
linear-stability equations then also form out of secondary multiple
solutions of such 2D equations. The small- and large-$G$ types of
3D behavior are thus reconciled via a hysteresis region between
them. Such a region arises from collision of the isolated
solution contours with the 3D boundary of primary
instability upon the respective reorganization of
the 2D marginal-stability solution structure.

Such a strongly nonlinear behavior of solutions of the linear
oscillatory-instability equations is an outcome of the broken symmetry
and the resulting universality of oscillatory manifestation of the effect of
different boundary conditions. For $\eta=\chi=0$ and small $\pi-\theta$, such a
universality transforms the solution of the linear steady-instability equations
for $\eta=1$ and $\chi=0$ into a small-frequency solution of such equations for
oscillatory perturbation. This furnishes conditions for the solution multiplicity
of the latter 3D equations. Utilizing the physics of such multiple solutions,
behavior of the resulting hysteresis region is interpreted based on the
above analogy between the 3D effect of $G=k_{y}/k$ and the 2D effect
of $|\tan\theta|$ in the respective linear stability equations. Such
interpretation leads to the differential properties of this region
with respect to $k_{z}$ and $\theta$ that are consistent
with the numerical data.

For $\eta=0$ and $\chi=1$,
abrupt marginal-stability changes arise when
$\theta\in(\pi/2,\pi)$ [$\theta\in(0,\pi/2)$]. One their class comes
from such a 2D change. Its 2D nature is due to an irreconcilableness
between the effects of along-slot and across-slot gravity. The failure
of the former to accommodate the latter is related to a phase shift between
the component and flow perturbations. Such a shift arises from the opposition
generated by the across-slot gravity at the inverse (distinction) boundary
to steady cell rotation. Above certain $k_{y}$, the convection pattern
thus switches from that largely driven by the along-slot gravity to a
slower-propagating (like nearly-steady) one driven mainly by the across-slot
gravity. The latter pattern is localized at the distinction (inverse)
boundary. According to the sensitivity to the phase shift, the
respective change in the 2D marginal-stability curve arises
at the longer wavelength the larger $|\theta-\pi/2|$ is.

Due to the effect of $G=k_{y}/k$($<1$) in the 3D linear-stability equations,
the 3D analogue of an above 2D abrupt change for $\eta=0$ and $\chi=1$ shifts
to smaller $k_{y}$ as $k_{z}$ increases from $0$. When such a 2D change is
steep enough, as for $|\theta-\pi/2|\geq 3\pi/8$, this shift leads to a small
region isolated from $k_{y}=0$ where 3D disturbances are most unstable.
Another region of 3D most unstable disturbances arises then near $k_{y}=0$.
Being due to conditions I and II at $\theta=\pi$ ($\theta=0$), it is
maintained for any $\theta\in(\pi/2,\pi)$ [$\theta\in(0,\pi/2)$]
by a sufficiently small $G$. This is allowed by virtue of an
appropriate weakness of the 2D effect of along-slot gravity
with respect to that of across-slot gravity near $k=0$.

3D marginally unstable disturbances arising
when $\theta\in(\pi/2,\pi)$ [$\theta\in(0,\pi/2)$] for
$\eta=0$ and $\chi=1$ from $k_{y}=0$ exhibit another class
of abrupt changes. Being essentially of a 3D nature, these changes however
also imply switching between a pattern driven by the across-slot gravity
and that driven by the along-slot gravity. The former is characterized
by slow-propagating convection cells localized at the distinction
(inverse) boundary while the latter by traveling convection cells
occupying the whole slot width. Differential behavior of such a
transition with respect to $k_{z}$ and $\theta$ is described
with the above analogy between the 3D effect of $G$
and the 2D effect of $|\tan\theta|$ in the
respective marginal-stability equations.

Steady linear instability at $\theta=\pi$
arises for any pair of $\eta$ and $\chi$ due to the mechanism of
differential gradient diffusion identified in \cite{r:twh,r:tbc}. When the slot
boundaries have dissimilar velocity conditions and $\eta=\chi$, such a mechanism
triggers the instability for the stress-free condition at the distinction boundary
before that for the no-slip one. This takes place because the stress-free condition
accommodates the formation of along-slot density differences infinitesimally close
to its boundary. Such an effect could have manifestations beyond the instability for
which it was exposed. Although differential gradient diffusion for $\eta=0$ and
$\chi=1$ at $\theta=\pi$ ($\theta=0$) acts (inversely) at both boundaries, the
steady perturbation is driven at the distinction (inverse) boundary. The
steady instability for the single stress-free velocity condition at
this boundary thus still precedes that for the single no-slip one.

For $\eta=\chi$, the distinction boundary alone could also
give rise to finite-amplitude steady instability at $\theta=0$. Compared to
$\eta=1$ and $\chi=0$ \cite{r:tpla}, such an instability is less subcritical
for $\mu<1$. Its onset at $\eta=\chi=1$ still precedes that for $\eta=1$ and $\chi=0$,
for the second boundary in \cite{r:tpla} is also a source of stabilizing differential
diffusion. When both such linear steady instabilities set in at infinite $Ra$ with
$\mu\geq 1$, the hysteresis for $\eta=\chi=1$ is thus even more prominent.
Still well-pronounced at $\eta=\chi\approx 0.41$, it practically vanishes
for the no-slip boundaries at $\eta=\chi=0$. The convective scales of both
components are then reduced too much for their disparity to give rise to
such a subcriticality. The hysteresis at $\eta=\chi<0.41$ could also
be more prominent for stress-free slot boundaries. The possibility
that abrupt climate change be largely controlled by such a
finite-amplitude instability is thus discussed.
\subsection{\label{s:cc}Main conclusions}
Some consequences of the broken symmetry
are associated with 2D small-amplitude convection.
At $\theta=0$, in particular, component diffusion at the
similarity boundary for $\eta=\chi=1$ is found to enhance efficiency
of the viscous standing-wave instability compared to $\eta=\chi=0$. The key
outcome in viscous fluid is however the universally oscillatory manifestation
of small-amplitude convection. Such a universality stems from the preference
given by the broken symmetry to one of the two counter-traveling waves over
the other. At $\eta=0$ and $\chi=1$, it also involves an abrupt change in
the 2D marginal-stability curve for an inclined slot, when dissimilar
oscillatory patterns are switched. In inviscid fluid, such a change 
comes with an interval of 2D wave numbers where $Ra_{c}=\omega_{c}=0$.
Also arising at $\theta=\pi/2$, such an interval is then a separate
cause for three-dimensionality of most unstable disturbances
near $\theta=\pi/2$.

The broken symmetry also generally enhances the role
of 3D small-amplitude disturbances. Some of them arise
from conditions I and II. Such a mechanism is also relevant
when the nature of instability changes from steady to oscillatory with
the parameter value at which conditions I and II hold being perturbed.
Besides the second dissipation mechanism, the 3D effect of condition II arises 
from solute diffusion at the similarity boundary for $\eta=\chi=1$,
both at $\theta=0$ and at $\theta=\pi$, and from differential gradient
diffusion at the inverse (distinction) boundary for $\eta=0$ and $\chi=1$
at $\theta=\pi$ ($\theta=0$). For $\theta=0$, the former case is also
relevant to inviscid fluid. Unlike the other 3D effects coming
from conditions I and II, such 3D inviscid disturbances at
$\eta=\chi=1$ are dominant only for small $\theta>0$.
This results from the 2D along-slot gravity near
$k=0$ being then relatively strong with
respect to the across-slot gravity.

3D disturbances also arise from or involve
mechanisms other than that specified by conditions I and II. For
$\eta=0$ and $\chi=1$, their dominance when $\theta\in(0,\pi/2)$
[$\theta\in(\pi/2,\pi)$] is associated with abrupt changes in the 3D
marginal-stability curves both in inviscid and in viscous fluid. In inviscid
fluid, in particular, such abrupt changes arise with $Ra_{c}=\omega_{c}=0$.
In viscous fluid for $\eta=\chi=0$, such a three-dimensionality stems from the
disparity between the 2D behavior of $Ra_{c}$ near $k_{y}=0$ at $\theta=\pi$ and
that at $\theta\in(\pi/2,\pi)$. This disparity also leads to multiplicity and 
isolated existence of as well as hysteresis between solutions of the 3D linear
oscillatory-instability equations. The 3D behavior of the inviscid and viscous
abrupt marginal-stability changes for $\eta=0$ and $\chi=1$ as well as of
the hysteresis region for $\eta=\chi=0$ follows from the
analogy between the 3D effect of $G=k_{y}/k$ and
the 2D effect of $|\tan\theta|$. 

The broken symmetry also has other important implications.
In the context of linear steady instability at $\theta=\pi$,
it exposes a higher effectiveness of differential gradient diffusion
at a stress-free fluid boundary than at a no-slip one. It also has a
two-fold effect on the mechanism of finite-amplitude steady instability
at $\theta=0$. Eliminating for $\eta=\chi$ one boundary at which gradient
disparity could form for $\eta=1$ and $\chi=0$, such a symmetry breaking
also reduces the stabilizing effect of differential gradient diffusion on
account of this boundary. The convection hysteresis then effectively arises
for $\eta=\chi=1$. Although such a hysteresis could practically vanish when
the convective scales of both components are reduced at $\eta=\chi=0$,
it remains significant when $\eta=\chi$ is closer to $0$ than to $1$.
Interpretation of abrupt climate change as being largely
underlain by such a finite-amplitude convective
instability is thus discussed.

\newpage
\clearpage
\newpage
\clearpage
\newpage
\clearpage
\newpage
\addtocounter{figure}{-1}
\newpage
\begin{figure}
\centerline{\psfig{file=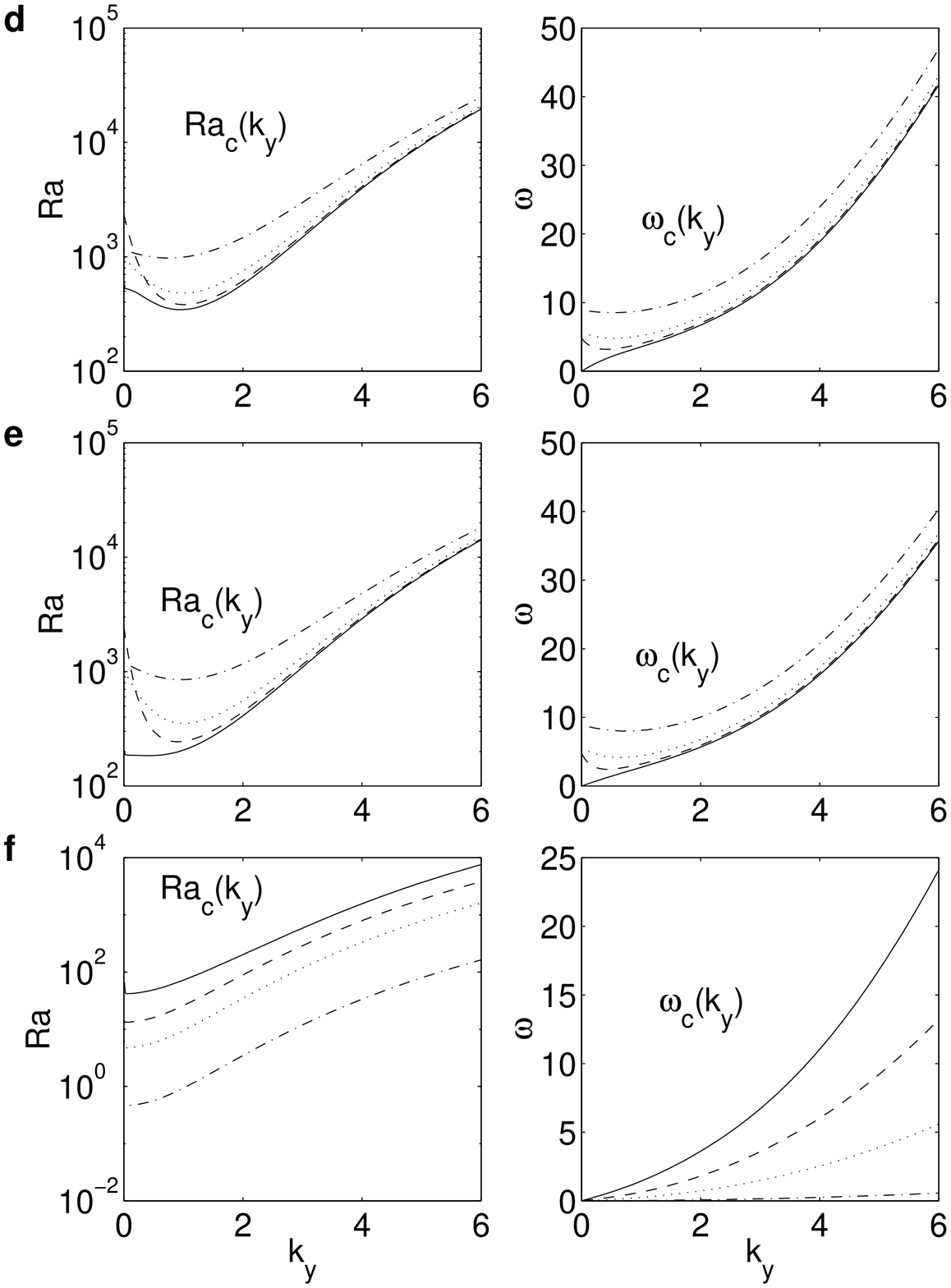,width=14.5cm}}
\caption{Inviscid fluid; $\eta=\chi=1$. 
Curves of the marginal linear stability to 3D ($k_{z}\geq 0$)
oscillatory disturbances for different $k_{z}$, $Ra_{c}(k_{y})$ and
$\omega_{c}(k_{y})$; $\mu=1$, $Le=1$. For (a)---(e), the solid lines:
$k_{z}=0$, the dashed lines: $k_{z}=0.5$, the dotted lines: $k_{z}=1$,
the dash-dot lines: $k_{z}=2$. (a) $\theta=0$; (b) $\theta=0.05\pi/2$;
(c) $\theta=0.1\pi/2$; (d) $\theta=0.15\pi/2$; (e) $\theta=\pi/8$;
(f) $k_{z}=0$, the solid lines: $\theta=\pi/4$, the dashed lines:
$\theta=3\pi/8$, the dotted lines: $\theta=0.9\pi/2$,
the dash-dot lines: $\theta=0.99\pi/2$.}
\end{figure}
\clearpage
\clearpage
\newpage
\begin{figure}
\vspace*{-0.8cm}
\centerline{\psfig{file=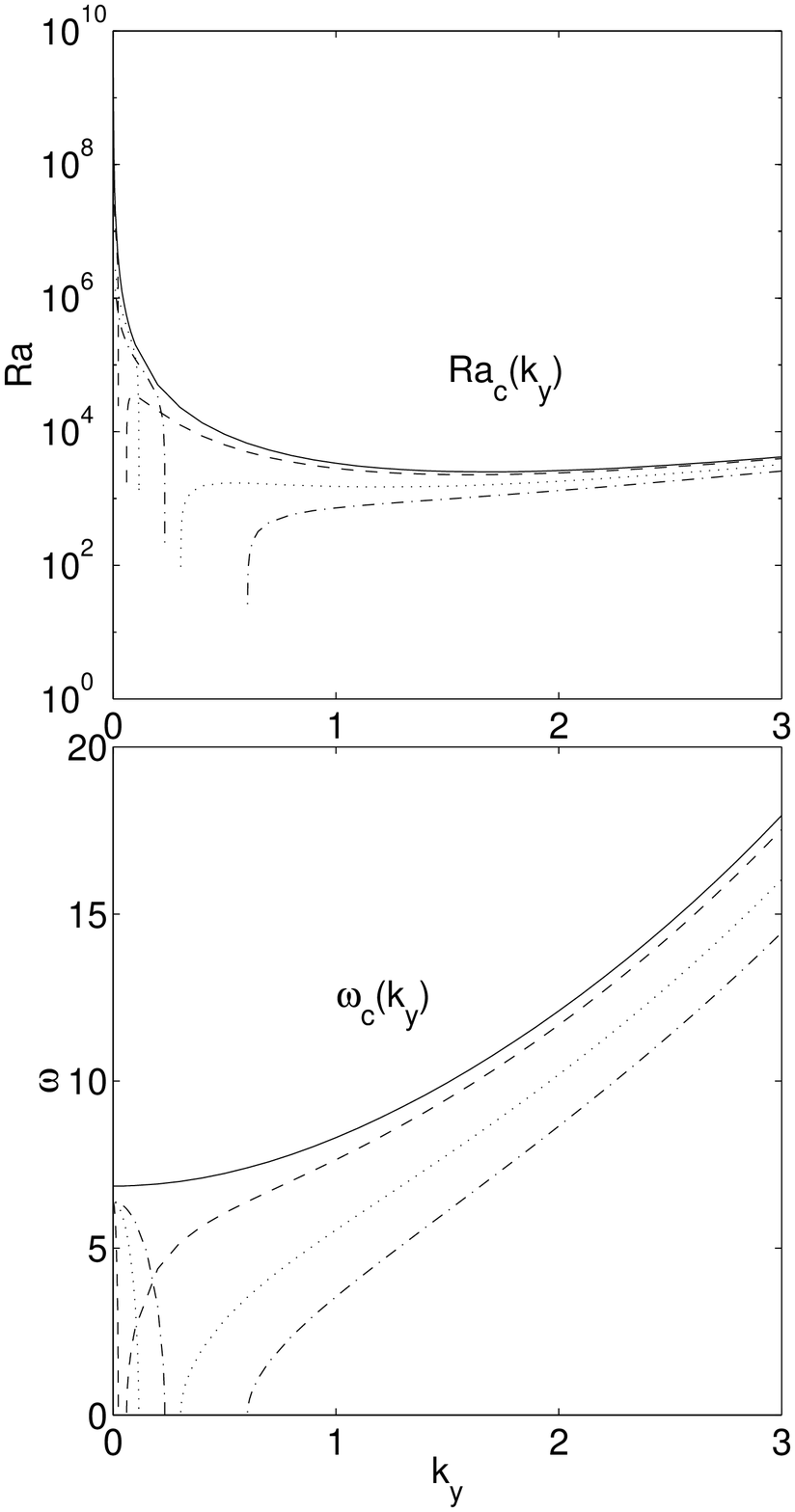,width=11.5cm}}
\vspace*{-0.4cm}
\caption{Inviscid fluid; $\eta=0$, $\chi=1$.
Curves of the marginal linear stability to 2D ($k_{z}=0$)
oscillatory disturbances, $Ra_{c}(k_{y})$ and $\omega_{c}(k_{y})$;
$\mu=1$, $Le=1$. The nearly vertical curves of $Ra_{c}(k_{y})$ are
expected to go to $Ra_{c}(k_{y})=0$. The solid lines: $\theta=(1\mp 1)\pi/2$, the
dashed lines: $\theta=(1\mp 0.99)\pi/2$, the dotted lines: $\theta=(1\mp 0.95)\pi/2$,
the dash-dot lines: $\theta=(1\mp 0.9)\pi/2$. The two values of $\theta$
are relevant in light of transformation (\ref{eq:tt01}).}
\label{f:msco01i2D}
\end{figure}
\clearpage
\newpage
\begin{figure}
\vspace*{-0.5cm}
\centerline{\psfig{file=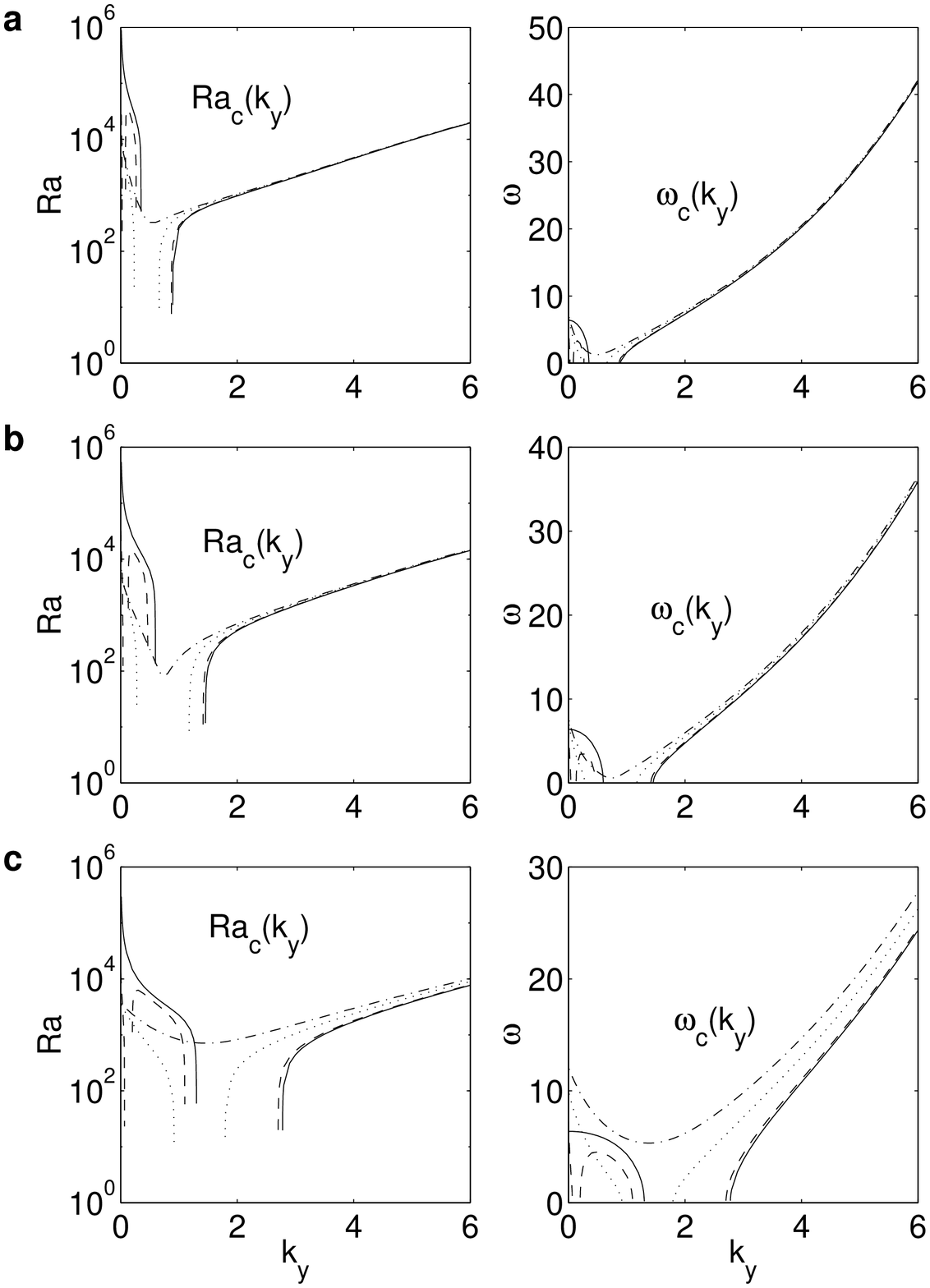,width=14cm}}
\vspace*{-0.3cm}
\caption{Inviscid fluid; $\eta=0$, $\chi=1$.
Curves of the marginal linear stability to 3D ($k_{z}\geq 0$)
oscillatory disturbances for different $k_{z}$, $Ra_{c}(k_{y})$
and $\omega_{c}(k_{y})$; $\mu=1$, $Le=1$. The nearly vertical 
curves of $Ra_{c}(k_{y})$ are expected to go to $Ra_{c}(k_{y})=0$.
(a) $\theta=(1\mp 0.85)\pi/2$; (b) $\theta=(1\mp 0.75)\pi/2$; (c)
$\theta=(1\mp 0.5)\pi/2$; (d) $\theta=(1\mp 0.25)\pi/2$; (e)
$\theta=(1\mp 0.1)\pi/2$; (f) $\theta=\pi/2$. The two
values of $\theta$ in (a)---(e) are relevant in light
of transformation (\ref{eq:tt01}).
The solid lines: $k_{z}=0$; the dashed lines:
(a) $k_{z}=0.15$, (b) $k_{z}=0.25$, (c) $k_{z}=0.5$,
(d) $k_{z}=1.3$, (e) $k_{z}=1.9$, (f) $k_{z}=1$;
the dotted lines: (a) $k_{z}=0.4$, (b) $k_{z}=0.6$, (c) $k_{z}=1.5$,
(d) $k_{z}=2.2$, (e) $k_{z}=3$, (f) $k_{z}=1.5$;
the dash-dot lines: (a) $k_{z}=0.5$, (b) $k_{z}=0.77$, (c) $k_{z}=2$,
(d) $k_{z}=2.8$, (e) $k_{z}=4.2$, (f) $k_{z}=2$.
Additional data are provided
in Table \ref{t:kc}.}
\label{f:msco01i3D}
\end{figure}
\clearpage
\newpage
\addtocounter{figure}{-1}
\begin{figure}
\vspace*{-0.5cm}
\centerline{\psfig{file=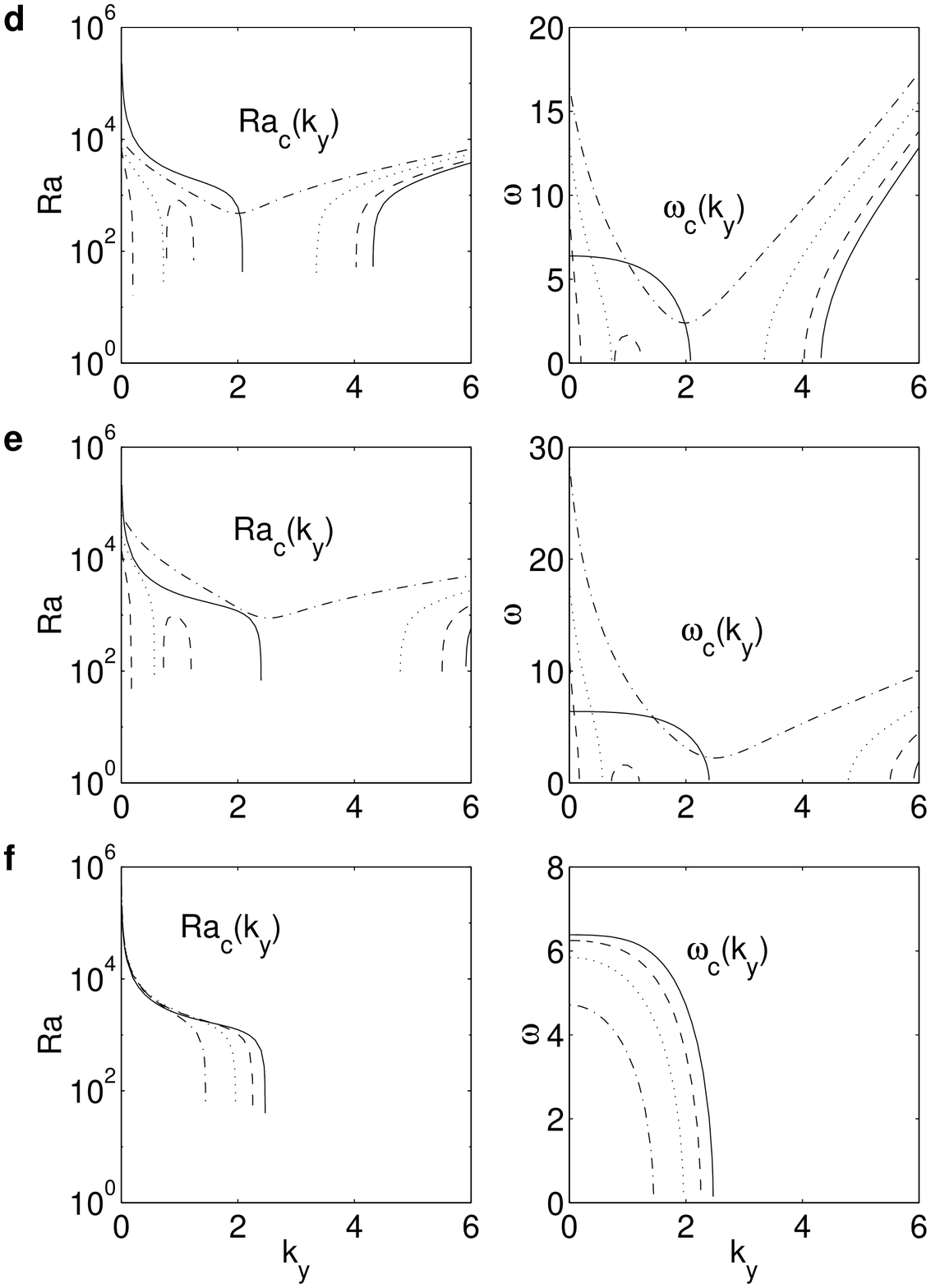,width=14cm}}
\vspace*{-0.3cm}
\caption{Inviscid fluid; $\eta=0$, $\chi=1$.
Curves of the marginal linear stability to 3D ($k_{z}\geq 0$)
oscillatory disturbances for different $k_{z}$, $Ra_{c}(k_{y})$
and $\omega_{c}(k_{y})$; $\mu=1$, $Le=1$. The nearly vertical 
curves of $Ra_{c}(k_{y})$ are expected to go to $Ra_{c}(k_{y})=0$.
(a) $\theta=(1\mp 0.85)\pi/2$; (b) $\theta=(1\mp 0.75)\pi/2$; (c)
$\theta=(1\mp 0.5)\pi/2$; (d) $\theta=(1\mp 0.25)\pi/2$; (e)
$\theta=(1\mp 0.1)\pi/2$; (f) $\theta=\pi/2$. The two
values of $\theta$ in (a)---(e) are relevant in light
of transformation (\ref{eq:tt01}).
The solid lines: $k_{z}=0$; the dashed lines:
(a) $k_{z}=0.15$, (b) $k_{z}=0.25$, (c) $k_{z}=0.5$,
(d) $k_{z}=1.3$, (e) $k_{z}=1.9$, (f) $k_{z}=1$;
the dotted lines: (a) $k_{z}=0.4$, (b) $k_{z}=0.6$, (c) $k_{z}=1.5$,
(d) $k_{z}=2.2$, (e) $k_{z}=3$, (f) $k_{z}=1.5$;
the dash-dot lines: (a) $k_{z}=0.5$, (b) $k_{z}=0.77$, (c) $k_{z}=2$,
(d) $k_{z}=2.8$, (e) $k_{z}=4.2$, (f) $k_{z}=2$.
Additional data are provided
in Table \ref{t:kc}.}
\end{figure}
\clearpage
\begin{figure}
\vspace*{-0.5cm}
\centerline{\psfig{file=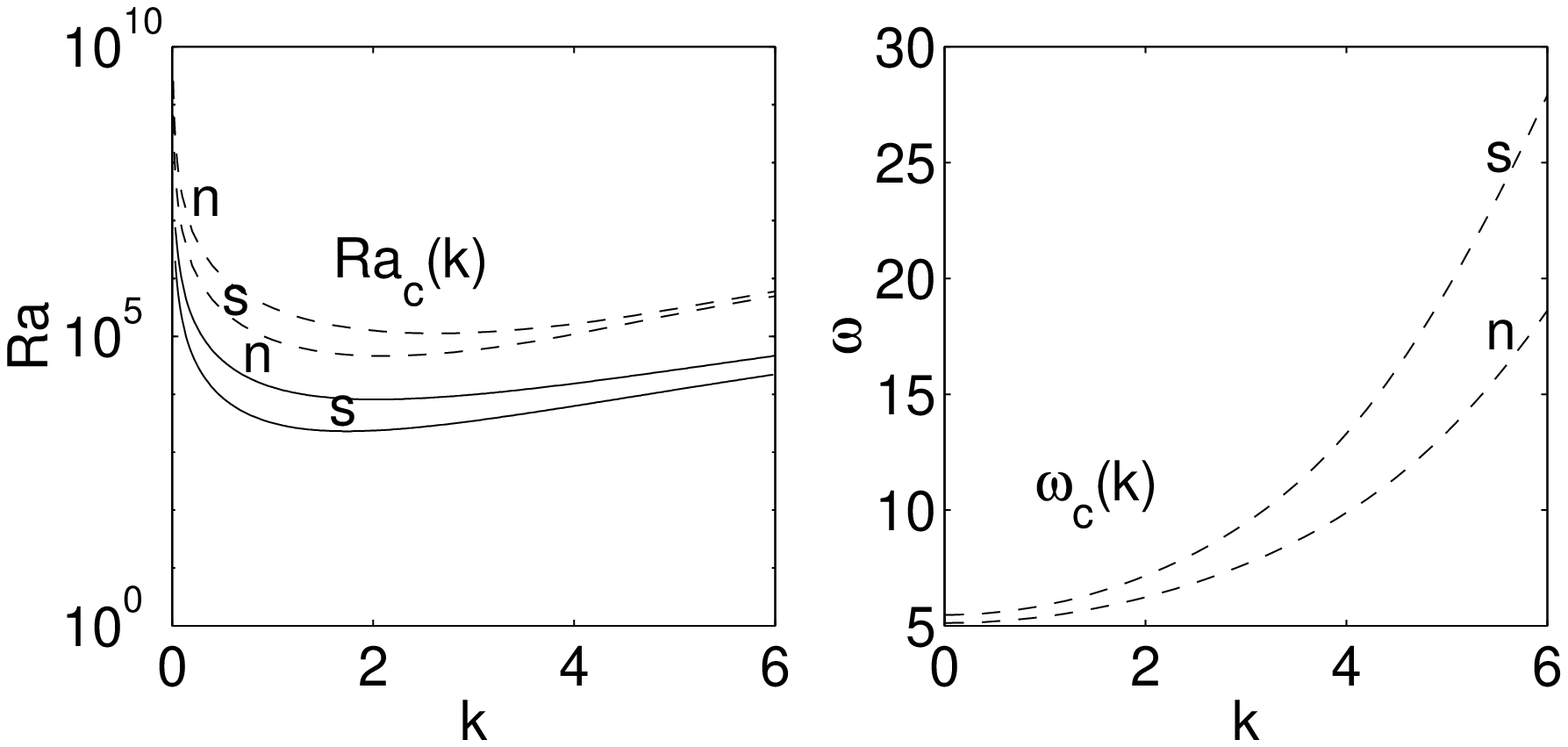,width=14.5cm}}
\caption{$\theta=0$. Viscous fluid; $\eta=0$, $\chi=1$;
$\mu=1$, $Pr=6.7$, $Le=1$. In light of transformation (\ref{eq:tt01}),
these results apply to $\theta=\pi$ as well. Curves of the marginal
linear stability to 2D steady (the solid lines) and oscillatory
(the dashed lines) disturbances, $Ra_{c}(k)$ and $\omega_{c}(k)$,
for stress-free ($\bf{s}$) and no-slip ($\bf{n}$) slot boundary
conditions. Notations $\bf{s}$ and $\bf{n}$ refer to the respective
closest curves below them. (The analogue of the standing-wave
oscillatory instabilities described in this figure is not
reported for the inclined slot with viscous fluid
considered herein below. It is then preceded by
two types of 2D traveling-wave instability,
one of which is related to the steady
instabilities described in
this figure.)}
\label{f:mscsov01}
\end{figure}
\clearpage
\newpage
\begin{figure}
\vspace*{-0.5cm}
\centerline{\psfig{file=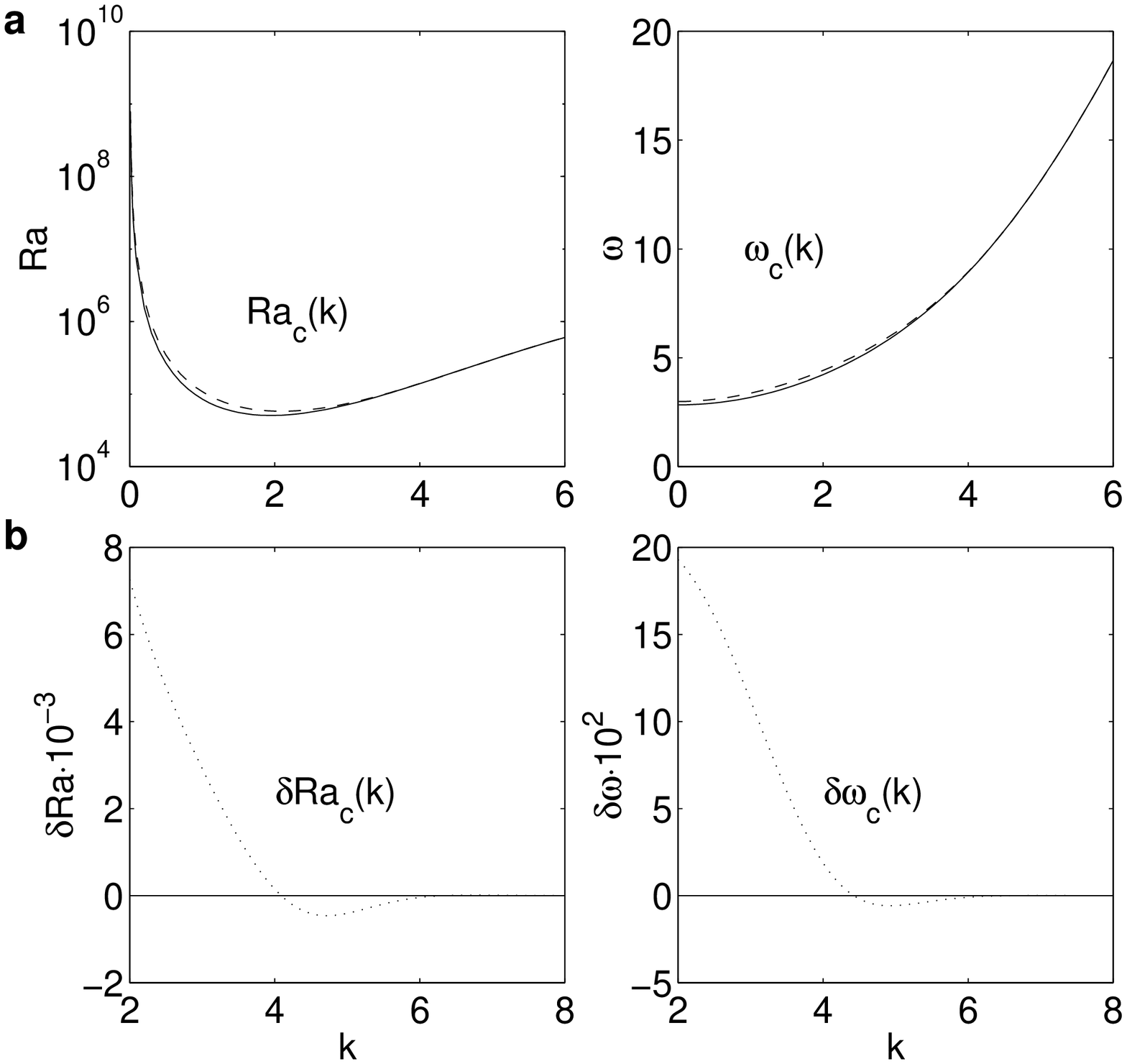,width=14.5cm}}
\caption{$\theta=0$. Viscous fluid and no-slip slot boundaries;
$\mu=1$, $Pr=6.7$, $Le=1$. (a) Curves of the marginal linear stability
to 2D oscillatory disturbances, $Ra_{c}(k)$ and $\omega_{c}(k)$; the solid
lines: $\eta=\chi=1$, the dashed lines: $\eta=\chi=0$. (b) The dotted lines:
$\delta Ra_{c}(k)\equiv{Ra_{c}(k)}_{|\eta=\chi=0}-{Ra_{c}(k)}_{|\eta=\chi=1}$ and
$\delta\omega_{c}(k)\equiv{\omega_{c}(k)}_{|\eta=\chi=0}-{\omega_{c}(k)}_{|\eta=\chi=1}$,
the solid lines designate the respective zero values.}
\label{f:mscot0}
\end{figure}
\clearpage
\newpage
\begin{figure}
\centerline{\psfig{file=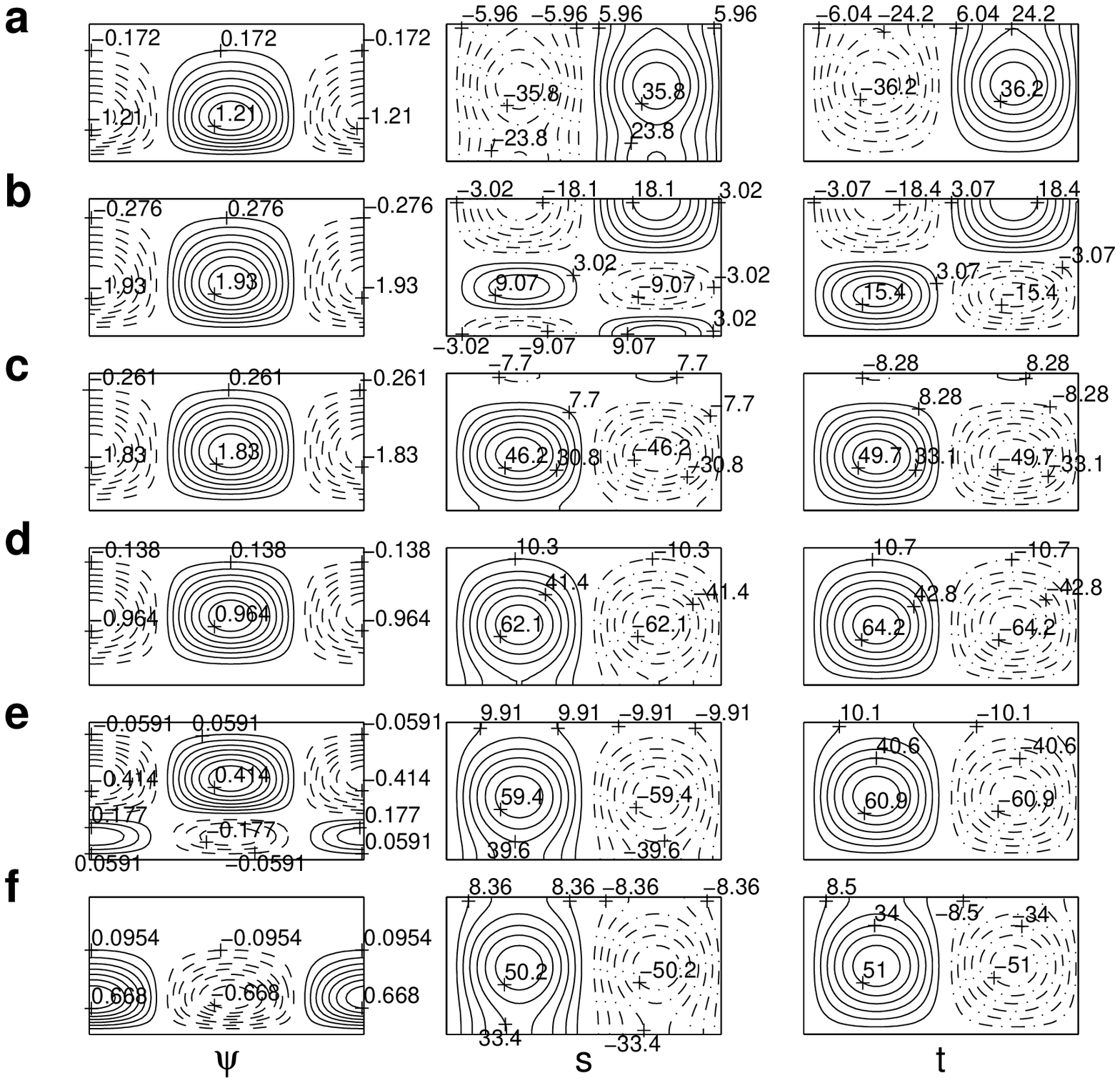,width=14.5cm}}
\vspace*{0.5cm}
\caption{$\theta=0$. Viscous fluid
and no-slip slot boundaries; $\eta=\chi=0$.
Perturbation temporal behavior throughout a half of the oscillation
period $\tau_{p}\approx18\delta\tau$ ($\delta\tau=0.05$) just beyond the onset of 2D
oscillatory instability. It was obtained from the numerical simulation of evolution of
the linearized Eqs. (\ref{eq:ns1})---(\ref{eq:dxi}) in response to the initial disturbance
proportional to the background state after initial time $\tau_{i}\approx 340000$ has passed;
$\lambda=2$, $\mu=1$, $Ra=87340$, $Pr=6.7$, $Le=1$. With this $\tau_{i}$, all perturbation
modes other than the unstable mode ($\tau_{p}\approx 18\delta\tau$) are practically negligible.
$\psi$: perturbation streamlines; $s$: isolines of solute concentration perturbation; $t$:
perturbation isotherms. The actual relative values of the streamfunction perturbation are
equal to $10^{-3}$ times the respective values in the figure. The solid and dashed
streamlines designate the clockwise and counterclockwise rotation and are equally
spaced within the positive and negative streamfunction intervals, respectively.
The solid and dash-dot isolines of the component perturbations are equally
spaced within the positive and negative component perturbation intervals,
respectively. (a) $\tau=\tau_{i}+\delta\tau$; (b) $\tau=\tau_{i}+3\delta\tau$;
(c) $\tau=\tau_{i}+5\delta\tau$; (d) $\tau=\tau_{i}+7\delta\tau$; (e)
$\tau=\tau_{i}+8\delta\tau$; (f) $\tau=\tau_{i}+9\delta\tau$.}
\label{f:perh00}
\end{figure}
\clearpage
\newpage
\begin{figure}
\centerline{\psfig{file=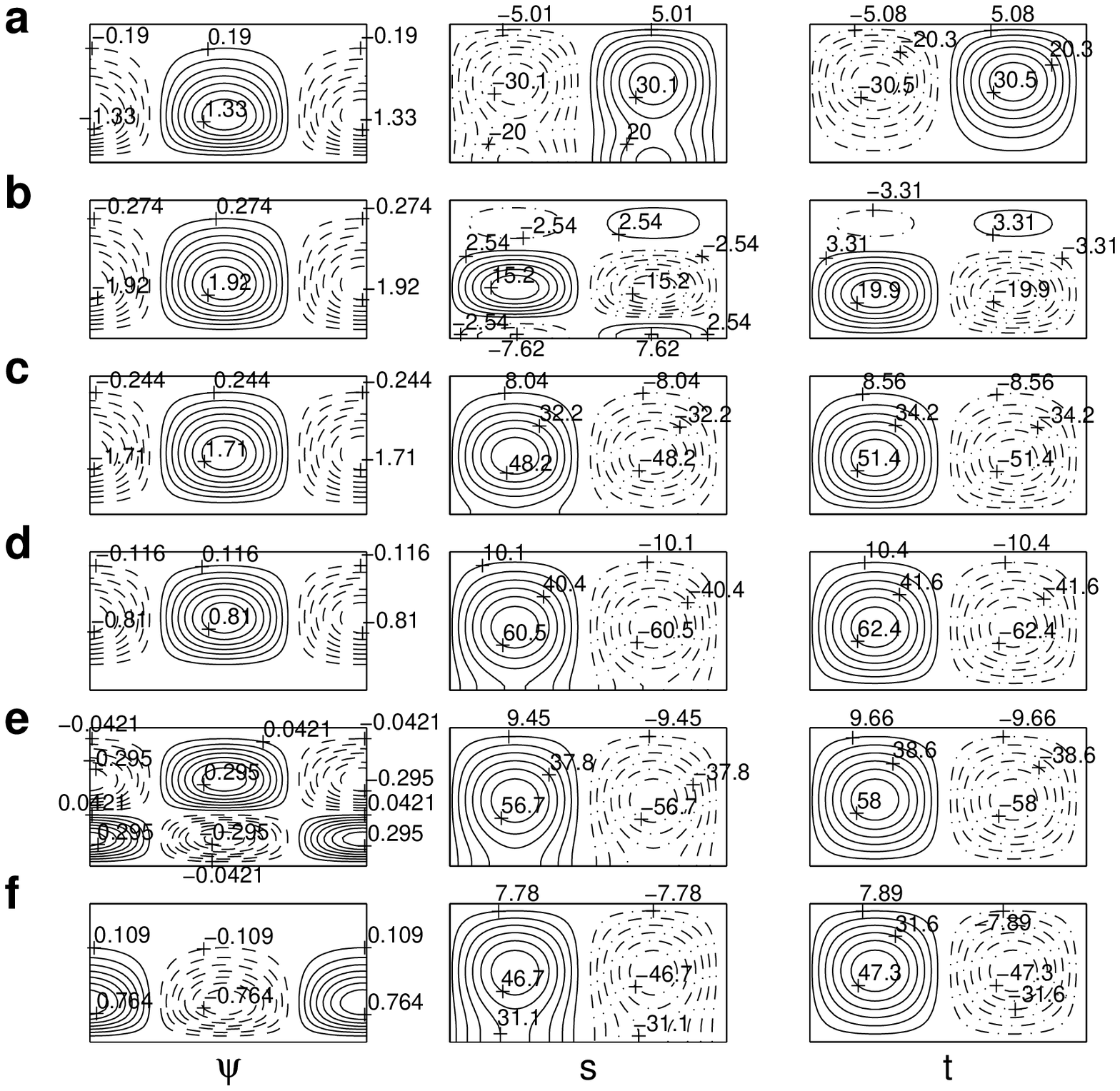,width=14.5cm}}
\vspace*{0.5cm}
\caption{$\theta=0$. Viscous fluid
and no-slip slot boundaries; $\eta=\chi=1$.
Perturbation temporal behavior throughout a half of the oscillation
period $\tau_{p}\approx18\delta\tau$ ($\delta\tau=0.05$) just beyond the onset of 2D
oscillatory instability. It was obtained from the numerical simulation of evolution of
the linearized Eqs. (\ref{eq:ns1})---(\ref{eq:dxi}) in response to the initial disturbance
proportional to the background state after initial time $\tau_{i}\approx 120000$ has passed;
$\lambda=2$, $\mu=1$, $Ra=85040$, $Pr=6.7$, $Le=1$. With this $\tau_{i}$, all perturbation
modes other than the unstable mode ($\tau_{p}\approx 18\delta\tau$) are practically negligible.
$\psi$: perturbation streamlines; $s$: isolines of solute concentration perturbation; $t$:
perturbation isotherms. The actual relative values of the streamfunction perturbation are
equal to $10^{-3}$ times the respective values in the figure. The solid and dashed
streamlines designate the clockwise and counterclockwise rotation and are equally
spaced within the positive and negative streamfunction intervals, respectively.
The solid and dash-dot isolines of the component perturbations are equally
spaced within the positive and negative component perturbation intervals,
respectively. (a) $\tau=\tau_{i}+\delta\tau$; (b) $\tau=\tau_{i}+3\delta\tau$;
(c) $\tau=\tau_{i}+5\delta\tau$; (d) $\tau=\tau_{i}+7\delta\tau$; (e)
$\tau=\tau_{i}+8\delta\tau$; (f) $\tau=\tau_{i}+9\delta\tau$.}
\label{f:perh11}
\end{figure}
\clearpage
\newpage
\begin{figure}
\vspace*{-0.5cm}
\centerline{\psfig{file=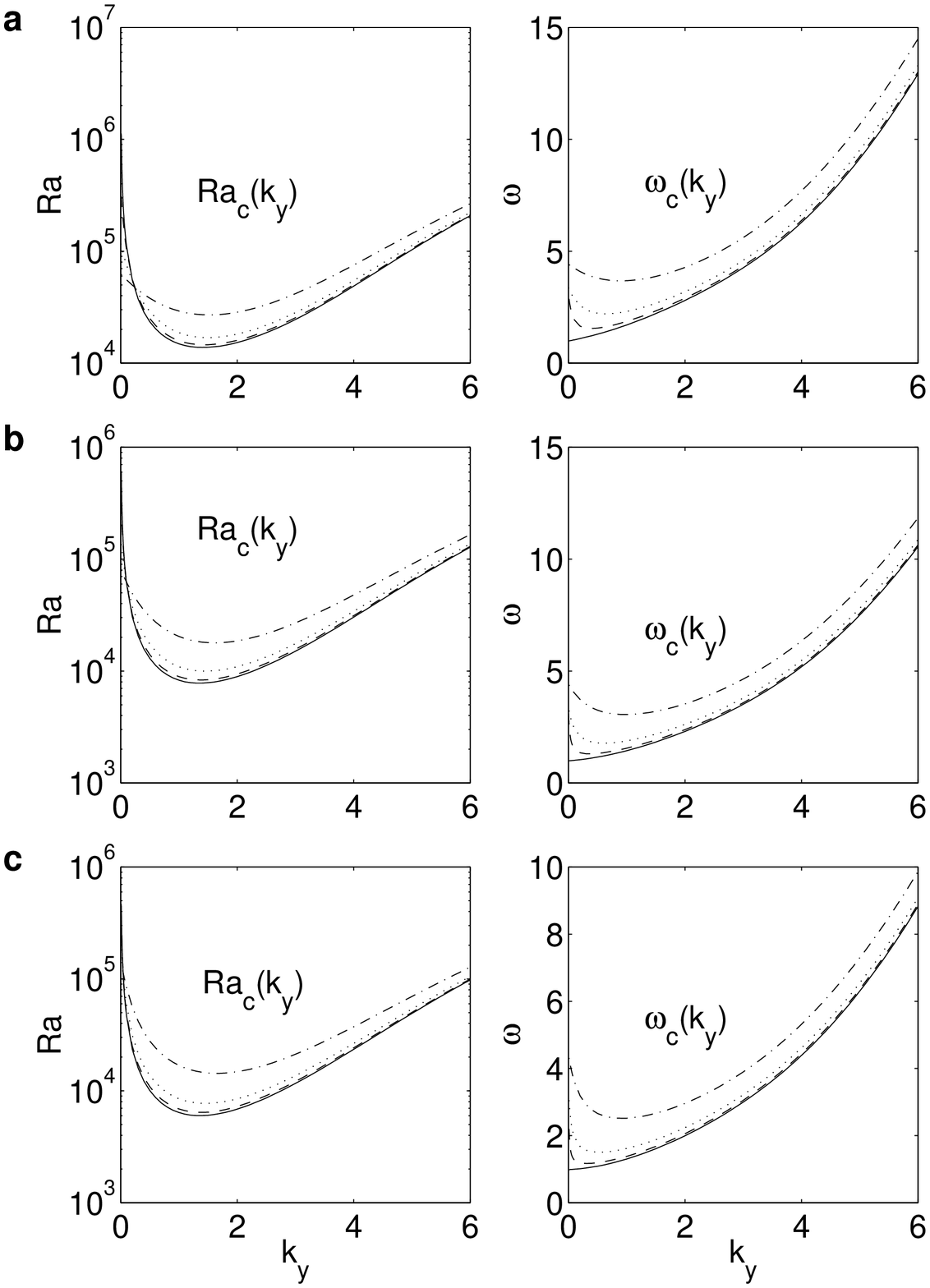,width=14.5cm}}
\caption{Viscous fluid and no-slip slot boundaries;
$\eta=\chi=0$. Curves of the marginal linear stability to 3D ($k_{z}\geq 0$)
oscillatory disturbances for different $k_{z}$, $Ra_{c}(k_{y})$ and $\omega_{c}(k_{y})$;
$\mu=1$, $Pr=6.7$, $Le=1$. The solid lines: $k_{z}=0$, the dashed lines:
$k_{z}=0.5$, the dotted lines: $k_{z}=1$, the dash-dot lines:
$k_{z}=2$. (a) $\theta=0.25\pi/2$; (b) $\theta=0.5\pi/2$;
(c) $\theta=0.75\pi/2$.}
\label{f:mscof}
\end{figure}
\clearpage
\newpage
\begin{figure}
\vspace*{-0.5cm}
\centerline{\psfig{file=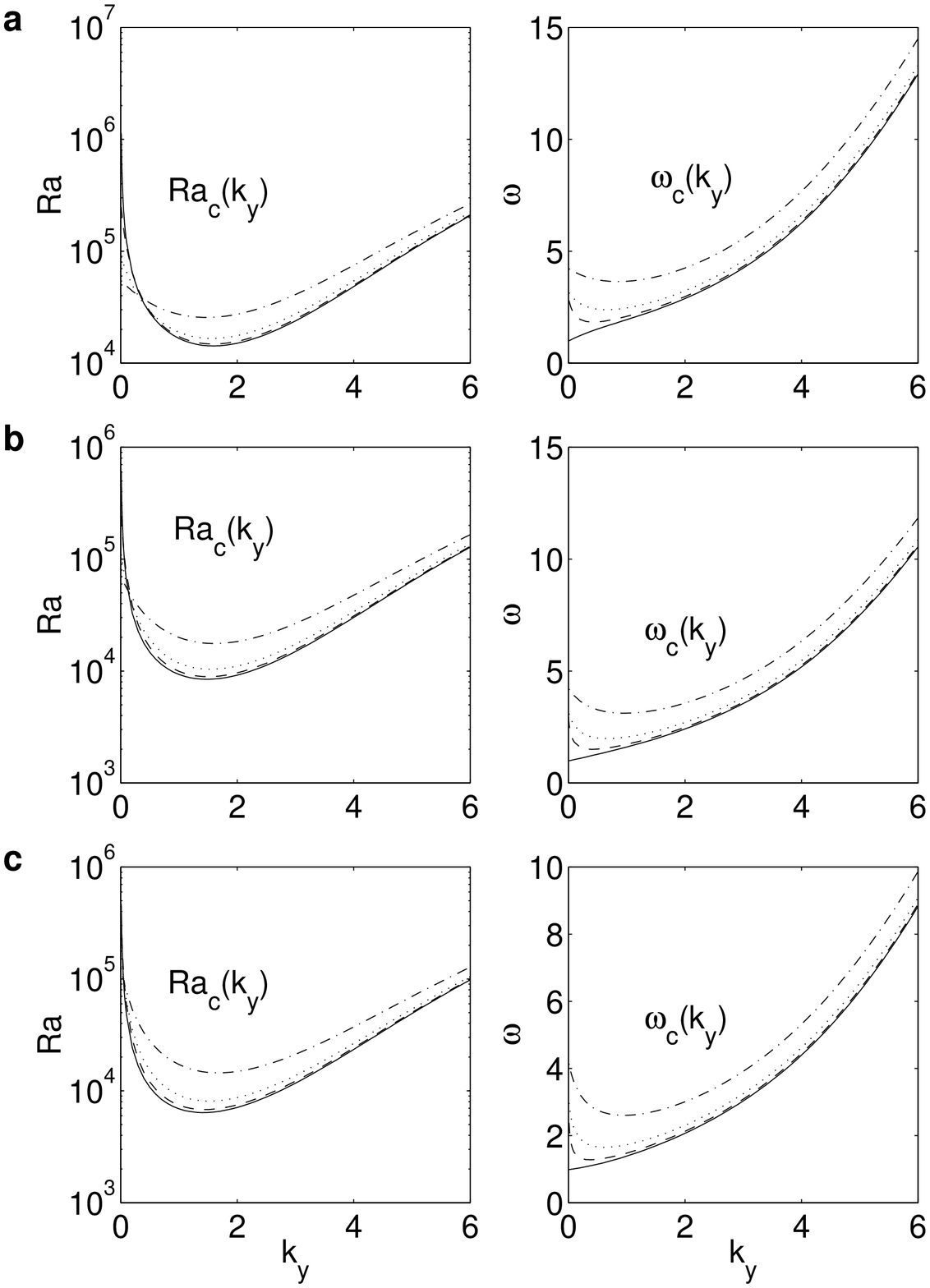,width=14.5cm}}
\caption{Viscous fluid and no-slip slot boundaries;
$\eta=\chi=1$. Curves of the marginal linear stability to 3D ($k_{z}\geq 0$)
oscillatory disturbances for different $k_{z}$, $Ra_{c}(k_{y})$ and $\omega_{c}(k_{y})$;
$\mu=1$, $Pr=6.7$, $Le=1$. The solid lines: $k_{z}=0$, the dashed lines:
$k_{z}=0.5$, the dotted lines: $k_{z}=1$, the dash-dot lines:
$k_{z}=2$. (a) $\theta=0.25\pi/2$; (b) $\theta=0.5\pi/2$;
(c) $\theta=0.75\pi/2$.}
\label{f:mscod}
\end{figure}
\clearpage
\newpage
\begin{figure}
\vspace*{-0.9cm}
\centerline{\psfig{file=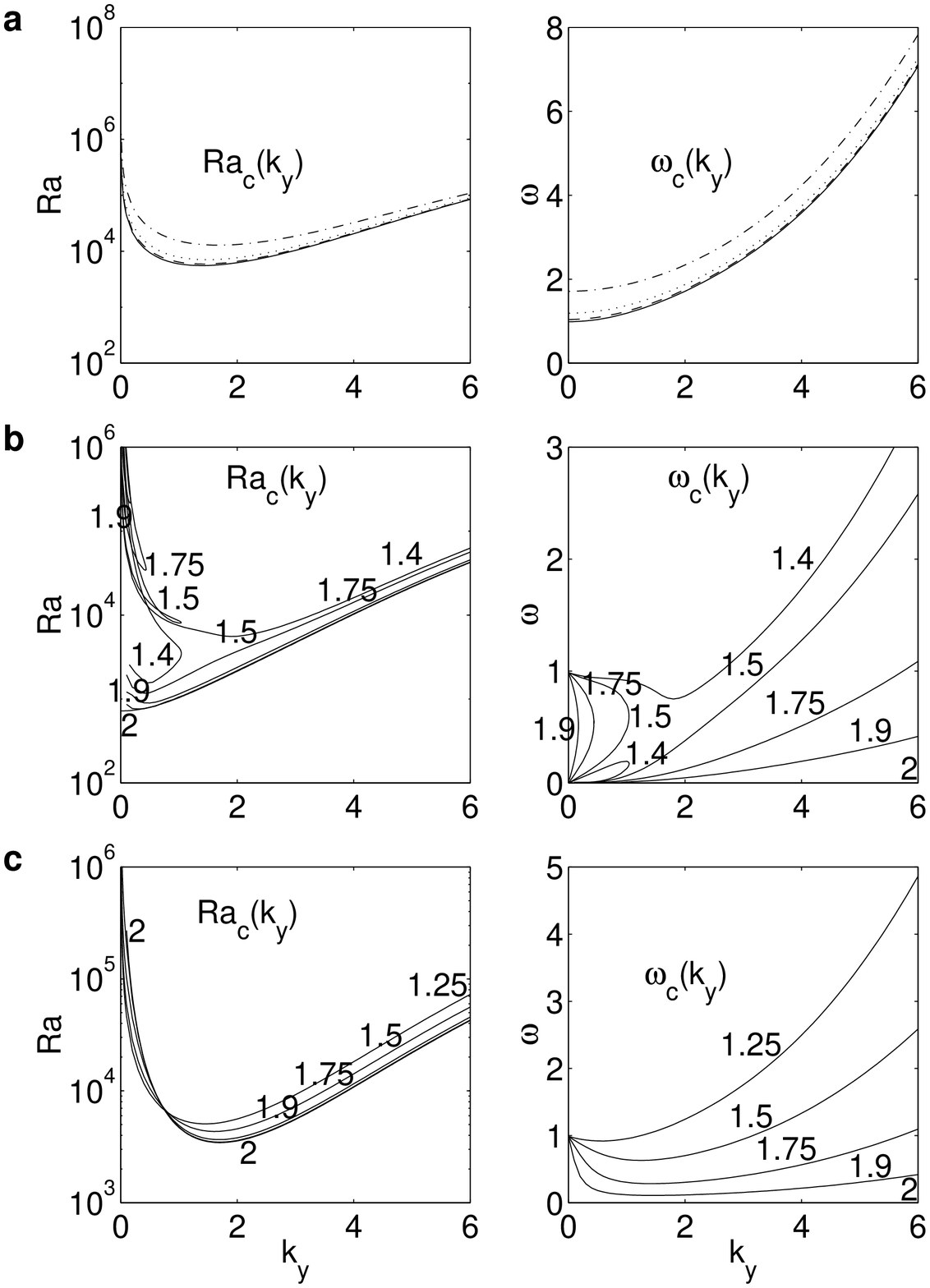,width=14.5cm}}
\vspace*{-0.45cm}
\caption{Viscous fluid and no-slip slot boundaries. 
Curves of the marginal linear stability to 3D ($k_{z}\geq 0$)
oscillatory disturbances for different $k_{z}$, $Ra_{c}(k_{y})$ and $\omega_{c}(k_{y})$;
$\mu=1$, $Pr=6.7$, $Le=1$. The solid lines: $k_{z}=0$, the dashed lines: $k_{z}=0.5$, the
dotted lines: $k_{z}=1$, the dash-dot lines: $k_{z}=2$. (a) $\theta=\pi/2$, $\eta=\chi=0$
and $\eta=\chi=1$; (b) $\eta=\chi=0$; (c) $\eta=\chi=1$.
In (b) and (c), $\theta=\vartheta\pi/2$, where the values of $\vartheta$ are given
in these figures next to their respective closest curves the numbers do
not intersect and for $\vartheta=2$, the instability is steady:
$\omega_{c}(k_{y})=0$. For the lowest $\omega_{c}(k_{y})>0$ in
(b) and for $\vartheta=2$ in (c), the data near $k_{y}=0$
(where such stable data were numerically difficult
to obtain) are not presented. For such data,
${Ra_{c}(k_{y})}_{|k_{y}\rightarrow 0}\rightarrow\infty$
are assumed.}
\label{f:mscoddff}
\end{figure}
\clearpage
\newpage
\begin{figure}
\vspace*{-0.5cm}
\centerline{\psfig{file=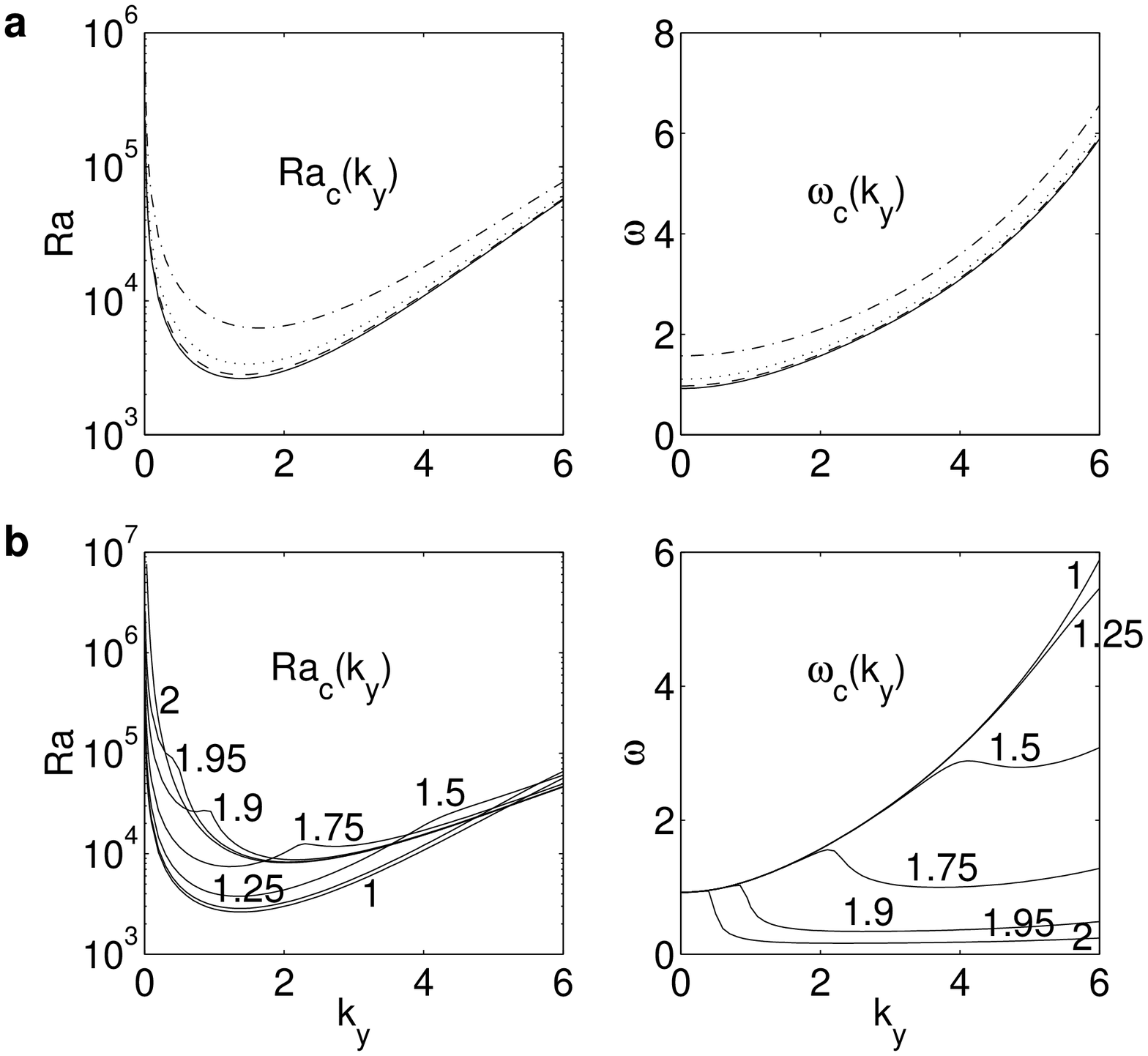,width=17cm}}
\vspace*{-0.3cm}
\caption{Viscous fluid and no-slip slot boundaries;
$\eta=0$, $\chi=1$. Curves of the marginal linear stability to 3D ($k_{z}\geq 0$)
oscillatory disturbances for different $k_{z}$, $Ra_{c}(k_{y})$ and $\omega_{c}(k_{y})$;
$\mu=1$, $Pr=6.7$, $Le=1$. The solid lines: $k_{z}=0$, the dashed lines: $k_{z}=0.5$,
the dotted lines: $k_{z}=1$, the dash-dot lines: $k_{z}=2$. (a) $\theta=\pi/2$;
(b) $\theta=\vartheta\pi/2$ and $\theta=(2-\vartheta)\pi/2$, where the values
of $\vartheta$ are given in the figure next to their respective closest curves
the numbers do not intersect and for $\vartheta=2$, the instability is steady:
$\omega_{c}(k_{y})=0$. The two values of $\theta$ in (b) are applicable in
light of transformation (\ref{eq:tt01}). For $\vartheta=2$ in (b),
the data near $k_{y}=0$ (where such stable data were numerically
difficult to obtain) are not presented.}
\label{f:mscodf}
\end{figure}
\clearpage
\newpage
\begin{figure}
\vspace*{-0.6cm}
\centerline{\psfig{file=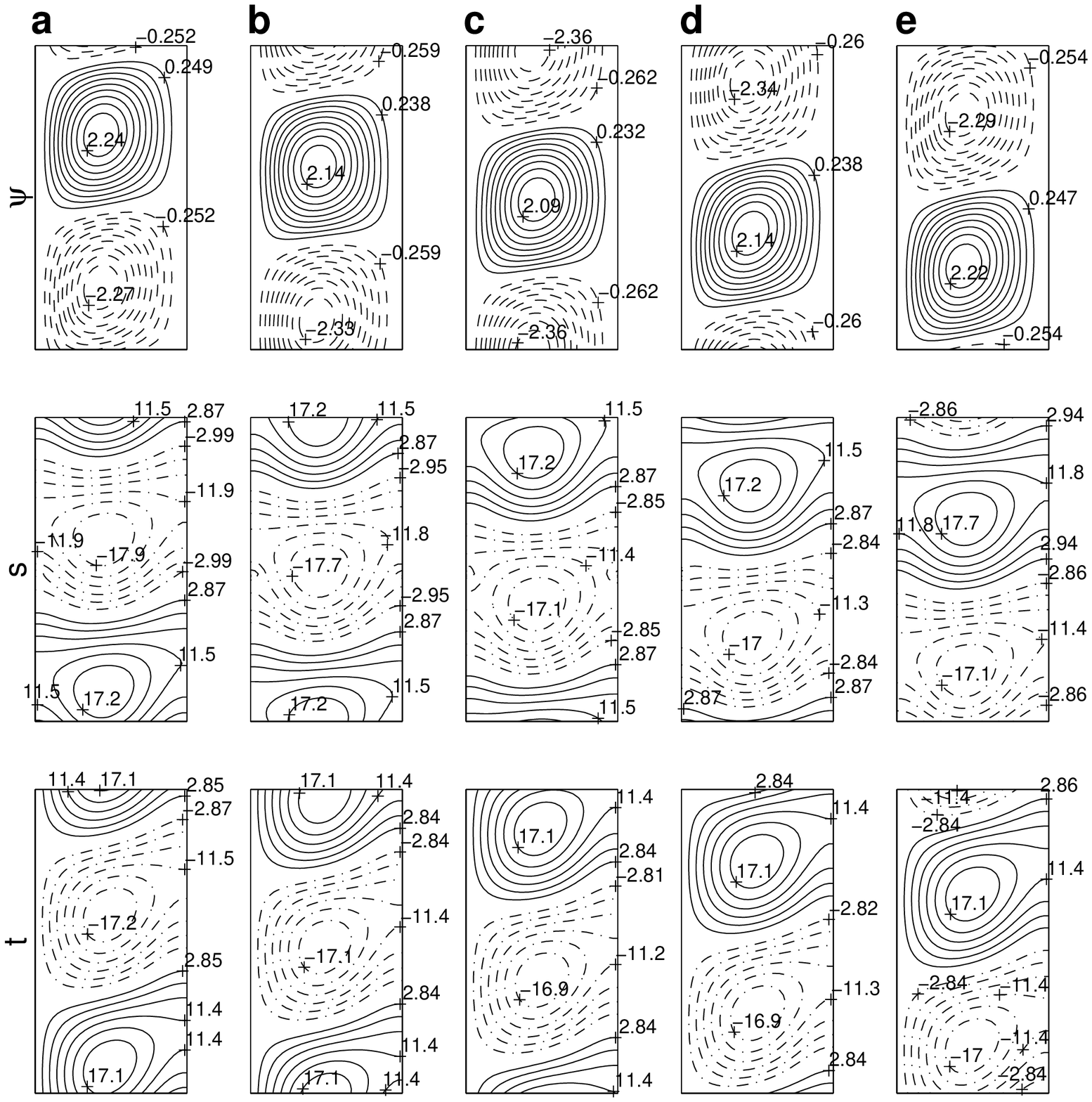,width=14.5cm}}
\caption{$\theta=\pi/2$. Viscous fluid and no-slip slot boundaries;
$\eta=\chi=0$. Perturbation temporal behavior throughout nearly a half of the oscillation
period $\tau_{p}\approx 46\delta\tau$ ($\delta\tau=0.05$) just beyond the onset of 2D
oscillatory instability. It was obtained from the numerical simulation of evolution of
the linearized Eqs. (\ref{eq:ns1})---(\ref{eq:dxi}) in response to the initial disturbance
proportional to the background state after initial time $\tau_{i}\approx 12000$ has passed;
$\lambda=2$, $\mu=1$, $Ra=11292$, $Pr=6.7$, $Le=1$. With this $\tau_{i}$, all perturbation
modes other than the unstable mode ($\tau_{p}\approx 46\delta\tau$) are practically negligible.
$\psi$: perturbation streamlines; $s$: isolines of solute concentration perturbation; $t$:
perturbation isotherms. The actual relative values of the streamfunction perturbation are
equal to $10^{-3}$ times the respective values in the figure. The solid and dashed
streamlines designate the clockwise and counterclockwise rotation and are equally
spaced within the positive and negative streamfunction intervals, respectively.
The solid and dash-dot isolines of the component perturbations are equally
spaced within the positive and negative component perturbation intervals,
respectively. Small asymmetries between the positive and negative phases
are due to the distortion introduced by the solute scale-fixing conditions.
The instability onset $Ra$ in the data underlying this figure is
different from that for Fig. \ref{f:perv11} only because the
scale-fixing conditions make the numerical formulation used
for these data slightly different from that used for
the data in Fig. \ref{f:perv11}.
(a) $\tau=\tau_{i}+\delta\tau$; (b) $\tau=\tau_{i}+6\delta\tau$; (c)
$\tau=\tau_{i}+11\delta\tau$; (d) $\tau=\tau_{i}+16\delta\tau$; (e)
$\tau=\tau_{i}+21\delta\tau$.}
\label{f:perv00}
\end{figure}
\clearpage
\newpage
\begin{figure}
\centerline{\psfig{file=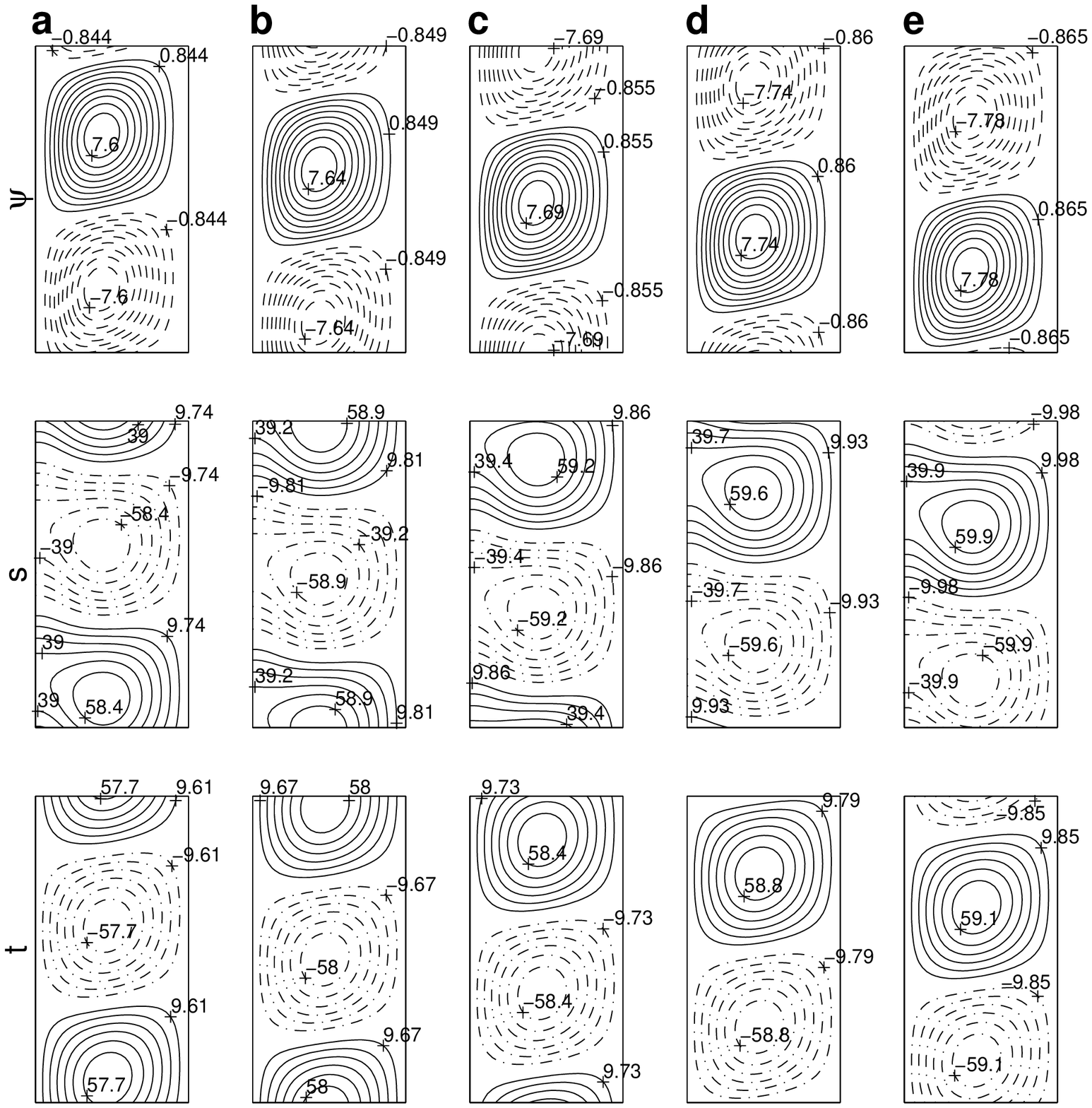,width=14.5cm}}
\caption{$\theta=\pi/2$. Viscous fluid and no-slip slot boundaries;
$\eta=\chi=1$. Perturbation temporal behavior throughout nearly a half of the oscillation
period $\tau_{p}\approx 46\delta\tau$ ($\delta\tau=0.05$) just beyond the onset of 2D
oscillatory instability. It was obtained from the numerical simulation of evolution of
the linearized Eqs. (\ref{eq:ns1})---(\ref{eq:dxi}) in response to the initial disturbance
proportional to the background state after initial time $\tau_{i}\approx 1200$ has passed;
$\lambda=2$, $\mu=1$, $Ra=11226$, $Pr=6.7$, $Le=1$. With this $\tau_{i}$, all perturbation
modes other than the unstable mode ($\tau_{p}\approx 46\delta\tau$) are practically negligible.
$\psi$: perturbation streamlines; $s$: isolines of solute concentration perturbation; $t$:
perturbation isotherms. The actual relative values of the streamfunction perturbation are
equal to $10^{-3}$ times the respective values in the figure. The solid and dashed
streamlines designate the clockwise and counterclockwise rotation and are equally
spaced within the positive and negative streamfunction intervals, respectively.
The solid and dash-dot isolines of the component perturbations are equally
spaced within the positive and negative component perturbation intervals,
respectively. The instability onset $Ra$ in the data underlying this
figure is different from that for Fig. \ref{f:perv00} only because
a slightly different numerical formulation is used.
(a) $\tau=\tau_{i}+\delta\tau$; (b) $\tau=\tau_{i}+6\delta\tau$; (c)
$\tau=\tau_{i}+11\delta\tau$; (d) $\tau=\tau_{i}+16\delta\tau$; (e)
$\tau=\tau_{i}+21\delta\tau$.}
\label{f:perv11}
\end{figure}
\clearpage
\newpage
\begin{figure}
\centerline{\psfig{file=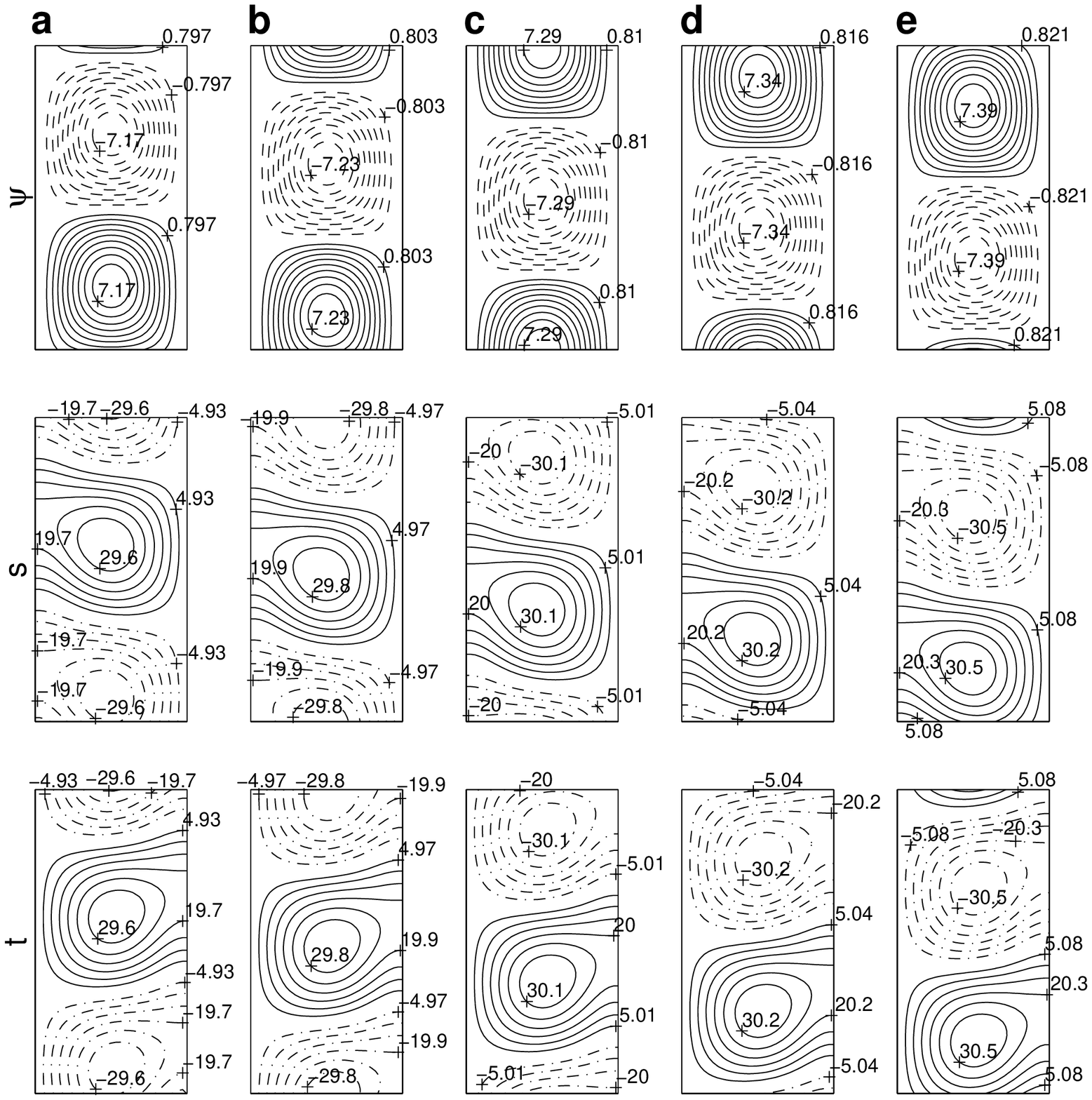,width=14.5cm}}
\vspace*{0.5cm}
\caption{$\theta=\pi/2$. Viscous fluid and no-slip slot boundaries; $\eta=0$,
$\chi=1$. Perturbation temporal behavior throughout nearly a half of the oscillation
period $\tau_{p}\approx 51\delta\tau$ ($\delta\tau=0.05$) just beyond the onset of 2D
oscillatory instability. It was obtained from the numerical simulation of evolution of
the linearized Eqs. (\ref{eq:ns1})---(\ref{eq:dxi}) in response to the initial disturbance
proportional to the background state after initial time $\tau_{i}\approx 1100$ has passed;
$\lambda=2$, $\mu=1$, $Ra=5586$, $Pr=6.7$, $Le=1$. With this $\tau_{i}$, all perturbation
modes other than the unstable mode ($\tau_{p}\approx 51\delta\tau$) are practically negligible.
$\psi$: perturbation streamlines; $s$: isolines of solute concentration perturbation; $t$:
perturbation isotherms. The actual relative values of the streamfunction perturbation are
equal to $10^{-3}$ times the respective values in the figure. The solid and dashed
streamlines designate the clockwise and counterclockwise rotation and are equally
spaced within the positive and negative streamfunction intervals, respectively.
The solid and dash-dot isolines of the component perturbations are equally
spaced within the positive and negative component perturbation intervals,
respectively. (a) $\tau=\tau_{i}+2\delta\tau$; (b) $\tau=\tau_{i}+7\delta\tau$;
(c) $\tau=\tau_{i}+13\delta\tau$; (d) $\tau=\tau_{i}+18\delta\tau$; (e)
$\tau=\tau_{i}+23\delta\tau$.}
\label{f:perv01}
\end{figure}
\clearpage
\newpage
\begin{figure}
\centerline{\psfig{file=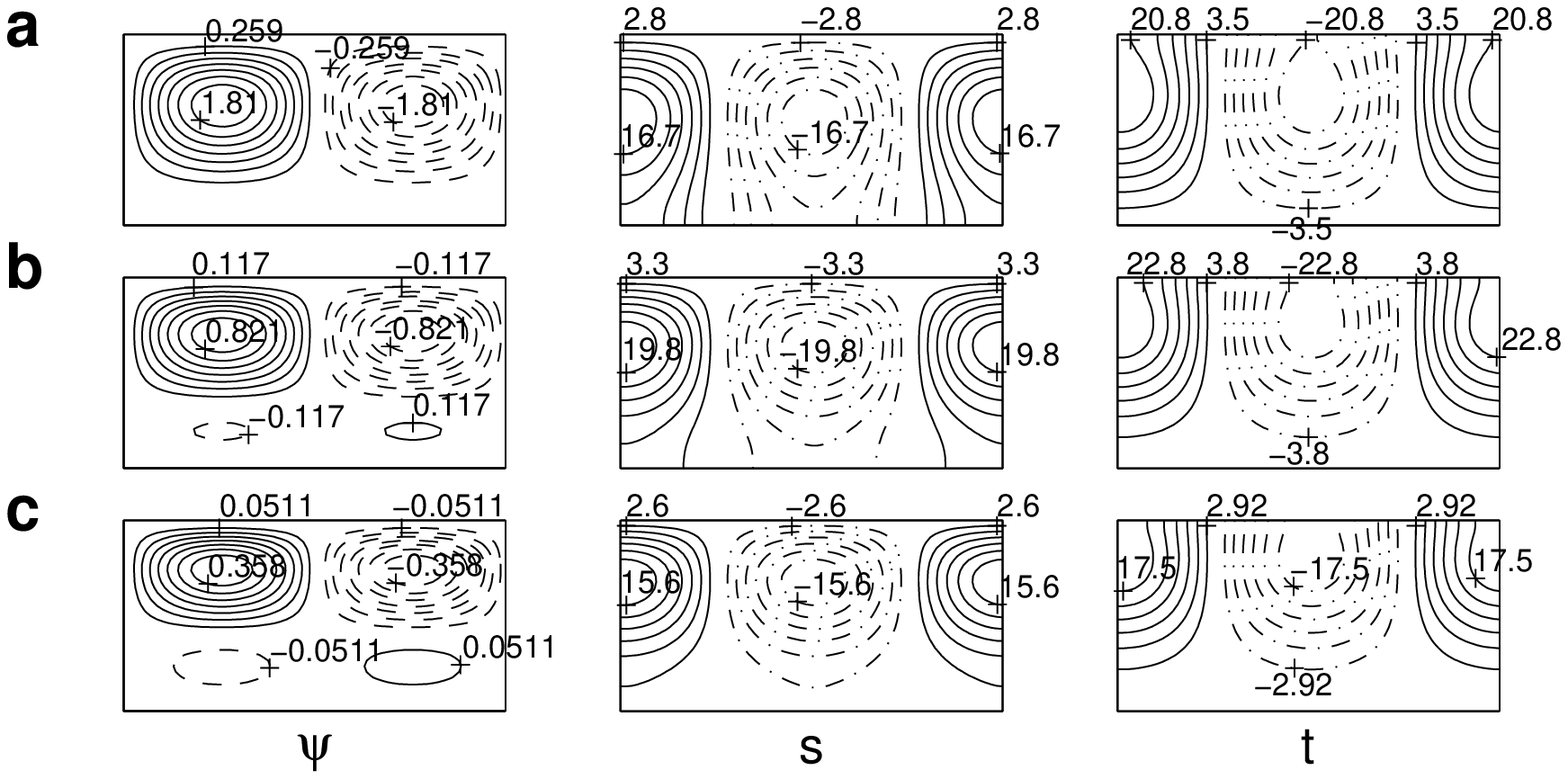,width=14.5cm}}
\vspace*{0.5cm}
\caption{$\theta=0$. Viscous fluid and no-slip slot
boundaries; $\eta=0$, $\chi=1$. [In light of transformation
(\ref{eq:tt01}), these results also apply to $\theta=\pi$ if
$Ra_{c}\mapsto Ra_{c}^{s}$, $Ra^{s}\mapsto Ra$, and thus $\mu\mapsto 1/\mu$.]
2D singular eigenvectors corresponding to the wavelength $\lambda=2$ at the onset
of small-amplitude steady convection; $Le=1$. $\psi$: perturbation streamlines;
$s$: isolines of solute concentration perturbation; $t$: perturbation isotherms.
The variables are nondimensionalized as in Eqs. (\ref{eq:ns1})---(\ref{eq:dxi}).
The actual relative values of the streamfunction perturbation are equal to
$10^{-3}$ times the respective values in the figure. The solid and dashed
streamlines designate the clockwise and counterclockwise rotation and
are equally spaced within the positive and negative streamfunction
intervals, respectively. The solid and dash-dot isolines of the
component perturbations are equally spaced within the positive
and negative component perturbation intervals, respectively.
(a) $\mu=1$, $Ra_{c}=9787$; (b) $\mu=1.3$, $Ra_{c}=25478$;
(c) $\mu=1.5$, $Ra_{c}=46908$.}
\label{f:perhs01}
\end{figure}
\clearpage
\newpage
\begin{figure}
\centerline{\psfig{file=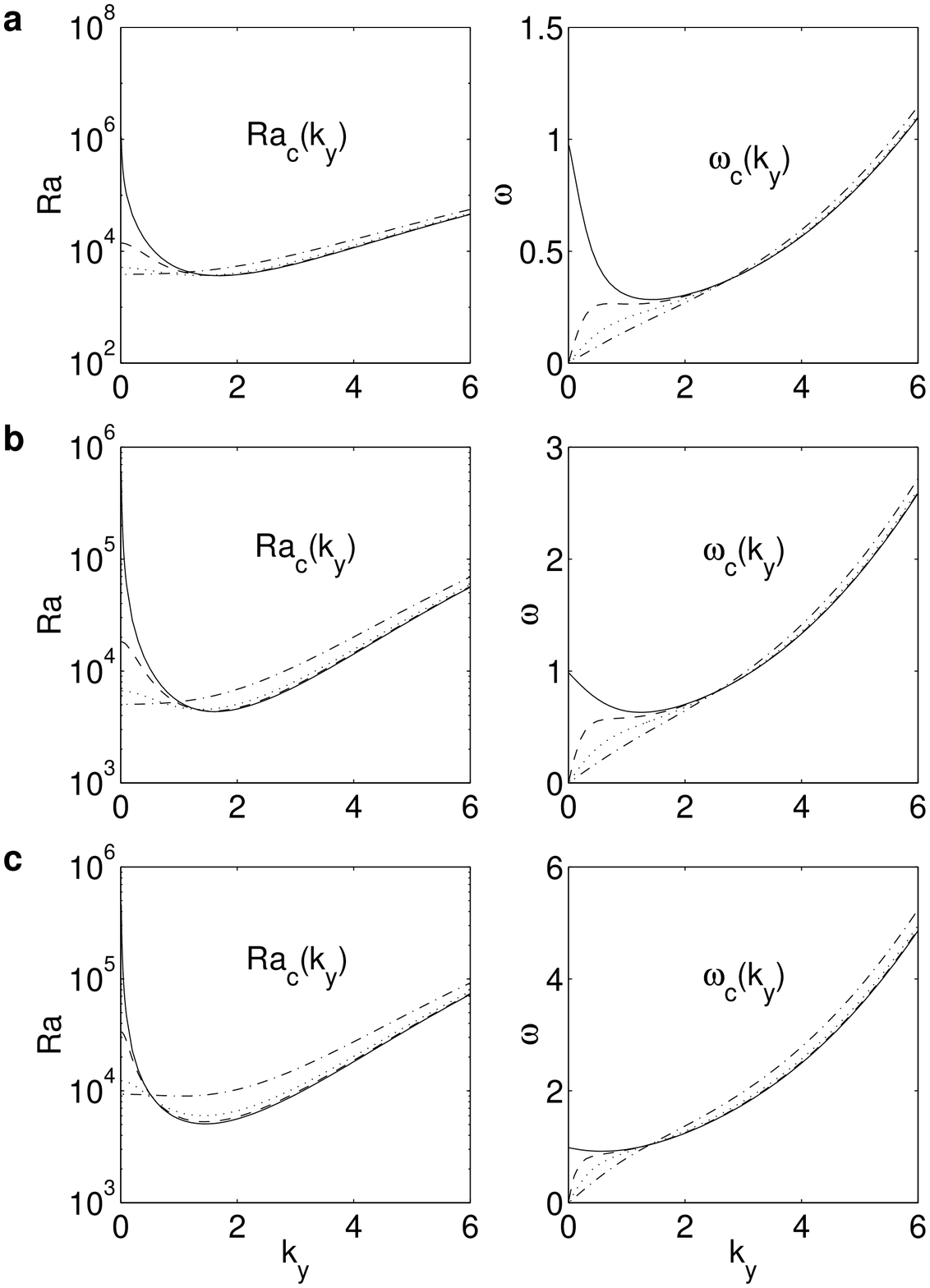,width=14.5cm}}
\vspace*{0.5cm}
\caption{Viscous fluid and no-slip slot boundaries;
$\eta=\chi=1$. Curves of the marginal linear stability to 3D ($k_{z}\geq 0$)
oscillatory disturbances for different $k_{z}$, $Ra_{c}(k_{y})$ and $\omega_{c}(k_{y})$;
$\mu=1$, $Pr=6.7$, $Le=1$. The solid lines: $k_{z}=0$, the dashed lines:
$k_{z}=0.5$, the dotted lines: $k_{z}=1$, the dash-dot lines:
$k_{z}=2$. (a) $\theta=1.75\pi/2$; (b) $\theta=1.5\pi/2$;
(c) $\theta=1.25\pi/2$.}
\label{f:ttp11}
\end{figure}
\clearpage
\newpage
\begin{figure}
\centerline{\psfig{file=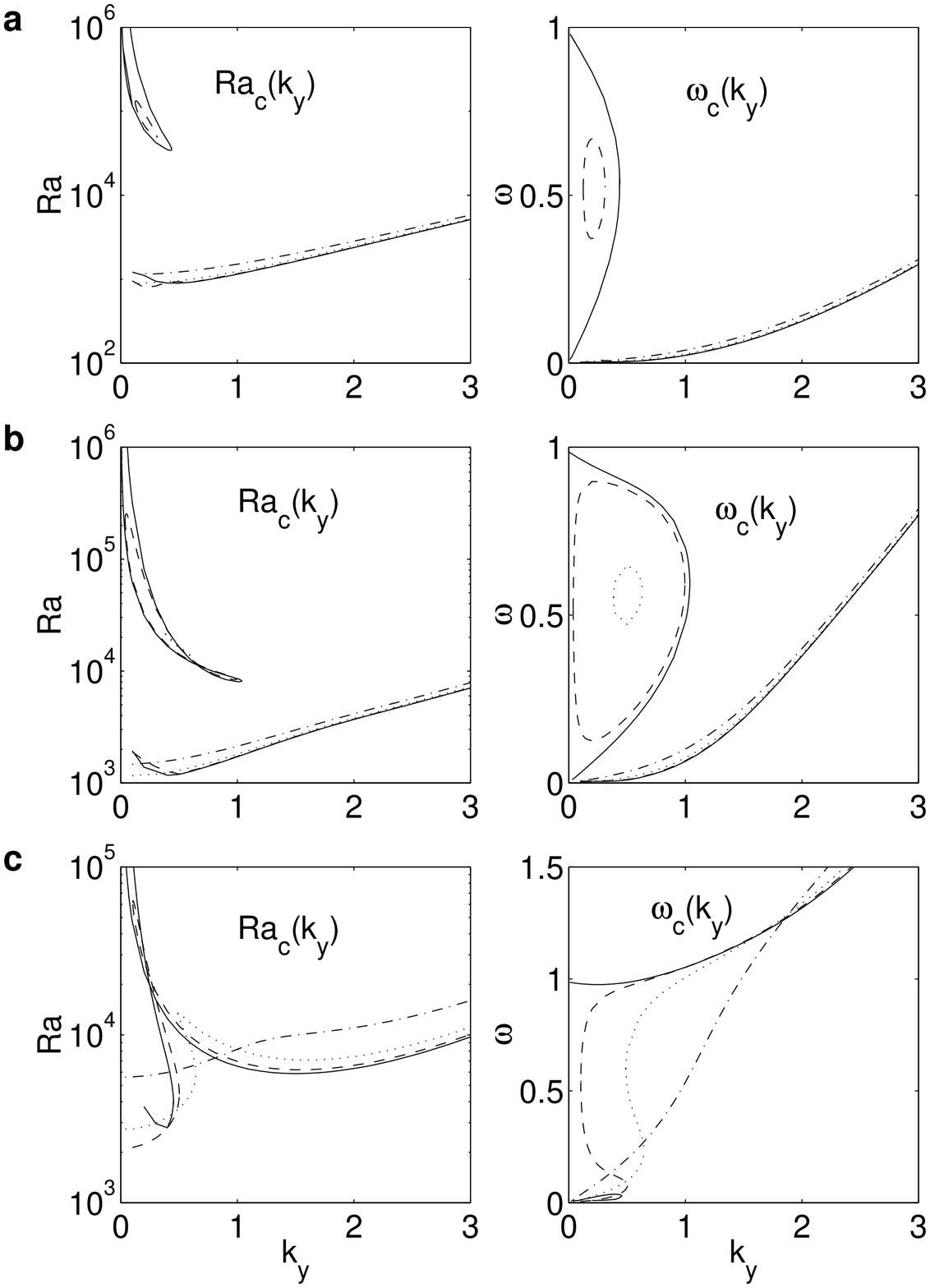,width=14.5cm}}
\vspace*{-0.4cm}
\caption{Viscous fluid and no-slip slot boundaries;
$\eta=\chi=0$. Curves of the marginal linear stability to 3D
($k_{z}\geq 0$) oscillatory disturbances for different $k_{z}$, $Ra_{c}(k_{y})$
and $\omega_{c}(k_{y})$; $\mu=1$, $Pr=6.7$, $Le=1$. The solid lines: $k_{z}=0$,
the dashed lines: $k_{z}=0.2$ in (a) and (b) and $k_{z}=0.5$ in (c), the dotted
lines: $k_{z}=0.5$ in (a) and (b) and $k_{z}=1$ in (c), the dash-dot lines: $k_{z}=1$
in (a) and (b) and $k_{z}=2$ in (c). (a) $\theta=1.75\pi/2$; (b) $\theta=1.5\pi/2$;
(c) $\theta=1.25\pi/2$. The lowest-$\omega_{c}(k_{y})$ data near $k_{y}=0$
(where such stable data were numerically difficult to obtain) are not
presented. Additional quantitative details are given
in Table \ref{t:lul}.}
\label{f:ttp00}
\end{figure}
\clearpage
\newpage
\begin{figure}
\vspace*{-0.5cm}
\centerline{\psfig{file=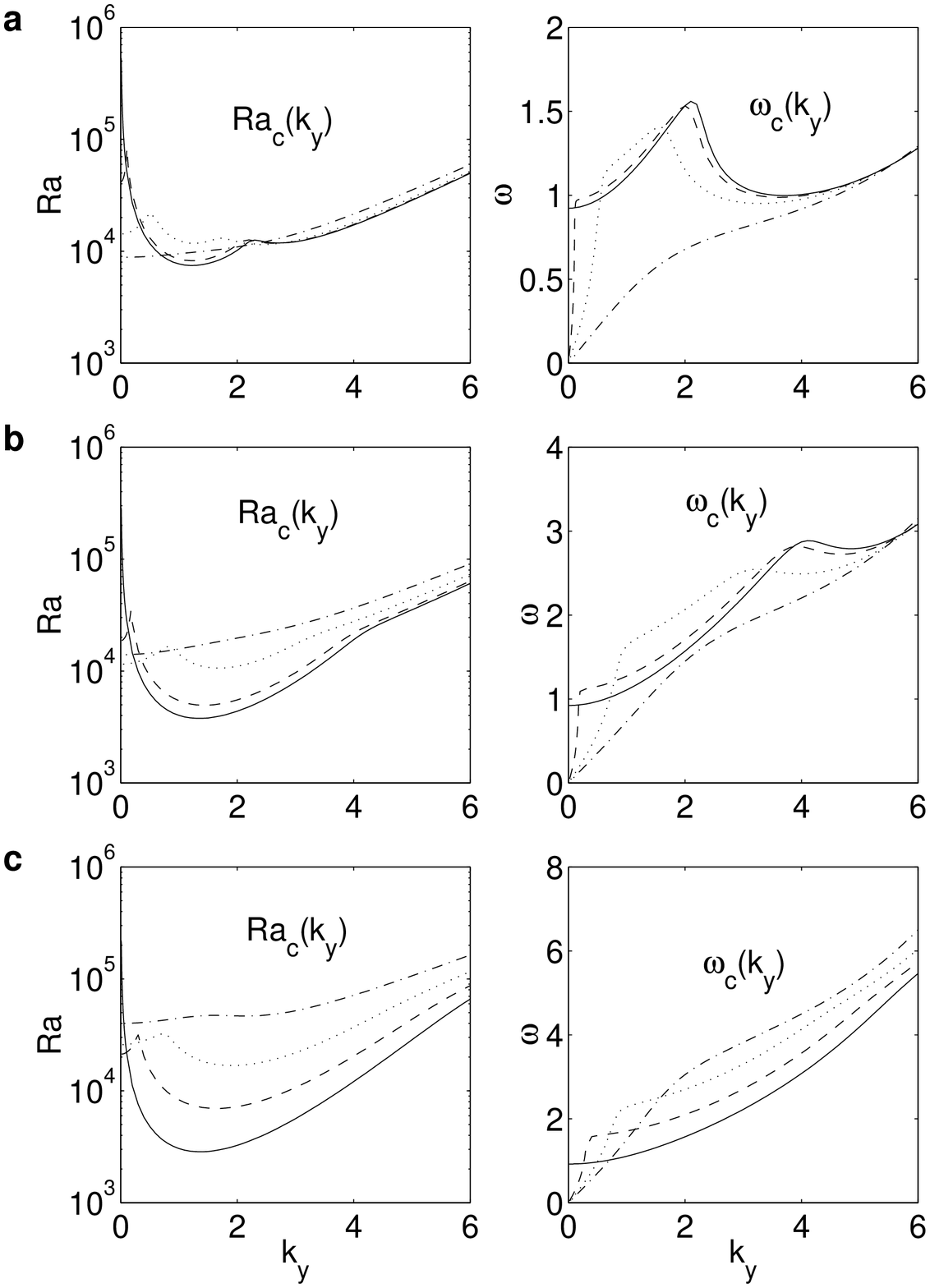,width=14.5cm}}
\vspace*{-0.3cm}
\caption{Viscous fluid and no-slip slot boundaries;
$\eta=0$, $\chi=1$. Curves of the marginal linear stability to 3D ($k_{z}\geq 0$)
oscillatory disturbances for different $k_{z}$, $Ra_{c}(k_{y})$ and $\omega_{c}(k_{y})$;
$\mu=1$, $Pr=6.7$, $Le=1$. (a) $\theta=(1\pm0.75)\pi/2$; (b) $\theta=(1\pm0.5)\pi/2$; (c)
$\theta=(1\pm0.25)\pi/2$. The two values of $\theta$ are relevant in light of transformation
(\ref{eq:tt01}). The solid lines: $k_{z}=0$; the dashed lines: (a) $k_{z}=0.5$, (b) $k_{z}=1$,
(c) $k_{z}=2$; the dotted lines: (a) $k_{z}=1$, (b) $k_{z}=2$, (c) $k_{z}=3$;
the dash-dot lines: (a) $k_{z}=2$, (b) $k_{z}=3$, (c) $k_{z}=4$.
Additional quantitative details are given in Table \ref{t:jk}.}
\label{f:ttp01}
\end{figure}
\clearpage
\newpage
\begin{figure}
\vspace*{-0.8cm}
\centerline{\psfig{file=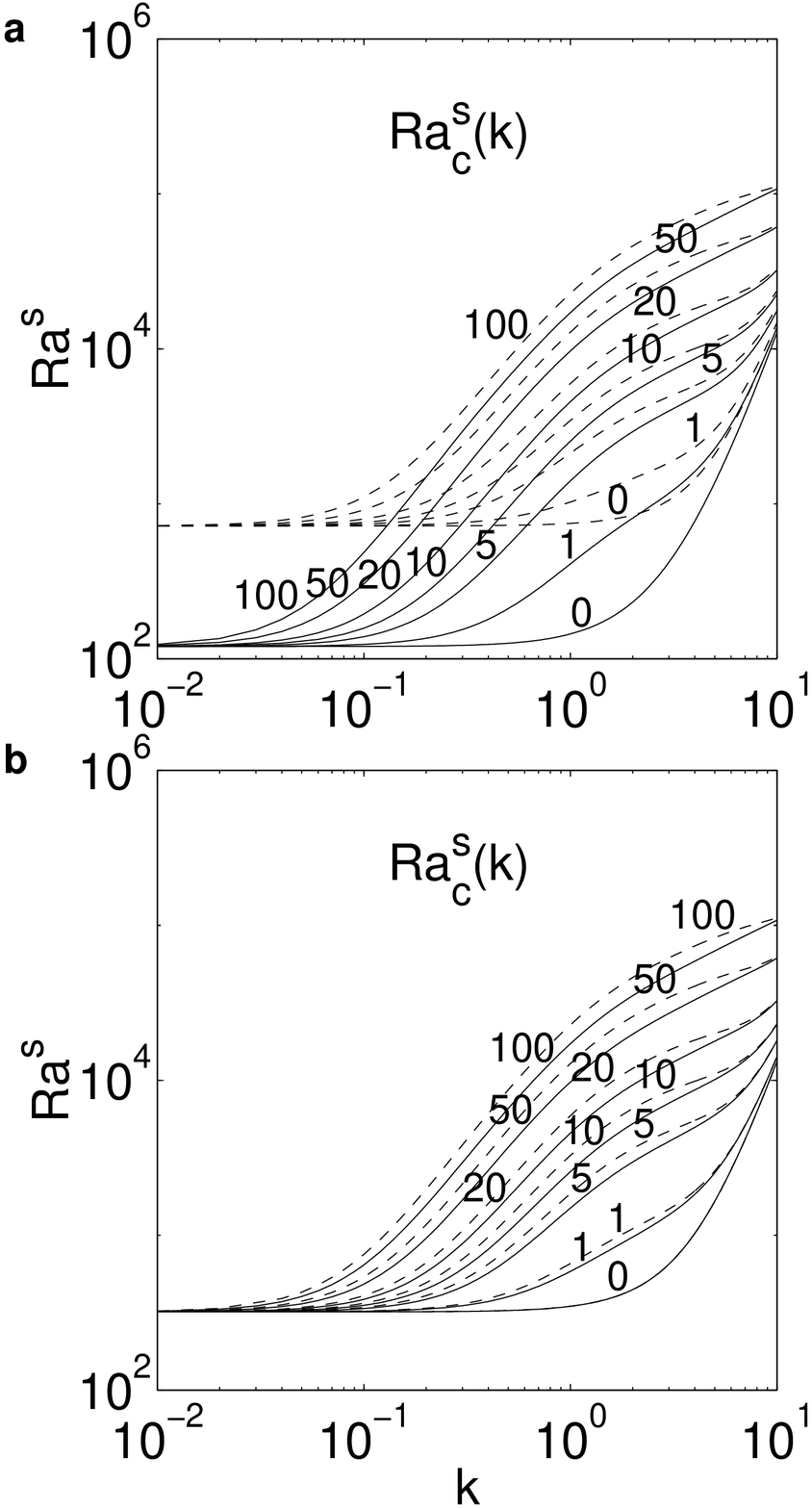,width=11.4cm}}
\vspace*{-0.5cm}
\caption{$\theta=\pi$. Viscous fluid; $\eta=\chi=0$. Curves of the marginal linear stability
to steady disturbances, $Ra_{c}^{s}(k)$, for different $Ra$; $Le=1$. The numbers multiplied
by $10^{3}$ give the values of $Ra$ for which the respective closest lines below them were
computed. (a) the solid lines: $\gamma_{\pm}=0$, the dashed lines: $\gamma_{\pm}=1$;
(b) the solid lines: $\gamma_{-}=0$ and $\gamma_{+}=1$, the dashed lines:
$\gamma_{-}=1$ and $\gamma_{+}=0$. The most unstable wave numbers
are described in Table \ref{t:km}.}
\label{f:sp00}
\end{figure}
\clearpage
\newpage
\begin{figure}
\vspace*{-0.8cm}
\centerline{\psfig{file=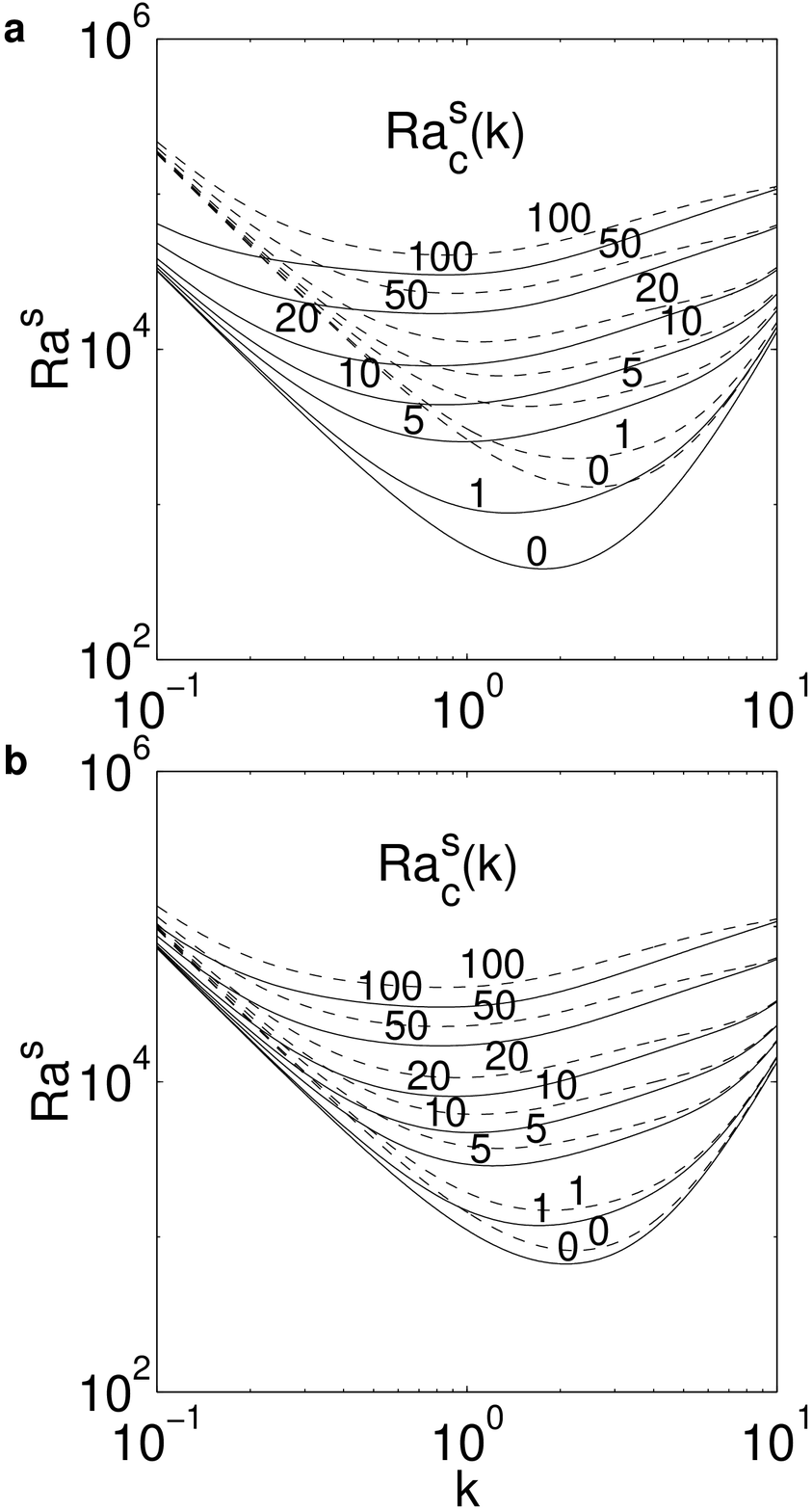,width=11.4cm}}
\vspace*{-0.5cm}
\caption{$\theta=\pi$. Viscous fluid; $\eta=\chi=1$. Curves of the marginal linear stability
to steady disturbances, $Ra_{c}^{s}(k)$, for different $Ra$; $Le=1$. The numbers multiplied
by $10^{3}$ give the values of $Ra$ for which the respective closest lines below them were
computed. (a) the solid lines: $\gamma_{\pm}=0$, the dashed lines: $\gamma_{\pm}=1$;
(b) the solid lines: $\gamma_{-}=0$ and $\gamma_{+}=1$, the dashed lines:
$\gamma_{-}=1$ and $\gamma_{+}=0$. The most unstable wave numbers
are described in Table \ref{t:km}.}
\label{f:sp11}
\end{figure}
\clearpage
\newpage
\begin{figure}
\vspace*{-0.8cm}
\centerline{\psfig{file=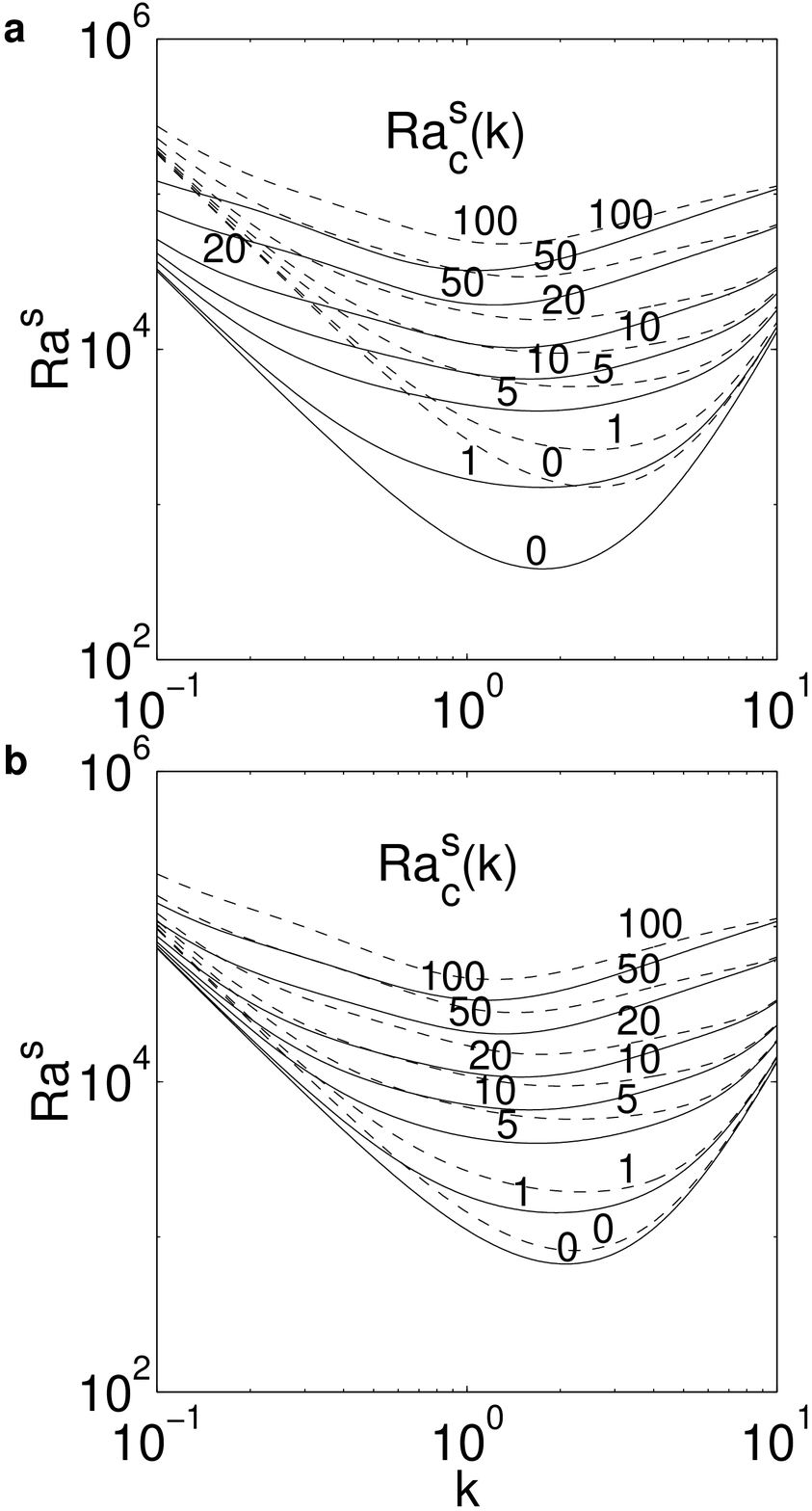,width=11.4cm}}
\vspace*{-0.5cm}
\caption{$\theta=\pi$. Viscous fluid;
$\eta=0$, $\chi=1$. [These data also apply to $\theta=0$ if transformation
(\ref{eq:tt01}) along with $Ra\mapsto Ra^{s}$, $Ra_{c}^{s}\mapsto Ra_{c}$, and
$\gamma_{\pm}\mapsto\gamma_{\mp}$ are allowed for.] Curves of the marginal linear stability
to steady disturbances, $Ra_{c}^{s}(k)$, for different $Ra$; $Le=1$. The numbers multiplied
by $10^{3}$ give the values of $Ra$ for which the respective closest lines below them were
computed. (a) the solid lines: $\gamma_{\pm}=0$, the dashed lines: $\gamma_{\pm}=1$;
(b) the solid lines: $\gamma_{-}=0$ and $\gamma_{+}=1$, the dashed lines:
$\gamma_{-}=1$ and $\gamma_{+}=0$. The most unstable wave numbers
are described in Table \ref{t:km}.\newline}
\label{f:sp01}
\end{figure}
\clearpage
\newpage
\begin{figure}
\centerline{\psfig{file=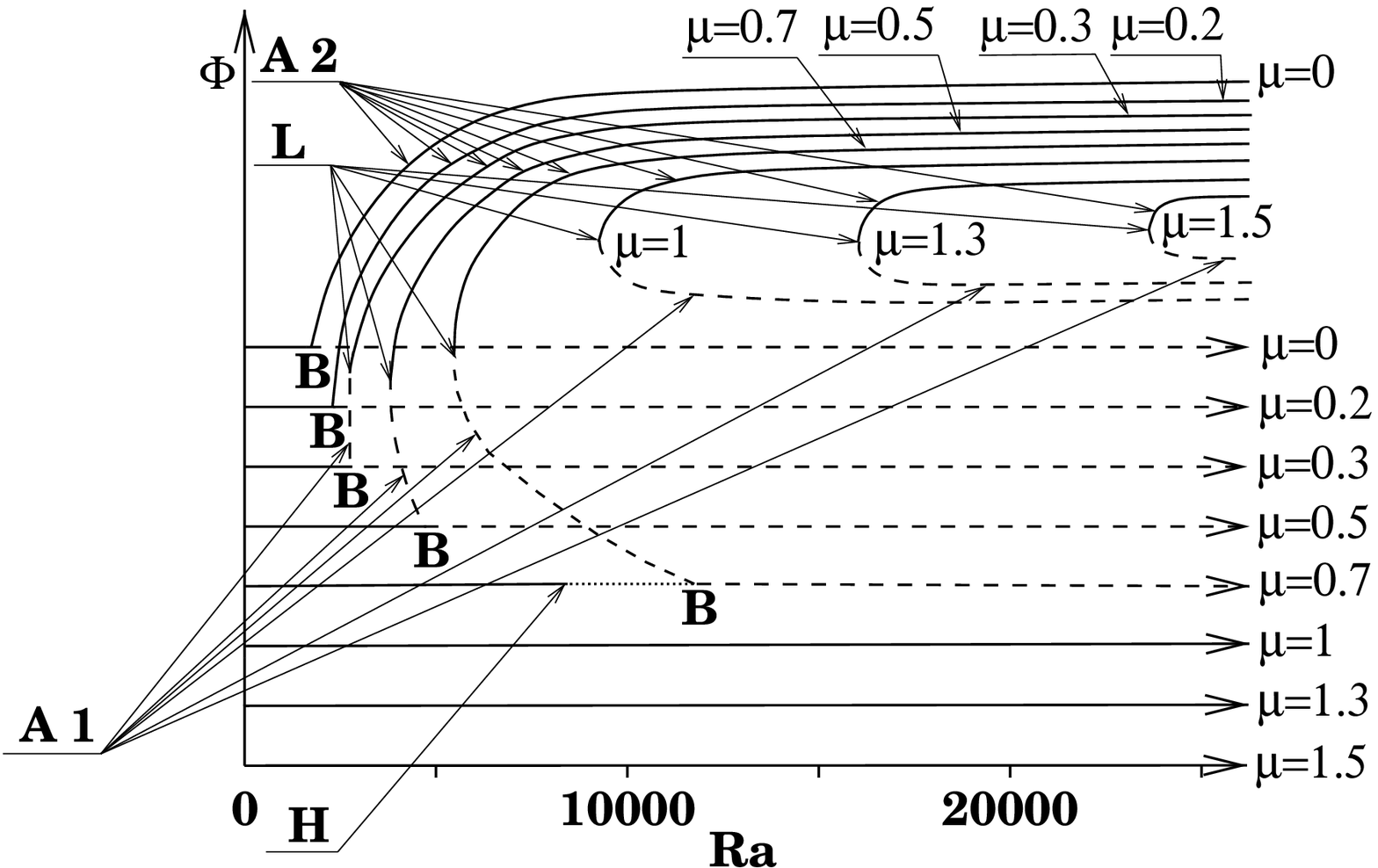,width=14.5cm}}
\vspace*{0.5cm}
\caption{$\theta=0$. Viscous fluid and no-slip slot
boundaries; $\eta=\chi=1$. Schematic structures of the
2D steady flows with minimal along-slot period $\lambda=2$
for $\mu\in[0,1.5]$; $Pr=6.7$, $Le=1$. $\Phi$ is an abstract
measure of the steady flows that distinguishes between different
solutions, specifies the location of the singularities (limit points
and symmetry-breaking bifurcations), and represents the flows arising
from a symmetry-breaking bifurcation as a single branch. The background
states are depicted by the horizontal lines with arrows (for $\mu=1.5$,
this is the coordinate axis). The solid lines stand for the solutions
being stable to the disturbances associated with the eigenvalues that
give rise to the steady instability of the conduction state. The dashed
lines represent the flows being unstable to either steady or both steady
and oscillatory disturbances. The dotted lines stand for the solutions
being unstable to oscillatory disturbances alone. Secondary bifurcations,
if any, are not shown. $B$ is the symmetry-breaking bifurcation standing
for the steady linear stability boundary for wave number $k=\pi$
($\lambda=2$). Its criticality changes at $\mu$ just below $0.3$.
$L$ is the limit point. $A1$ and $A2$ are the lower- and
higher-amplitude branches associated with the limit point,
respectively. H is a \mbox{Hopf} bifurcation. For $\mu=1$,
it arises just below $Ra\approx 85040$
(Fig. \ref{f:perh11}).}
\label{f:bd}
\end{figure}
\clearpage
\newpage
\begin{figure}
\centerline{\psfig{file=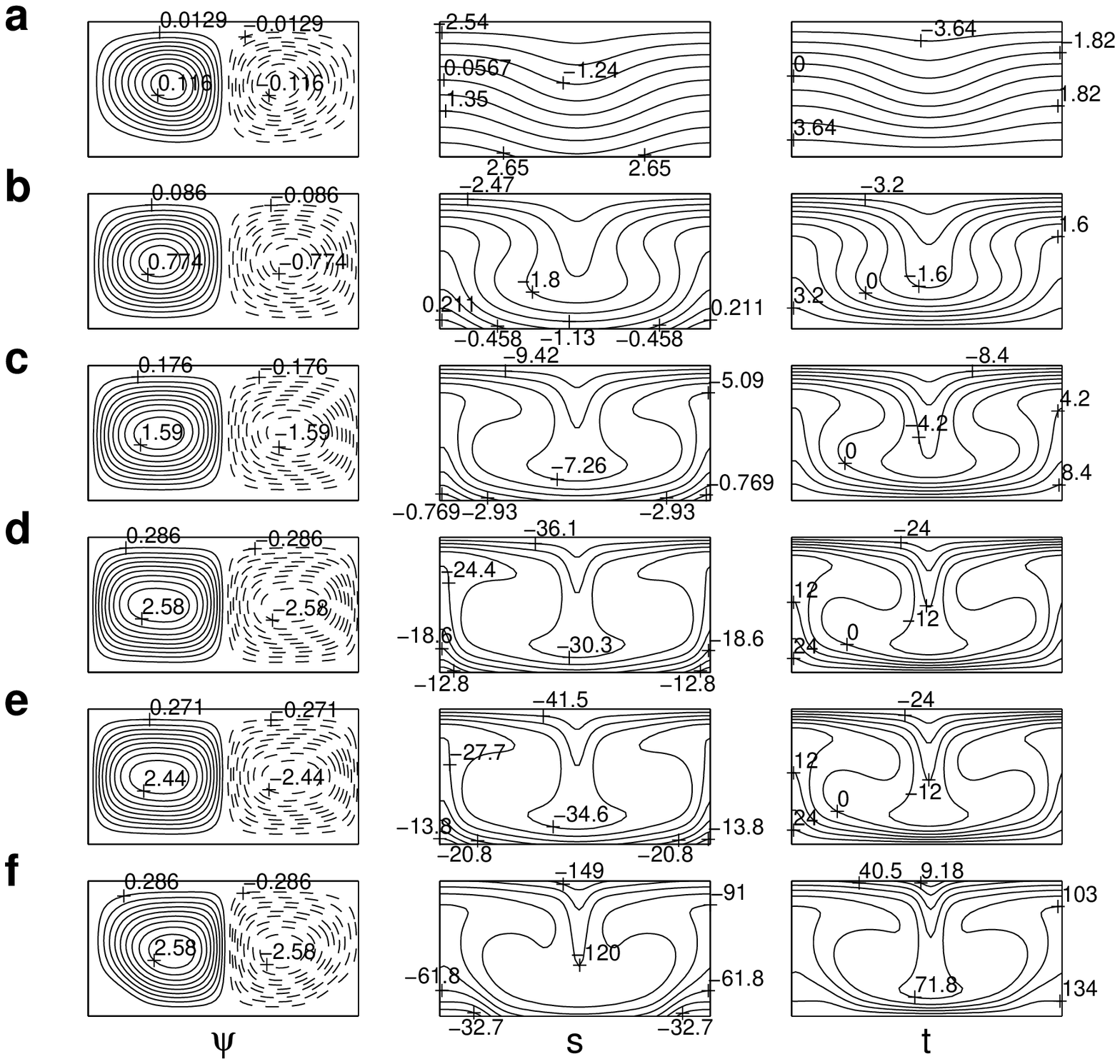,width=14.5cm}}
\vspace*{0.5cm}
\caption{$\theta=0$. Viscous fluid and no-slip slot boundaries;
$Pr=6.7$, $Le=1$, $\lambda=2$. 2D convective steady flows. $\psi$:
streamlines; $s$: isolines of solute concentration; $t$: isotherms. The
solid and dashed streamlines represent the clockwise and counterclockwise 
rotation and are equally spaced within the positive and negative streamfunction
intervals, respectively. The actual values of $s$ and $t$ are equal to $10^3$
times the respective values in the figure. (a)---(e): $\eta=\chi=1$. (a) $\mu=0.7$,
$Ra=9107$, branch $A1$; (b) $\mu=0.7$, $Ra=8000$, branch $A2$; (c) $\mu=1$,
$Ra=24000$, branch $A2$; (d) $\mu=1.3$, $Ra=60000$, branch $A2$; (e) 
$\mu=1.5$, $Ra=60000$, branch $A2$; (f) $\eta=\chi\approx 0.41$,
$\mu=1$, $Ra=300000$, branch $A2$.}
\label{f:ff11}
\end{figure}
\clearpage
\newpage
\begin{table}
\vspace*{-0.5cm}
\caption{Inviscid fluid; $\eta=0$, $\chi=1$; $\mu=1$, $Le=1$.
$k_{y}^{2l}(k_{z})$, $k_{y}^{2u}(k_{z})$, $k_{y}^{3l}(k_{z})$,
and $k_{y}^{3u}(k_{z})$ are approximate values of wave number
$k_{y}$ at which $Ra_{c}(k_{y})$ and $\omega_{c}(k_{y})$ are
expected to decrease to zero for different $\theta$ and $k_{z}$.
$k_{y}^{2l}(k_{z})$ and $k_{y}^{2u}(k_{z})$ originate at
$k_{z}=0$, whereas $k_{y}^{3l}(k_{z})$ and $k_{y}^{3u}(k_{z})$ arise only at $k_{z}>0$.
For $k_{z}\geq 0$, $q_{y}^{l}(k_{z})\equiv[{k_{y}^{2l}(0)}^2-k_{z}^{2}]^{1/2}$
and $q_{y}^{u}(k_{z})\equiv[{k_{y}^{2u}(0)}^2-k_{z}^{2}]^{1/2}$
are provided for comparison with $k_{y}^{2l}(k_{z})$ and
$k_{y}^{2u}(k_{z})$, respectively. The two values of
$\theta$ are relevant due to transformation
(\ref{eq:tt01}).}
\label{t:kc}
\vspace*{0.5cm}
\begin{tabular}{|c|c|c|c|c|c|c|c|} \hline
$\theta$    & $k_{z}$ &  $k_{y}^{3l}$ & $k_{y}^{3u}$ & $k_{y}^{2l}$ & $k_{y}^{2u}$ & $q_{y}^{l}$ & $q_{y}^{u}$ \\ \hline
            & $0.0$   &  Not found    & Not found    & $0.3500$     & $0.8954$     & $0.3500$  & $0.8954$         \\ \cline{2-8}
            & $0.1$   &  $0.0109$     & $0.0312$     & $0.3189$     & $0.8844$     & $0.3354$  & $0.8898$ \\ \cline{2-8}
            & $0.15$  &  $0.0249$     & $0.0840$     & $0.2660$     & $0.8704$     & $0.3162$  & $0.8827$ \\ \cline{2-8} 
$(1\mp 0.85)\pi/2$ & $0.2$   &  $0.0453$     & Not found    & Not found    & $0.8496$     &           & $0.8728$ \\ \cline{2-8}
            & $0.3$   &  $0.1105$     & Not found    & Not found    & $0.7847$     &           & $0.8436$ \\ \cline{2-8}
            & $0.4$   &  $0.2327$     & Not found    & Not found    & $0.6622$     &           & $0.8011$ \\ \cline{2-8}
            & $0.5$   &  Not found    & Not found    & Not found    & Not found    &           &          \\ \hline
            & $0.0$   &  Not found    & Not found    & $0.5959$     & $1.4562$     & $0.5959$  & $1.4562$         \\ \cline{2-8}
            & $0.1$   &  $0.0063$     & $0.0169$     & $0.5789$     & $1.4495$     & $0.5874$  & $1.4528$ \\ \cline{2-8}
            & $0.2$   &  $0.0254$     & $0.0753$     & $0.5206$     & $1.4303$     & $0.5613$  & $1.4424$ \\ \cline{2-8}
            & $0.25$  &  $0.0401$     & $0.1322$     & $0.4639$     & $1.4154$     & $0.5409$  & $1.4346$ \\ \cline{2-8}
            & $0.3$   &  $0.0585$     & $0.2741$     & $0.3221$     & $1.3966$     & $0.5149$  & $1.4250$ \\ \cline{2-8}
$(1\mp 0.75)\pi/2$ & $0.4$   &  $0.1079$     &  Not found   & Not found    & $1.3465$     &           & $1.4001$ \\ \cline{2-8}
            & $0.5$   &  $0.1780$     &  Not found   & Not found    & $1.2781$     &           & $1.3677$ \\ \cline{2-8}
            & $0.6$   &  $0.2784$     &  Not found   & Not found    & $1.1741$     &           & $1.3268$ \\ \cline{2-8}
            & $0.75$  &  $0.6027$     &  Not found   & Not found    & $0.8483$     &           & $1.2482$ \\ \cline{2-8}
            & $0.76$  &  $0.6910$     &  Not found   & Not found    & $0.7600$     &           & $1.2421$ \\ \cline{2-8}
            & $0.77$  &  Not found    &  Not found   & Not found    & Not found    &           &          \\ \hline
            & $0.0$   &  Not found    &  Not found   &  $1.3006$    & $2.7783$     &  $1.3006$ & $2.7783$         \\ \cline{2-8}
            & $0.5$   &  $0.0665$     &  $0.2019$    &  $1.1006$    & $2.7039$     &  $1.2007$ & $2.7329$ \\ \cline{2-8}
            & $0.6$   &  $0.0971$     &  $0.3266$    &  $0.9756$    & $2.6697$     &  $1.1539$ & $2.7127$ \\ \cline{2-8}
 $(1\mp 0.5)\pi/2$ & $0.7$   &  $0.1344$     &  Not found   & Not found    & $2.6284$     &           & $2.6887$ \\ \cline{2-8}
            & $1.0$   &  $0.2954$     &  Not found   &  Not found   & $2.4523$     &           & $2.5921$ \\ \cline{2-8}
            & $1.5$   &  $0.9178$     &  Not found   &  Not found   & $1.7937$     &           & $2.3386$ \\ \cline{2-8}
            & $1.55$  &  $1.0848$     &  Not found   &  Not found   & $1.6222$     &           & $2.3057$ \\ \cline{2-8}
            & $1.6$   &  Not found    &  Not found   &  Not found   & Not found    &           &          \\ \hline
\end{tabular}
\end{table}
\clearpage
\newpage
\addtocounter{table}{-1}
\begin{table}
\vspace*{-1cm}
\caption{(Continued)}
\vspace*{0.5cm}
\begin{tabular}{|c|c|c|c|c|c|c|c|} \hline
$\theta$   & $k_{z}$ &  $k_{y}^{3l}$ & $k_{y}^{3u}$ & $k_{y}^{2l}$ & $k_{y}^{2u}$ & $q_{y}^{l}$ & $q_{y}^{u}$ \\ \hline
           & $0.0$   &  Not found    &  Not found   &   $2.0795$   &  $4.3196$    &  $2.0795$ & $4.3196$         \\ \cline{2-8}
           & $1.0$   &  $0.1136$     &  $0.3379$    &   $1.7081$   &  $4.1517$    &  $1.8233$ & $4.2023$ \\ \cline{2-8}
           & $1.3$   &  $0.2004$     &  $0.7769$    &   $1.2421$   &  $4.0289$    &  $1.6231$ & $4.1193$ \\ \cline{2-8}
           & $1.33$  &  $0.2108$     &  $0.9239$    &   $1.0932$   &  $4.0140$    &  $1.5986$ & $4.1097$ \\ \cline{2-8}
$(1\mp 0.25)\pi/2$& $1.4$   &  $0.2365$     &  Not found   &   Not found  &  $3.9787$    &           & $4.0864$ \\ \cline{2-8}
           & $1.5$   &  $0.2767$     &  Not found   &   Not found  &  $3.9236$    &           & $4.0508$ \\ \cline{2-8}
           & $2.0$   &  $0.5583$     &  Not found   &   Not found  &  $3.5556$    &           & $3.8287$ \\ \cline{2-8}
           & $2.2$   &  $0.7264$     &  Not found   &   Not found  &  $3.3475$    &           & $3.7174$ \\ \cline{2-8}
           & $2.7$   &  $1.6070$     &  Not found   &   Not found  &  $2.3569$    &           & $3.3718$ \\ \cline{2-8}
           & $2.8$   &  Not found    &  Not found   &   Not found  &  Not found   &           &          \\ \hline
           & $0.0$   &  Not found    &  Not found   &    $2.3988$  &  $5.9121$    &  $2.3988$ & $5.9121$         \\ \cline{2-8}
           & $1.0$   &  $0.0429$     &  $0.1166$    &    $2.1658$  &  $5.8015$    &  $2.1804$ & $5.8269$ \\ \cline{2-8}
           & $1.5$   &  $0.1025$     &  $0.3052$    &    $1.8162$  &  $5.6620$    &  $1.8720$ & $5.7186$ \\ \cline{2-8}
           & $1.7$   &  $0.1356$     &  $0.4432$    &    $1.5889$  &  $5.5887$    &  $1.6924$ & $5.6624$ \\ \cline{2-8}
           & $1.9$   &  $0.1752$     &  $0.7245$    &    $1.2013$  &  $5.5035$    &  $1.4643$ & $5.5985$ \\ \cline{2-8}
$(1\mp 0.1)\pi/2$ & $1.93$  &  $0.1818$     &  $0.8369$    &    $1.0709$  &  $5.4891$    &  $1.4245$ & $5.5882$ \\ \cline{2-8}
           & $1.95$  &  $0.1862$     &  Not found   &    Not found &  $5.4804$    &      & $5.5813$ \\ \cline{2-8}
           & $2.0$   &  $0.1977$     &  Not found   &    Not found &  $5.4565$    &      & $5.5635$ \\ \cline{2-8}
           & $3.0$   &  $0.5673$     &  Not found   &    Not found &  $4.7860$    &           & $5.0944$ \\ \cline{2-8}
           & $3.5$   &  $0.9303$     &  Not found   &    Not found &  $4.2388$    &           & $4.7648$ \\ \cline{2-8}
           & $4.1$   &  $2.0274$     &  Not found   &    Not found &  $2.9020$    &           & $4.2595$ \\ \cline{2-8}
           & $4.2$   &  Not found    &  Not found   &    Not found &  Not found   &           &          \\ \hline
           & $0.0$   &  Not found    &  Not found   &   $2.4677$   &  $\infty$    &  $2.4677$ & $\infty$         \\ \cline{2-8}
           & $0.5$   &  Not found    &  Not found   &   $2.4165$   &  $\infty$    &  $2.4165$ & $\infty$         \\ \cline{2-8}
$\pi/2$    & $1.0$   &  Not found    &  Not found   &   $2.2560$   &  $\infty$    &  $2.2560$ & $\infty$         \\ \cline{2-8}
           & $1.5$   &  Not found    &  Not found   &   $1.9595$   &  $\infty$    &  $1.9595$ & $\infty$         \\ \cline{2-8}
           & $2.0$   &  Not found    &  Not found   &   $1.4455$   &  $\infty$    &  $1.4455$ & $\infty$         \\ \hline
\end{tabular}
\end{table}
\begin{table}
\caption{Viscous fluid and no-slip slot boundaries; $\eta=\chi=0$; $\mu=1$,
$Pr=6.7$, $Le=1$. $k_{y}^{u}(k_{z})$ and $k_{y}^{l}(k_{z})$ are approximate
locations of the limit points of the upper and lower branches,
respectively, of $Ra_{c}(k_{y},k_{z})$ and $\omega_{c}(k_{y},k_{z})$
for different $k_{z}>0$ and such $\theta$ as the limit points are
connected to intervals of the primary-instability boundary
[Fig. \ref{f:ttp00}(c)]. Sign $\approx$ implies that the
accuracy with which the number was obtained is smaller
than that of the other numbers, due to very small
values of $\omega_{c}$ near such a limit point.
For $\theta=(1+0.35)\pi/2$, the limit points
vanish just below $k_{z}=1$.}
\label{t:lul}
\vspace*{0.5cm}
\begin{tabular}{|c|c|c|c|c|c|c|c|} \hline
\multicolumn{2}{|c|}{$k_{z}$} & $0.1$ & $0.2$ & $0.3$ & $0.5$ & $1$ & $2$ \\ \hline
$\theta=(1+0.35)\pi/2$ & $k_{y}^{u}$ & $0.0057$  & $0.0230$  & $0.0531$  & $0.1620$ & Not found & Not found \\ \cline{2-8} 
                       & $k_{y}^{l}$ & $0.7656$  & $0.7768$  & $0.7953$  & $0.8546$ & Not found & Not found \\ \hline
$\theta=(1+0.25)\pi/2$ & $k_{y}^{u}$ & $0.0038$  & $0.0154$  & $0.0353$  & $0.1031$ & $0.4895$ & Not found \\ \cline{2-8} 
                       & $k_{y}^{l}$ & $\approx0.4550$  & $\approx0.4603$  & $0.4703$  & $0.5023$ & $0.6442$ & Not found \\ \hline
$\theta=(1+0.15)\pi/2$ & $k_{y}^{u}$ & $0.0022$  & $0.0089$  & $0.0203$  & $0.0581$ & $0.2454$ & Not found \\ \cline{2-8} 
                       & $k_{y}^{l}$ & $\approx0.2469$  & $\approx0.2520$  & $0.2572$  & $0.2740$ & $0.3473$ & Not found \\ \hline
\end{tabular}
\end{table}
\begin{table}
\caption{Viscous fluid and no-slip slot boundaries; $\eta=0$, $\chi=1$; $\mu=1$,
$Pr=6.7$, $Le=1$. $k_{y}^{G}(k_{z})$ and $k_{y}^{\theta}(k_{z})$ are approximate
values of wave number $k_{y}$ for different $k_{z}$ at a given $\theta$ where 
relatively abrupt changes in $\partial Ra_{c}/\partial k_{y}$ are distinguishable.
(Such respective changes in $\partial\omega_{c}/\partial k_{y}$, in particular in
Fig. \ref{f:ttp01}, could be most pronounced at slightly shifted values of $k_{y}$.)
$k_{y}^{\theta}(k_{z})$ originate at $k_{z}=0$, whereas $k_{y}^{G}(k_{z})$ arise only
at $k_{z}>0$. For $k_{z}\geq 0$,
$q_{y}^{\theta}(k_{z})\equiv[{k_{y}^{\theta}(0)}^2-k_{z}^{2}]^{1/2}$
are provided for comparison with $k_{y}^{\theta}(k_{z})$. For $\theta=(1\pm0.5)\pi/2$,
$k_{y}^{\theta}(k_{z})$ stands for the values of $k_{y}$ where the respective relatively
abrupt change in $\partial\omega_{c}/\partial k_{y}$ takes place at a given $k_{z}$,
since such a change in $\partial Ra_{c}/\partial k_{y}$ is then little
distinguishable. The two values of $\theta$ are relevant
due to transformation (\ref{eq:tt01}).}
\label{t:jk}
\vspace*{0.5cm}
\begin{tabular}{|c|c|c|c|c|} \hline
$\theta$         & $k_{z}$ &  $k_{y}^{G}$ &  $k_{y}^{\theta}$  & $q_{y}^{\theta}$   \\ \hline
                 & $0.0$   &  Not found   &  $2.3$             & $2.3$         \\ \cline{2-5}
$(1\pm0.75)\pi/2$& $0.5$   &  $0.107$     &  $2.2$             & $2.24$        \\ \cline{2-5}
                 & $1.0$   &  $0.5$       &  $1.7$             & $2.07$        \\ \cline{2-5}
                 & $2.0$   &  Not found   &  Not found         & $1.14$        \\ \hline
                 & $0.0$   &  Not found   &  $4.15$            & $4.15$        \\ \cline{2-5}
                 & $0.5$   &  $0.04$      &  $4.10$            & $4.12$        \\ \cline{2-5}
$(1\pm0.5)\pi/2$ & $1.0$   &  $0.175$     &  $3.90$            & $4.03$        \\ \cline{2-5}
                 & $2.0$   &  $0.8$       &  $3.20$            & $3.64$        \\ \cline{2-5}
                 & $3.0$   &  Not found   &  Not found         & $2.87$        \\ \hline
                 & $0.0$   &  Not found   &  Not found         &               \\ \cline{2-5}
                 & $0.5$   &  $0.0175$    &  Not found         &               \\ \cline{2-5}
                 & $1.0$   &  $0.07$      &  Not found         &               \\ \cline{2-5}
$(1\pm0.25)\pi/2$& $2.0$   &  $0.3$       &  Not found         &               \\ \cline{2-5}
                 & $3.0$   &  $0.7$       &  Not found         &               \\ \cline{2-5}
                 & $4.0$   &  $1.6$       &  Not found         &               \\ \cline{2-5}
                 & $5.0$   &  Not found   &  Not found         &               \\ \hline
\end{tabular}
\end{table}
\begin{table}
\caption{$\theta=\pi$. Viscous fluid; $Le=1$. Approximate values
of the most unstable wave number, $k_{c}$, and of the respective
$Ra_{c}^{s}(k_{c})$ for the linear steady instability at the
values of $Ra$ and $\gamma_{\pm}$ used for Figs. \ref{f:sp00},
\ref{f:sp11}, and \ref{f:sp01}. For $\eta=0$ and $\chi=1$, these
data also apply to $\theta=0$ if transformation (\ref{eq:tt01})
along with $Ra\mapsto Ra^{s}$, $Ra_{c}^{s}\mapsto Ra_{c}$,
and $\gamma_{\pm}\mapsto\gamma_{\mp}$ are allowed for.}
\label{t:km}
\vspace*{0.5cm}
\begin{tabular}{|c|c|c|c|c|c|c|c|c|c|} \hline
\multicolumn{3}{|c|}{$Ra$} & $0$ & $1000$ & $5000$ & $10000$ & $20000$ & $50000$ & $100000$ \\ \hline
                     &    $\gamma_{\pm}=0$ & $k_{c}$    & $0$    & $0$    & $0$   & $0$   & $0$   &
		       $0$     & $0$             \\ \cline{3-10} 
$\eta=\chi=0$        &        & $Ra_{c}^{s}(k_{c})$ & $120$  & $120$  & $120$  & $120$  & $120$  &
		       $120$  & $120$                     \\ \cline{2-10}
[Fig. \ref{f:sp00}(a)] & $\gamma_{\pm}=1$ & $k_{c}$    & $0$    & $0$    & $0$   & $0$   & $0$   &
                       $0$     & $0$              \\ \cline{3-10} 
                     &        & $Ra_{c}^{s}(k_{c})$ & $720$  & $720$  & $720$  & $720$  & $720$  &
		       $720$  & $720$   \\ \hline
                     & $\gamma_{-}=0$,    & $k_{c}$    & $0$    & $0$    & $0$   & $0$   & $0$   &
		       $0$     & $0$              \\ \cline{3-10} 
$\eta=\chi=0$        & $\gamma_{+}=1$    & $Ra_{c}^{s}(k_{c})$ & $320$  & $320$  & $320$  & $320$ &
		       $320$  & $320$   & $320$   \\ \cline{2-10}
[Fig. \ref{f:sp00}(b)] & $\gamma_{-}=1$,  & $k_{c}$    & $0$    & $0$    & $0$   & $0$   & $0$   &
		       $0$     & $0$              \\ \cline{3-10} 
                     & $\gamma_{+}=0$    & $Ra_{c}^{s}(k_{c})$ & $320$  & $320$  & $320$  & $320$ &
		       $320$  & $320$   & $320$   \\ \hline
                     & $\gamma_{\pm}=0$  & $k_{c}$    & $1.76$ & $1.36$ & $0.97$ & $0.85$ & $0.77$&
		       $0.79$ & $0.85$            \\ \cline{3-10}
$\eta=\chi=1$        &                   & $Ra_{c}^{s}(k_{c})$ & $385$  & $882$  & $2539$ & $4392$ &
		       $7825$ & $17007$  & $30186$ \\ \cline{2-10}
[Fig. \ref{f:sp11}(a)] & $\gamma_{\pm}=1$& $k_{c}$    & $2.55$ & $2.21$ & $1.54$ & $1.28$ & $1.09$&
                       $0.91$ & $0.85$            \\ \cline{3-10}
                     &                   & $Ra_{c}^{s}(k_{c})$ & $1296$ & $1980$ & $4284$ & $6744$ &
		       $11167$ & $22973$ & $40590$ \\ \hline
                     & $\gamma_{-}=0$,    & $k_{c}$    & $2.09$ & $1.71$ & $1.22$ & $1.04$ & $0.91$&
		       $0.82$ & $0.83$            \\ \cline{3-10}
$\eta=\chi=1$        & $\gamma_{+}=1$    & $Ra_{c}^{s}(k_{c})$ & $669$  & $1183$ & $2870$ & $4704$ &
		       $8048$ & $17024$ & $30241$  \\ \cline{2-10}
[Fig. \ref{f:sp11}(b)] & $\gamma_{-}=1$,  & $k_{c}$    & $2.21$ & $1.84$ & $1.27$ & $1.07$ & $0.93$ &
		       $0.83$ & $0.84$            \\ \cline{3-10}
                     & $\gamma_{+}=0$    & $Ra_{c}^{s}(k_{c})$ & $817$  & $1482$ & $3717$ & $6159$ &
		       $10630$ & $22706$ & $40574$ \\ \hline
                     & $\gamma_{\pm}=0$  & $k_{c}$    & $1.76$ & $1.75$ & $1.69$ & $1.59$ & $1.44$ &
		       $1.22$ & $1.07$            \\ \cline{3-10}
$\eta=0$, $\chi=1$ &                  & $Ra_{c}^{s}(k_{c})$ & $385$  & $1289$ & $4003$ & $6437$ &
		       $10246$ & $19327$ & $32200$  \\ \cline{2-10}
[Fig. \ref{f:sp01}(a)] & $\gamma_{\pm}=1$& $k_{c}$    & $2.55$ & $2.51$ & $2.20$ & $1.96$ & $1.76$ &
		       $1.51$ & $1.32$             \\ \cline{3-10}
                     &                   & $Ra_{c}^{s}(k_{c})$ & $1296$ & $2255$ & $5745$ & $9445$ &
		       $15500$ & $29334$ & $47784$  \\ \hline
                     & $\gamma_{-}=0$,    & $k_{c}$    & $2.09$ & $1.94$ & $1.68$ & $1.59$ & $1.49$ &
		       $1.31$ & $1.16$             \\ \cline{3-10}
$\eta=0$, $\chi=1$ & $\gamma_{+}=1$   & $Ra_{c}^{s}(k_{c})$ & $669$  & $1431$ & $4008$ & $6584$ &
		       $10718$ & $20322$ & $33526$  \\ \cline{2-10}
[Fig. \ref{f:sp01}(b)] & $\gamma_{-}=1$,  & $k_{c}$    & $2.21$ & $2.35$ & $2.20$ & $1.96$ & $1.72$ &
			$1.43$ & $1.23$             \\ \cline{3-10}
                     & $\gamma_{+}=0$    & $Ra_{c}^{s}(k_{c})$ & $817$  & $1949$ & $5733$ & $9372$ &
                       $15004$ & $27897$ & $45606$  \\ \hline
\end{tabular}
\end{table}

\begin{table}
\caption{$\theta=0$. Viscous fluid; $\lambda=2$, $Pr=6.7$, $Le=1$.
Obtained from the numerical steady solutions of Eqs. (\ref{eq:ns1})---(\ref{eq:dxi}),
approximate values of $Ra$ characterizing limit point $L$ and symmetry-breaking
bifurcation point $B$ (Fig. \ref{f:bd}) for $\eta=\chi=1$ at $\gamma_{\pm}=1$
and such singularities for some $\eta=\chi<1$ at $\gamma_{\pm}=1$ and
$\gamma_{\pm}=0$ in the range of $\mu\leq 1$.}
\label{t:hys}
\vspace*{0.5cm}
\begin{tabular}{|c|c|c|c|c|c|c|c|} \hline
\multicolumn{2}{|c|}{$\mu$} & $0.3$ & $0.4$ & $0.5$ & $0.6$ & $0.7$ & $1$ \\ \hline
$\eta=\chi=1$ & $L$ & $2726$& $3261$& $3872$& $4581$& $5417$& $9071$      \\ \cline{2-8} 
$\gamma_{\pm}=1$& $B$ & $2733$& $3419$& $4547$& $6694$& $11834$& $\infty$   \\ \hline
$\eta=\chi\approx 0.64$&$L$ & Not found & Not found & $3682$& $4855$& $6563$& $25107$ \\ \cline{2-8}
$\gamma_{\pm}=1$ & $B$ & $2299$ & $2844$& $3716$& $5300$& $8800$& $\infty$  \\ \hline
$\eta=\chi\approx 0.41$&$L$ & Not found & Not found & Not found & $4983$& $7713$& $269057$ \\ \cline{2-8}
$\gamma_{\pm}=1$ & $B$ & $2194$ & $2708$& $3524$& $4992$& $8184$ & $\infty$ \\ \hline
$\eta=\chi\approx 0.41$&$L$ & Not found & Not found & $1616\div1617$ & $2274$& $3343$& $96701$ \\ \cline{2-8}
$\gamma_{\pm}=0$ & $B$ & $980$ & $1223$& $1616\div1617$& $2342$& $3938$ & $\infty$ \\ \hline
\end{tabular}
\end{table}

\end{document}